\documentclass[prd,preprint,onecolumn,superscriptaddress,nofootinbib]{revtex4-2}

\selectfont

\usepackage{placeins}
\usepackage{mathtools}
\usepackage{amssymb}
\usepackage{amsfonts}
\usepackage{fdsymbol}
\usepackage{soul}
\usepackage{color,xcolor}
\usepackage{braket}
\usepackage{hyperref}
\usepackage{url}
\usepackage{multirow}
\usepackage{relsize}
\usepackage{fullpage}
\usepackage{makecell}
\usepackage{fullpage}
\usepackage[normalem]{ulem}
\usepackage[T1]{fontenc}
\usepackage[utf8]{inputenc}

\usepackage[]{hyperref}
\usepackage[]{changes} 

\usepackage{tabularx}
\usepackage{array}
 
\makeatother
\hypersetup{
 unicode=false,          
 pdftoolbar=true,        
 pdfmenubar=true,        
 pdffitwindow=true,     
 pdfstartview={FitH},    
 pdfsubject={Neutrino mixing-schemes},   
 pdfnewwindow=true,      
 pdfcreator={RevTeX},
 colorlinks=true,       
 linkcolor=blue,
 citecolor=blue,        
 filecolor=black,      
 urlcolor=blue,           
}

\begin{document}

\title{
    Correlations in  oscillation data and their impact \\ on neutrino mixing models
}
\author{Janusz Gluza}
\affiliation{Institute of Physics, University of Silesia, Katowice, Poland}
\author{Biswajit Karmakar}
\affiliation{Institute of Physics, University of Silesia, Katowice, Poland}
\author{Szymon Zięba}
\affiliation{Institute of Physics, University of Silesia, Katowice, Poland}  

\date{\today}


\begin{abstract}
With increasing precision of neutrino experiments, complete correlation among the oscillation parameters will scrutinize theoretical neutrino mass and mixing models.  
Using available 1D--3D global-fit datasets, we construct a multidimensional $\tilde{\chi}^2$ function that provides a conservative estimate of the full six-dimensional parameter correlations. In our approach, we delineate the model-constrained parameter space by combining the interpolated experimental data with the analytical relations implied by a given mixing scheme. We identify $\tilde{\chi}^2_{\text{min}}$, corresponding to the optimal agreement attainable for parameter points satisfying the model. Applying this framework to several schemes  ($\text{TM}_{1,2}$, $\mu1,2$--$\tau1,2$) yields generally better fits for IO than for NO, with the $\mu1$--$\tau1$ scheme under IO providing the best overall fit. While  $\mu1,2$--$\tau1,2$ and $\text{TM}_1$ schemes perform significantly better than $\text{TM}_2$, the latter is excluded at $>3\sigma$ C.L. using NuFIT~v6.1 data. We further evaluate the implications of these correlation-respecting constraints on the effective neutrino mass observables relevant to tritium $\beta$-decay ($m_{\beta}$) and neutrinoless double $\beta$-decay ($m_{\beta\beta}$). The proposed approach can be readily extended to test any theoretical framework that predicts algebraic relations among neutrino oscillation parameters, and we provide a GitHub starter kit.
\end{abstract}
\maketitle  
\par\noindent\ignorespaces
\section{Introduction}
\label{section:intro}
{Neutrino oscillation experiments have firmly established the existence of non-zero neutrino masses and flavor mixing, while achieving remarkable precision in the determination of the associated oscillation parameters. These achievements have ushered neutrino physics into an era of precision measurements, driven by the wealth of high-quality data collected by experiments such as Super-Kamiokande~\cite{Super-Kamiokande:1998kpq}, SNO~\cite{SNO:1999crp}, T2K~\cite{T2K:2019bcf}, Daya Bay~\cite{DayaBay:2012fng}, RENO~\cite{RENO:2012mkc}, and Double Chooz~\cite{DoubleChooz:2011ymz}.

Neutrino oscillation data implies} that three types of known flavor neutrinos $|\nu_\alpha \rangle$, ($\alpha =e,\mu,\tau$) mix, in a minimal setup, with three massive states $|\nu_i\rangle$, $(i=1,2,3$), i.e. $|\nu_\alpha \rangle = {\rm U}_{\rm PMNS} |\nu_i\rangle$.
The standard parametrization of the unitary Pontecorvo–Maki–Nakagawa–Sakata (PMNS) mixing matrix is ~\cite{Pontecorvo:1957qd,Maki:1962mu, Kobayashi:1973fv}
\begin{eqnarray}
{\rm U}_{\rm PMNS} =
\begin{pmatrix}
    c_{12}c_{13} & s_{12}c_{13} & s_{13} e^{-i\delta_{\rm CP}} \\
    -s_{12}c_{23} -c_{12} s_{13} s_{23} e^{i\delta_{\rm CP}} & 
    c_{12} c_{s3} -s_{12} s_{13} s_{23} e^{i\delta_{\rm CP}}  & 
    c_{13} s_{23} \\
    s_{12}s_{23} - c_{12} s_{13} c_{23} e^{i\delta_{\rm CP}} & 
    -c_{12}s_{23} -s_{12} s_{13} c_{23} e^{i\delta_{\rm CP}}& 
    c_{13} c_{23}
\end{pmatrix}
{\rm U}_{\rm M}
\label{upmnsmatrix}
\end{eqnarray}
where $s(c)_{12,13,23} \equiv \sin(\cos)\theta_{12,13,23}$ are three mixing angles and {$\delta_{\rm CP}$} is the Dirac CP phase. The matrix 
${\rm U}_{\rm M} \equiv \rm{diag} (e^{i \alpha_1}, e^{i \alpha_2},1)$ is associated with additional CP phases  $\alpha_{1,2}$ in case neutrinos are self-conjugate Majorana particles (oscillation experiments are not sensitive to $\alpha_{1,2}$).
 \color{black}
\newpage
 In this framework, neutrino  oscillations are described by six independent parameters: the three mixing angles $\theta_{12}$, $\theta_{13}$, and $\theta_{23}$, the Dirac CP-violating phase
$\delta_{\rm CP}$, and two independent mass-squared differences,
$\Delta m_{21}^2$ and $\Delta m_{31}^2$ ($\Delta m_{32}^2$) for normal ordering (NO)
(inverted ordering (IO)). Over the past few decades, a wide range of neutrino oscillation experiments using solar, atmospheric, reactor, and accelerator neutrinos have been dedicated to measuring these parameters with increasing precision. The complementarity in neutrino oscillation data implies that different classes of neutrino oscillation experiments, classified by their source and baseline, are sensitive to distinct oscillation channels and parameter sets. For instance, solar neutrino experiments (such as SNO~\cite{SNO:1999crp} and Super-Kamiokande~\cite{Super-Kamiokande:1998kpq}) and medium-baseline reactor experiments (such as KamLAND~\cite{KamLAND:2002uet}) are uniquely positioned to pin down the "solar" parameters, $\theta_{12}$ and $\Delta m^2_{21}$. Conversely, short-baseline reactor experiments (such as Daya Bay~\cite{DayaBay:2012fng}, RENO~\cite{RENO:2012mkc}, and Double Chooz~\cite{DoubleChooz:2011ymz}) provide clean measurements of the "reactor" mixing angle $\theta_{13}$ by observing electron antineutrino disappearance. Meanwhile, atmospheric and long-baseline accelerator experiments (such as T2K~\cite{T2K:2019bcf}, NO$\nu$A~\cite{NOvA:2019cyt}, and the upcoming DUNE~\cite{DUNE:2020jqi} and Hyper-Kamiokande~\cite{Hyper-Kamiokande:2018ofw}) probe the "atmospheric" parameters ($\theta_{23}$ and $|\Delta m^2_{3\ell}|$) and search for CP violation through the Dirac phase $\delta_{\rm CP}$ via $\nu_\mu \to \nu_e$ appearance. Combining these different experimental datasets into unified global fits is therefore essential to reconstruct the complete multidimensional correlation matrix across all six parameters. In this work,  we use the NuFIT global fit data\footnote{For other analysis and global fits, see Capozzi et al~\cite{Capozzi:2021fjo} and de Salas et al.  ~\cite{deSalas:2020pgw,10.5281/zenodo.4726908}.}. The NuFIT data \cite{Esteban:2024eli} are summarized at the 3$\sigma$ level (1dof, $\Delta\chi^2 \leq 9$) in Table \ref{table:nufit60}.  

\color{black}
\begin{table}[h!]
 \centering
 \scriptsize
    \begin{tabular}{c c c c c c c}
    \hline 
         \multicolumn{7}{c}{ NuFIT 6.0~~~~~~~~~~~~~~~~~~~~~~ $3\sigma$ ranges~~~($\Delta\chi^2 \leq 9$)}
        \\
         \hline 
        Ordering
        &$\theta_{13}/^{\circ}$
        &
        $\theta_{12}/^{\circ}$
        &
        $\theta_{23}/^{\circ}$
        &
        $\delta_{\rm CP}/^{\circ}$
        &
        $\Delta m^2_{21}/{10^{-5} \rm{eV}^2}$
        &
         $\Delta m^2_{3\ell}/{10^{-3} \rm{eV}^2}$
         \\
         \hline
        NO
        &$8.19 \xrightarrow{} 8.89$
        &$31.63 \xrightarrow{} 35.95$
        &$41.3 \xrightarrow{} 49.9$
        &$124 \xrightarrow{} 364$
        &$6.92 \xrightarrow{} 8.05$
        &$2.451 \xrightarrow{} 2.578$
        \\
        IO
        &$8.25 \xrightarrow{} 8.93$
        &$31.63 \xrightarrow{} 35.95$
        &$41.5 \xrightarrow{} 49.8$
        &$201 \xrightarrow{} 335$
        &$6.92 \xrightarrow{} 8.05$
        &$-2.547 \xrightarrow{} -2.421$
         \\
        \hline \hline
    \end{tabular}
    \caption{
    NuFIT 6.0 with SK/IC atmospheric data  \cite{Esteban:2024eli,NuFIT6.0data} oscillation parameter $3\sigma$ ranges, $\Delta m^2_{3\ell} = \Delta m^2_{31(32)}$ for NO (IO).
    }
\label{table:nufit60}
\end{table}

Recently, the first physics results from the Jiangmen Underground Neutrino Observatory (JUNO)~\cite{JUNO:2025gmd} have further advanced the precision neutrino program by significantly improving the determination of the solar oscillation parameters $\theta_{12}$ and $\Delta m^2_{21}$. Remarkably, using only 59 days of data, JUNO has already achieved uncertainties substantially smaller than those obtained in previous global analyses, yielding updated estimates
  
\begin{equation}
\sin^2\theta_{12} \simeq 0.3092 \pm 0.0087,
\qquad
\Delta m^2_{21} \simeq (7.5 \pm 0.12)\times10^{-5}\ {\rm eV}^2,
\label{eq:juno}
\end{equation}
corresponding to precisions of about $2.4\%$ and $1.3\%$, respectively. Subsequently, NuFIT also included this result ~\cite{NuFIT6.0data} in the global analysis as summarized in Table \ref{table:nufit61}. 
\begin{table}[h!]
 \centering
 \scriptsize
    \begin{tabular}{c c c c c c c}
    \hline 
         \multicolumn{7}{c}{ NuFIT 6.1~~~~~~~~~~~~~~~~~~~~~~ $3\sigma$ ranges~~~($\Delta\chi^2 \leq 9$)}
        \\
         \hline 
        Ordering
        &$\theta_{13}/^{\circ}$
        &
        $\theta_{12}/^{\circ}$
        &
        $\theta_{23}/^{\circ}$
        &
        $\delta_{\rm CP}/^{\circ}$
        &
        $\Delta m^2_{21}/{10^{-5} \rm{eV}^2}$
        &
         $\Delta m^2_{3\ell}/{10^{-3} \rm{eV}^2}$
         \\
         \hline
        NO
        &$8.26 \xrightarrow{} 8.95$
        &$32.54 \xrightarrow{} 35.03$
        &$41.27 \xrightarrow{} 49.86$
        &$125 \xrightarrow{} 365$
        &$7.236 \xrightarrow{} 7.823$
        &$2.450 \xrightarrow{} 2.576$
        \\
        IO
        &$8.32 \xrightarrow{} 8.99$
        &$32.54 \xrightarrow{} 35.03$
        &$41.51 \xrightarrow{} 49.83$
        &$203 \xrightarrow{} 335$
        &$7.236 \xrightarrow{} 7.822$
        &$-2.547 \xrightarrow{} -2.421$
         \\
        \hline \hline
    \end{tabular}
    \caption{
    NuFIT 6.1 with SK/IC atmospheric data and JUNO 59.1 days dataset \cite{Esteban:2024eli,NuFIT6.1data} oscillation parameter $3\sigma$ ranges, $\Delta m^2_{3\ell} = \Delta m^2_{31(32)}$ for NO (IO).
    }
\label{table:nufit61}
\end{table}
As expected, with Eq.~(\ref{eq:juno}), the main difference between the two fits is connected with $\sin^2\theta_{12}$ and $\Delta m^2_{21}$.
With experimental improvements, further increase in precision is expected in the coming years \cite{Song:2020nfh}.
 
From a theoretical perspective, understanding the origin of the observed pattern of neutrino mixing remains one of the central challenges in particle physics, and substantial effort has been devoted over the past few decades to address this question\footnote{For an alternative description of neutrino mixing, see the recent work Ref.~\cite{Duda:2025vmc} and further references therein.}. Symmetry-based approaches have attracted particular attention owing to their strong predictive power. Among them, models based on non-Abelian discrete flavor symmetries, including modular symmetries, constitute some of the most compelling frameworks for explaining the structure of the lepton mixing matrix and the origin of CP violation (see, e.g., Refs.~\cite{Altarelli:2010gt,King:2013eh,Petcov:2017ggy,Chauhan:2023faf}). Following the precise measurement of the nonzero reactor mixing angle, $\theta_{13}$~\cite{An:2012eh,Ahn:2012nd}, scenarios such as trimaximal mixing ($\mathrm{TM}_1$ and $\mathrm{TM}_2$)~\cite{Albright:2010ap} and (partial) $\mu$--$\tau$ reflection symmetry~\cite{Harrison:2002et} have emerged as viable and predictive frameworks for describing the observed lepton mixing pattern.

These theoretical models are highly predictive, often establishing algebraic relations between otherwise independent oscillation parameters (e.g., relating the CP-phase $\delta_{\rm CP}$ to the mixing angles). Simultaneously, the experimental global fits~\cite{NuFIT6.1data}, while often presented as 1D or 2D parameter projections, contain rich, higher-dimensional correlation information.  
Because these distinct experimental methodologies yield highly specialized, decoupled parameter constraints, interpreting their synergistic bounds requires the sophisticated framework of global fits to properly reconstruct the underlying multi-dimensional correlation matrix.

Most theoretical studies of neutrino mixing schemes have traditionally assessed model predictions by comparing them with the allowed intervals of individual oscillation parameters, without fully accounting for the correlations among these parameters encoded in global fits to experimental data~\cite{Karmakar:2014dva,Borah:2017dmk}. Such an approach can lead to overly optimistic conclusions about model viability, since individually allowed parameter values may correspond to combinations that are strongly disfavored once the underlying correlations are taken into account.

In this work, we demonstrate how correlations among neutrino oscillation parameters can be systematically incorporated to explore and rigorously constrain realistic mixing schemes.
Unlike approaches that combine individual $\chi^2$ contributions~\cite{Petcov:2018snn, Zhang:2025jnn} from the solar, atmospheric, and reactor sectors under an assumption of parameter independence, our method preserves the genuine multi-dimensional shape of the experimental likelihood as reported by global-fit collaborations.  As mentioned earlier, correlations among selected sets of oscillation parameters, extracted from experimental data, are provided by 
NuFIT~\cite{Esteban:2024eli,NuFIT6.0data} and by de Salas et al.~\cite{deSalas:2020pgw,10.5281/zenodo.4726908} in the form of $\Delta\chi^2$ tables. However, complete correlation information for all oscillation parameters which can be utilized by the community is generally not available publicly. This may have severe consequences in neutrino model-building, as well as unified astroparticle frameworks~\cite{Chauhan:2023faf, Borah:2024gql, Chen:2026fod, Karmakar:2014dva, Karmakar:2015jza, Datta:2021zzf}, which try to explain neutrino mixing as well as cosmic phenomena such as matter-antimatter asymmetry, dark matter, deviations in the effective number of neutrino species or gravitational waves. To address this limitation, we develop a methodology in the following section to reconstruct the full set of up to six-dimensional (6D) correlations from available two- and three-dimensional (2D and 3D) correlation data. To illustrate the importance of the full correlations in neutrino oscillation data, we have employed our methodology on popular neutrino mixing schemes as an example to demonstrate its effects in neutrino model building, such as trimaximal mixing ($\mathrm{TM}_1$ and $\mathrm{TM}_2$) and (partial) $\mu$--$\tau$ reflection symmetry. Furthermore, as flavor-symmetric  models can easily demonstrate such mixing schemes and   typically formulate algebraic relations\footnote{{Specifically, such symmetry based models often yield `sum-rules' among neutrino mixing angles and CP phases~\cite{King:2007pr,Antusch:2007rk, Ballett:2014dua, Petcov:2014laa} as well as among the masses~\cite{Gehrlein:2016wlc, Gehrlein:2017ryu, Barry:2010yk, King:2013psa, Chauhan:2023faf}.}} connecting two or more oscillations
parameters, testing these relations against the actual correlated likelihood, rather
than against independently quoted 1D or 2D intervals,  is essential for a meaningful
assessment of their viability. 

Beyond testing the mixing schemes against correlated neutrino oscillation data, we translate the
resulting constraints into predictions for the effective Majorana mass $m_{\beta\beta}$ probed by
neutrinoless double beta decay and the effective electron-neutrino mass $m_{\beta}$ probed by
direct kinematic measurements, connecting oscillation-sector correlations to observables of direct relevance for ongoing and planned beta-decay and double-beta-
decay programs \cite{KATRIN:2001ttj, KATRIN:2021dfa,Project8:2013trt,nEXO:2017nam,nEXO:2021ujk,LEGEND:2017cdu,LEGEND:2021bnm,CUPID:2019imh}. While we focus on $\mathrm{TM}_1$, $\mathrm{TM}_2$, partial $\mu$--$\tau$ reflection symmetries  as illustrative examples, the method is
not tied to these particular schemes: any model predicting algebraic relations among a subset of the six oscillation parameters can be tested using the same $\chi^2$ construction. To facilitate such applications, we make the underlying Mathematica notebook and
interpolated correlation functions publicly available \cite{GitHubCorrelations}.   

This paper is organized as follows. In Section~\ref{chapt:method}, we describe the methodology developed in this work, which is based on the NuFIT global fit, and explain how the full multidimensional correlation structure is reconstructed from the available 2D and 3D data. In Section~\ref{section:method}, we first review several neutrino mixing schemes and confront them with the correlated neutrino oscillation data. We then present the resulting 2D and 1D projections of the correlated $\chi^2$ function and discuss their implications for $m_{\beta\beta}$ and $m_\beta$. In Section~\ref{section:nufit61}, we illustrate the impact of the first JUNO results, incorporated through NuFIT~6.1, on the present analysis. Finally, in Section~\ref{section:conc}, we summarize our main findings, discuss future prospects, and present our conclusions.
  
\section{Reconstructing Correlated Neutrino Oscillation Data\label{chapt:method}} 

In practice, neutrino oscillation experiments never measure the six parameters $\theta_{12}$, $ \theta_{13}$, $ \theta_{23}$, $ \delta_{\rm CP}$, $ \Delta m^2_{12}$ and $|\Delta m^2_{3\ell}|$ independently. Every observable is a function of an oscillation probability $P(\nu_{\alpha}\rightarrow \nu_{\beta})$ that, in general, depends on several of these parameters simultaneously through interference terms in the appearance and disappearance channels. As a result, the parameters extracted from any single class of experiments are already correlated with one another within a single data set, and these correlations become further entangled when independent data sets are combined in a global fit. This distinction is critical, and one must not confuse the `six independent parameters' of the theoretical framework with `six independently measured parameters' in a statistical context. While the standard three-flavor Lagrangian allows these parameters to vary freely, experimental measurements inherently tie them together.  To illustrate this, let us consider long-baseline \(\nu_\mu \to \nu_e\) measurements at ongoing T2K, NO\(\nu \)A and future DUNE, Hyper-Kamiokande, which also points towards crucial parameter degeneracies~\cite{Barger:2001yr, Agarwalla:2013ju, Sugama:2023sjf, T2K:2025wet}. At leading order, the appearance probability depends on \(\theta _{13}\), \(\theta _{23}\), \(\delta _{\mathrm{CP}}\), and the mass ordering \(|\Delta m^2_{3\ell}|\) entangled within comparable oscillatory phases and interference terms. This gives rise to the classic octant--\(\delta _{\mathrm{CP}}\) and mass-ordering--\(\delta _{\mathrm{CP}}\) degeneracies, where distinct parameter combinations yield identical event rates. Global fits cannot magically erase these degeneracies; they can only weight them using complementary data from reactor, solar, and atmospheric experiments. Consequently, the allowed six-dimensional parameter space is not a simple product of independent intervals, but a complex, curved structure featuring non-elliptical, multi-lobed projections. 
This observation has an immediate and important consequence for any theoretical mixing scheme that imposes an algebraic relation between two or more oscillation parameters and thereby selects a lower-dimensional hypersurface within the full six-dimensional parameter space. Whether that hypersurface passes through a region favored or disfavored by the data can only be judged correctly if the correlation structure of the data is known along that hypersurface. Testing the scheme's predictions against six independent 1$\sigma$ or 3$\sigma$ intervals, one parameter at a time, implicitly assumes a rectangular, uncorrelated allowed region, and can therefore both wrongly accept parameter combinations that lie outside the true correlated region and wrongly reject combinations that lie inside it. Hence, we aim to construct a practical, data-driven estimate of the correlated region using only the correlation information that global fits make publicly available. 

To enable a comprehensive assessment of how experimental constraints impact theoretical predictions, we develop a methodology to reconstruct the full correlation structure of neutrino oscillation parameters, extending up to six-dimensional (6D) correlations. In our analysis, we rely on the NuFIT data sets~\cite{NuFIT6.0data, NuFIT6.1data}, which provide the complete set of two-dimensional (2D) correlations for pairs of neutrino oscillation parameters, as well as a single three-dimensional (3D) correlation set for each mass ordering. These are, to our knowledge, the most recent complete correlation products publicly released by any global-fit collaboration to date, and they form the entire experimental input to the construction below. 
The tabulated NuFIT data are interpolated using third-degree polynomial functions implemented in Wolfram Mathematica~\cite{Mathematica14}. No extrapolation beyond the provided parameter ranges is performed to avoid introducing artificial distortions in the data. In the following, we refer to the interpolated correlation data as $\chi^2(X_1,\ldots,X_i)$ functions, with $i=2,3$, corresponding to the available 2D and 3D correlated datasets.

 \begin{figure}[!h]
    \centering
      \includegraphics[width=0.49\textwidth]{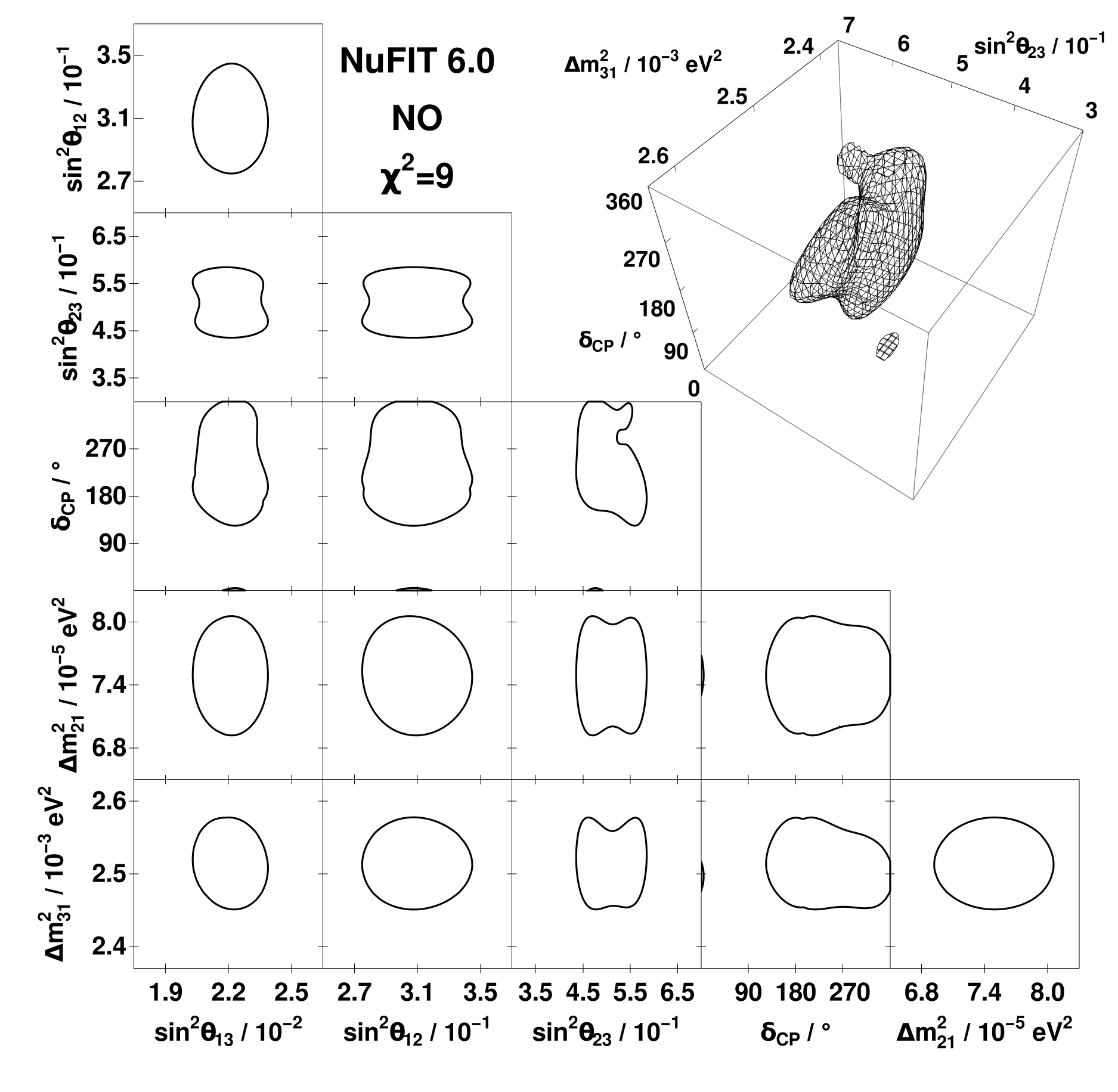}
      \includegraphics[width=0.49\textwidth]{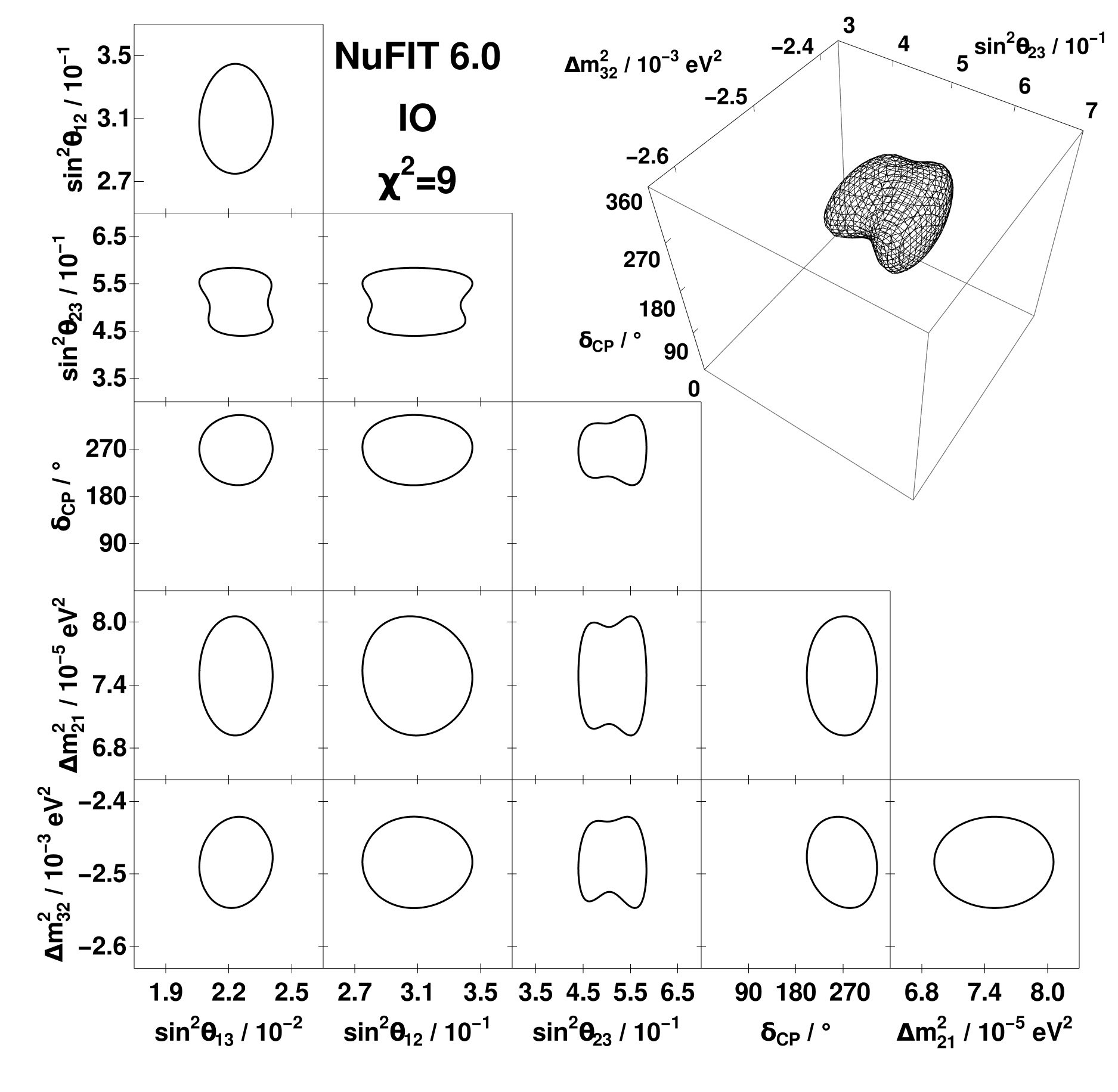}
     \caption{  
      Examples of two- and three-dimensional correlation contours for neutrino oscillation parameters at $\chi^2=9$, derived from the NuFIT 6.0 data sets (IC24 including SK atmospheric data)~\cite{NuFIT6.0data} for NO in the left panel and for IO in the right panel.
     }
    \label{figure:1}
\end{figure}

Fig.~\ref{figure:1} displays, for both the normal (left panel set) and inverted (right panel set) mass orderings, the full collection of two-dimensional NuFIT 6.0 correlation contours at $\chi^2=9$, arranged as a lower-triangular matrix of panels spanning all six oscillation parameters  $\sin^2\theta_{12}, \sin^2\theta_{13}, \sin^2\theta_{23}, \delta_{\rm CP}, \Delta m^2_{12}$ and $|\Delta m^2_{3\ell}|$, together with one representative three-dimensional $\chi^2=9$ surface, in the inset within each panel set. Each row and column of the matrix corresponds to one oscillation parameter, and each off-diagonal panel shows the two-dimensional boundary obtained after profiling over the remaining four parameters; the diagonal is omitted, since the corresponding one-dimensional information is already summarized in Table~\ref{table:nufit60}. 
The publicly available three-dimensional data set covers only one specific 3D ($\theta_{23}, |\Delta m^2_{3\ell}|, \delta_{\rm CP}$) of parameter set per mass ordering, rather than all possible correlations of six parameters; this is precisely the limitation that motivates the extension and reduction formulas introduced next, which combine the two- and three-dimensional information visible in Fig. 1 into estimates for parameter combinations of arbitrary dimensionality, up to the full six-dimensional case needed in Section \ref{section:method}.
 
To generalize the analysis beyond the three-dimensional correlation case, we define a $\chi^2$   for an arbitrary number of variables as the maximum $\chi^2$ value over all reduced subsystems obtained by omitting a single variable

\begin{equation}
\chi^2_{\rm ext}(X_1,...,X_n) \equiv   
\max\left\{\chi^2(X_1,...,X_{k-1},X_{k+1},...,X_n), ~~~ k \in \{1...n\}\right\}.
\label{eq:ext2}
\end{equation}

Since the $\chi^2$ function must be known for every subset appearing on the right-hand side of Eq.~\eqref{eq:ext2}, the procedure is applied recursively whenever necessary, terminating once the subsets coincide with the available 2D or 3D correlated likelihood contours.  The recursive construction is illustrated in Fig.~\ref{figure:1} and compensates for the fact that the full $n$-dimensional correlated likelihood is not directly available. Using Eq.~\eqref{eq:ext2}, one obtains, for example, the following expression in terms of the neutrino oscillation parameters
\begin{align}
\chi^2_{\rm ext}(\theta_{23},\theta_{13},\theta_{12},\delta_{\rm CP})
&\equiv
\max\big\{
    \chi^2(\theta_{23},\delta_{\rm CP}),
    \chi^2(\theta_{12},\delta_{\rm CP}),
    \chi^2(\theta_{13},\delta_{\rm CP}), \nonumber\\
&\qquad\qquad
    \chi^2(\theta_{13},\theta_{12}),
    \chi^2(\theta_{13},\theta_{23}),
    \chi^2(\theta_{12},\theta_{23})
\big\}\label{eq:ext1},
\\[1ex]
\chi^2_{\rm ext}(\theta_{23},\theta_{13},\Delta m_{31}^2,\delta_{\rm CP})
&\equiv
\max\big\{
    \chi^2(\theta_{23},\Delta m_{31}^2,\delta_{\rm CP}),
    \chi^2(\theta_{13},\Delta m_{31}^2), \nonumber\\
&\qquad\qquad
    \chi^2(\theta_{13},\delta_{\rm CP}),
    \chi^2(\theta_{13},\theta_{23})\label{eq:ext22}
\big\}.
\end{align}
{The representative example presented in Eq. (\ref{eq:ext1}), is directly relevant for testing neutrino mixing schemes where there exist algebraic relations among oscillation parameters, $\theta_{23},\theta_{13},\theta_{12}$ and $\delta_{\rm CP}$. The six terms on the right-hand side of Eq. (\ref{eq:ext1}) exhaust every two-dimensional correlation directly available from NuFIT among $\theta_{23},\theta_{13},\theta_{12},\delta_{\rm CP}$;  capturing both the leading-order parameter degeneracies and the subtle, residual correlations encoded within the global fit dataset. On the other hand, Eq. (\ref{eq:ext22}), illustrates a complementary case which includes a mass-squared difference appears in the reconstructed set. Such combinations are crucial to the mass-ordering dependence of $m_{\beta\beta}$ and $m_{\beta}$ in section \ref{section:mbb}. Such construction needs to be extended further for all parameters, as no six-parameter correlation of this kind is available directly from global analysis.} It has been checked for all  available 2D and 3D correlated datasets from NuFIT, that the extension formula Eq.\eqref{eq:ext2} holds, for both
mass orderings.  

To reduce the number of variables, we introduce the following projections
\begin{align}
\label{eq:red2}
 \chi^2_{\rm red} (X_1,...,X_{k-1},X_{k+1},...,X_n) &\equiv \min \left\{ \chi^2(X_1,...,X_{k}=x_k,...,X_n)  \text{ for any } x_k \right\}. 
\end{align}
{This helps us in profiling higher-dimensional $\chi^2$ down to a lower-dimensional one by minimizing over one of its parameters.}
{For instance,
\begin{equation}
\chi^2_{\rm red}(\theta_{23}, \theta_{13}, \delta_{\rm CP}) \equiv
   \min \left\{  \chi^2(\theta_{23}, \theta_{13}, \theta_{12}, \delta_{\rm CP}) \text{ for any } \theta_{12} \right\}, 
\end{equation}
which illustrates the four-parameter function $\chi^2(\theta_{23},\theta_{13},\theta_{12},\delta_{\rm CP})$ profiled over $\theta_{12}$ to give the reduced three-parameter function.}  
This procedure can be applied recursively, yielding, if needed, to the reduced one-dimensional estimates of $\chi^2$ profiles within a given mixing-scheme model. {Just like} in the extended case {in Eq. (\ref{eq:ext2})}, we have verified for all available correlated 2D and 3D NuFIT datasets that the reduction formula, Eq.~\eqref{eq:red2}, holds for both neutrino mass orderings. 
 
Together, Eqs. (\ref{eq:ext2}) and (\ref{eq:red2}) form a closed pair,  combining and projecting. That allows us to move between any subset of the six oscillation parameters using only the 2D and 3D data provided by global analyses. In the subsequent sections, we will use the notation $\tilde{\chi}^2$ whenever the extended or reduced $\chi^2$ formulas, $\chi^2_{\rm ext}$ and $\chi^2_{\rm red}$, are employed.
\\

\noindent 
{Regarding the construction presented here,} the following remarks are in order:

\begin{enumerate}

\item
The extension formula is constructed by taking the maximum of the $\chi^2$ values in Eq.~\eqref{eq:ext2}. For model testing, such underestimation is acceptable because it avoids excluding viable models.
An alternative approximation, based on summing $\chi^2$  for solar, atmospheric and
reactor data (see, e.g., Ref.~\cite{Petcov:2018snn}), may overestimate the  $\chi^2$, potentially leading to the unjustified rejection of otherwise acceptable models.
For instance, the considered later TM$_{1(2)}$ and partial $\mu$--$\tau$ symmetry {mixing schemes} strongly constrain the value of $\delta_{\rm CP}$. Consequently, neglecting correlations involving $\delta_{\rm CP}$ (e.g. see a recent discussion and a summary in Ref.~\cite{Zhang:2025jnn}) serves as a robust first approximation (underestimation of the corresponding $\chi^2$ values).
Therefore, a meaningful analysis requires a correlated $\chi^2$ including at least the parameters $\theta_{12}$, $\theta_{13}$, $\theta_{23}$, and $\delta_{\rm CP}$. The same conclusion applies to the TM$_{1(2)}$ analysis in the $\theta_{13}$--$\theta_{12}$ plane: although $\delta_{\rm CP}$ does not enter the analytical relations explicitly, it influences the results through parameter correlations.
Optionally, extending the analysis to include neutrino masses allows their values to be constrained through these correlations, even when they do not explicitly appear in the mixing-scheme relations. 

\item The constructed $\tilde{\chi}^2(P)$ provides a robust lower bound on the true $\chi^2(P)$. For any point $P$, the true $\chi^2(P)$ is necessarily greater than or equal to the $\chi^2$ value obtained from any projection onto a lower-dimensional subspace, since the latter is obtained by minimizing over the remaining parameters. Consequently, $\tilde{\chi}^2(P)$ captures the correlated experimental information against which theoretical models are tested. The resulting allowed parameter regions should therefore be regarded as conservative: the actual allowed regions cannot be larger and will generally be smaller than those shown here.
\end{enumerate} 

The full set of interpolated global fit functions, the implementation of  Eq. (\ref{eq:ext2}) and Eq. (\ref{eq:red2}) to obtain Fig. (\ref{figure:1}) we provide  an accompanying Mathematica notebook~\cite{GitHubCorrelations} to allow independent verification of the reconstruction and its direct application to the mixing schemes discussed in the following section.
 
\section{Correlation in neutrino oscillation data and its implications. }
\label{section:method}

The procedure introduced in the previous section for estimating the correlated ${\chi}^2$ of an arbitrary subset of oscillation parameters is entirely general and can be applied to any theoretical framework predicting algebraic relations among two or more of them. In this section, we put them to use with some representative examples to show tighter, correlation-respecting constraints than those obtainable from the individual 1D or 2D NuFIT intervals alone. 

As of now, models based on discrete flavor or modular or residual CP symmetries and/or their combinations are most frequently used to explain the observed pattern of neutrino mixing~\cite{Chauhan:2023faf,Ding:2024ozt,King:2013eh,Altarelli:2010gt,Petcov:2017ggy,Ishimori:2010au}.   Unsurprisingly, as mentioned earlier, these models are very predictive in nature, and often two or more oscillation parameters get related by closed-form algebraic relations commonly known as mixing {\it sum rules}. Such sum rules transform arbitrary parameter fits into testable structural relations. Generally, two broad classes of such sum rules exist in the literature. First, the mixing sum rules which yield algebraic relations among mixing angles and CP-phase~\cite{King:2007pr,Antusch:2007rk, Ballett:2014dua, Petcov:2014laa}, relating two or more neutrino oscillation parameters. The representative examples considered here, such as the trimaximal and $\mu-\tau$ reflection symmetries studied in the following section, are themselves concrete realizations of this class. On the other hand, mass sum rules,  relate not the mixing angles but the complex neutrino mass eigenvalues themselves, typically through linear or bilinear relations. It has been argued that such mass sum rules may originate from flavor symmetry constructions and provide testable predictions beyond fitting the
known mixing angles~\cite{Karmakar:2014dva, Karmakar:2015jza, Gehrlein:2016wlc, Gehrlein:2017ryu, Barry:2010yk, King:2013psa,  Gehrlein:2020jnr,Chauhan:2023faf, CentellesChulia:2023osj, Chulia:2025nsa}. {Both classes share the feature that enables the correlated-fit approach to accurately confront them, making it an efficient tool for surveying such a class of models.} 

In the following, we first review representative neutrino mixing schemes inspired by discrete flavor symmetries and the characteristic algebraic relations among their parameters. Next, by taking into account the relations among the neutrino mixing parameters of the model, we reconstruct the full correlation structure and derive the corresponding constraints imposed by the global neutrino oscillation fit. Finally, we employ this framework to investigate the impact of these correlations on the phenomenological predictions of the mixing schemes under consideration.

\subsection{Flavor symmetry mixing TM$_1$, TM$_2$ and partial $\mu$-$\tau$ reflection symmetry models}

Neutrino oscillation data exhibit a distinct mixing pattern in the lepton sector, in contrast to the near-diagonal structure of the quark sector.  Consequently, significant efforts have been devoted to understanding the neutrino mixing pattern, particularly models based non-Abelian discrete flavor symmetry groups~\cite{Altarelli:2010gt, King:2013eh, Petcov:2017ggy, Chauhan:2023faf}. Various `fixed' mixing schemes, such as bi-maximal~\cite{Vissani:1997pa,Fukugita:1998vn, Barger:1998ta, Baltz:1998ey}, tribimaximal (TBM)~\cite{Harrison:2002er, Harrison:2002kp}, hexagonal~\cite{Albright:2010ap}, and golden ratio ~\cite{Datta:2003qg, Kajiyama:2007gx, Everett:2008et, Rodejohann:2008ir, Adulpravitchai:2009bg, Ding:2011cm} mixing were proposed until 2012, when the measurement of a non-zero $\theta_{13}$~\cite{An:2012eh,Ahn:2012nd, DoubleChooz:2011ymz} ruled out all such possibilities. However, mixing schemes may still play a pivotal role in neutrino model building and can be considered as a first-order approximation of neutrino oscillation data. It is interesting to note that systematic approaches to revamp these schemes have been developed to align with updated experimental constraints. For a summary of all such possibilities, for example, see~\cite{ King:2013eh}.  In this regard,  continuous/discrete non-Abelian symmetries alongside modular symmetries continue to play a central role in current model-building, further contributing to the existing plethora of models~\cite{Ding:2023htn, King:2013eh, Ding:2024ozt, Altarelli:2010gt,Petcov:2017ggy, Chauhan:2023faf}. {Here, the aim is to propose a general framework that enables the confrontation of different classes of neutrino mixing schemes, or the associated sum rules relating two or more neutrino oscillation parameters, with correlated neutrino oscillation data, without relying on the details of construction of the underlying flavor model, e.g exact particle content \cite{Chauhan:2023faf}.} 

Among all the fixed mixing schemes mentioned earlier, the TBM mixing was probably the most popular due to its close proximity of predictions on the two large mixing angles:  $\sin^2{\theta_{12}}=1/3$ and $\sin^2{\theta_{23}}=1/2$. With $\sin\theta_{13}=0$, {the corresponding TBM mixing matrix} can be written as  
\begin{eqnarray}\label{eq:tbm}
|\rm{U_{{TBM}}}|=	
					\begin{pmatrix}
						\frac{2}{\sqrt{6}} &\frac{1}{\sqrt{3}} &0\\
						\frac{1}{\sqrt{6}} &\frac{1}{\sqrt{3}} &\frac{1}{\sqrt{2}}\\
						\frac{1}{\sqrt{6}} &\frac{1}{\sqrt{3}} &\frac{1}{\sqrt{2}}
					\end{pmatrix}. 
\end{eqnarray}
To address the observed non-zero $\theta_{13}$ (and CP violation), the trimaximal (TM) mixing generalizes TBM by relaxing some of its strict conditions while preserving certain symmetries~\cite{Albright:2008rp, Albright:2010ap, He:2011gb}. A straightforward generalization of TBM mixing is achieved by preserving a single column of the lepton mixing matrix in its TBM form given in Eq. (\ref{eq:tbm}). Preserving the first column of the TBM mixing matrix yields the TM$_1$ mixing pattern, while retaining the second column results in the TM$_2$ mixing pattern. In the $\rm{TM}_1$ and $\rm{TM}_2$ mixing-schemes,  
the mixing matrices  have the following structure 
\begin{eqnarray}
\label{u_tm1tm2}
\begin{array}{cc}
|\rm{U_{{TM}_1}}|=	
					\begin{pmatrix}
						\frac{2}{\sqrt{6}} &* &*\\
						\frac{1}{\sqrt{6}} &* &*\\
						\frac{1}{\sqrt{6}} &* &*
					\end{pmatrix},
     &
      |\rm{U_{{ TM}_2}}|=	
					\begin{pmatrix}
						* &\frac{1}{\sqrt{3}} &*\\
						* &\frac{1}{\sqrt{3}} &*\\
						* &\frac{1}{\sqrt{3}} &*
					\end{pmatrix}. 
\end{array}
\end{eqnarray}
Comparing the corresponding elements of the first (second) column of $\rm{U_{PMNS}}$ and $\rm{U_{{TM}_{1(2)}}}$, {given in Eq. (\ref{upmnsmatrix}) and Eq. (\ref{u_tm1tm2}) respectively,  one can easily obtain the algebraic} relations among the oscillation parameters {connecting the mixing angles and CP phase} (see e.g. \cite{deMedeirosVarzielas:2012apl,Chauhan:2023faf, Ganguly:2022qxj, Ganguly:2023jml}). Hence, the relation between  $\theta_{12}$ and $\theta_{13}$ {for the $\rm{TM}_1$ and $\rm{TM}_2$ mixing-schemes}  reads

\begin{equation}
     \label{eq:tms12s13}
     \theta_{12}^{\mbox{TM}_1}(\theta_{13}) :~~~  s_{12}^2 = \frac{1-3 s_{13}^2}{3 - 3 s_{13}^2}, 
    ~~~~~~~~~
     \theta_{12}^{\mbox{TM}_2}(\theta_{13}) :~~~ s_{12}^2 = \frac{1}{3 - 3 s_{13}^2}.
\end{equation}

\noindent Similarly, the relation between $\delta_{\rm CP}$, $\theta_{13}$ and $\theta_{23}$ ($\theta_{12}$ is expressed in term of $\theta_{13}$ via Eq.~\eqref{eq:tms12s13}) can be written as 
\begin{align}
     \label{eq:tmdcps23s13}
      \delta_{\rm CP}^{\mbox{TM}_1}(\theta_{13},\theta_{12}^{\mbox{TM}_1}(\theta_{13}),\theta_{23})  &:~~~  \cos \delta_{\rm CP} = \frac{
    (1-5s_{13}^2)(2s_{23}^2-1)
    }{
    4 s_{13}s_{23}\sqrt{2(1-3s_{13}^2)(1-s_{23}^2)}
    }, \\
     \delta_{\rm CP}^{\mbox{TM}_2}(\theta_{13},\theta_{12}^{\mbox{TM}_2}(\theta_{13}),\theta_{23})  &:~~~ \cos \delta_{\rm CP} = -\frac{
    (2-4s_{13}^2)(2s_{23}^2-1)
    }{
    4 s_{13}s_{23}\sqrt{(2-3s_{13}^2)(1-s_{23}^2)}
    }.\label{eq:tmdcps23s13-2}
\end{align}
{These sum rules clearly show that $\cos\delta_{\rm CP}$ cannot be chosen freely; its value is entirely dictated by the three mixing angles of the lepton sector. }

{In a  complementary, and more general, class of mixing scheme known as $\mu$-$\tau$ reflection symmetry~\cite{Harrison:2002et},  where elements of the second and third row of the lepton mixing matrix satisfy, $|U_{\mu i}|=|U_{\tau i}|$ where $i=1,2,3$.  In other words, this idea is based on the cumulative operation of $\mu$-$\tau$ flavor exchange and CP transformations~\cite{Harrison:2002et}. An immediate consequences of this equality of moduli are $\theta_{23}=45^{\circ}$ and $\delta_{\rm CP}=0^{\circ}$ or $270^{\circ}$. Further, this scheme allows $\theta_{13}\neq 0$, unlike the fixed-mixing schemes stated above or the simple $\mu$-$\tau$ permutation symmetry, where the reactor mixing angle is strictly zero. Although all these predictions are in agreement with the observed neutrino oscillation data, a sizeable range of these parameters is allowed at 3$\sigma$. In addition, the global fits also point to a non-maximal atmospheric mixing angle in either mass hierarchy. Hence, the partial $\mu$-$\tau$ reflection symmetry is assumed where the condition $|U_{\mu i}|=|U_{\tau i}|$ holds only for a single column of the lepton mixing matrix, $i.e.,$}   $ |U_{\mu 1}|=|U_{\tau 1}|$ and $|U_{\mu 2}|=|U_{\tau 2}|$\footnote{{If the equality holds for only for the third column, one can get maximal $\theta_{23}$ with unrestricted $\delta_{\rm CP}$. And if we apply the equality on any two columns simultaneously then by unitarity it holds for the third.} } (in this work, respectively denoted by \mbox{$\mu1$-$\tau1$} and \mbox{$\mu2$-$\tau2$}). 
{Now, again comparing \mbox{$\mu1$-$\tau1$} and \mbox{$\mu2$-$\tau2$} mixing schemes with $U_{\rm PMNS}$ one can obtain the relations among all three mixing angles}  $\theta_{23}$, $\theta_{13}$, $\theta_{12}$ {and the Dirac CP phase } as \cite{Ge:2011ih,Ge:2011qn, Chakraborty:2018dew} 
\begin{align}
     \label{eq:mutaudcps23s13s12}
      \delta_{\rm CP}^{\mbox{$\mu1$-$\tau1$}}(\theta_{13},\theta_{12},\theta_{23})  &:~~~  \cos \delta_{\rm CP} = \frac{
     (c_{23}^2-s_{23}^2)(c_{12}^2 s_{13}^2-s_{12}^2)
     }{
     4 c_{12}s_{12}c_{23}s_{23}s_{13}
     }, \\
     \delta_{\rm CP}^{\mbox{$\mu2$-$\tau2$}}(\theta_{13},\theta_{12},\theta_{23})  &:~~~ \cos \delta_{\rm CP} = \frac{
     (c_{23}^2-s_{23}^2)(c_{12}^2- s_{12}^2s_{13}^2)
     }{
     4 c_{12}s_{12}c_{23}s_{23}s_{13}
     }\label{eq:mutaudcps23s13s12-2}.
\end{align}
{The representative examples of algebraic relations (belonging to the broader class of mixing and CP-phase sum rules) in Eqs.~(\ref{eq:tms12s13}) - (\ref{eq:mutaudcps23s13s12-2}) clearly show an interdependence of neutrino oscillation parameters among each other. In the next subsection, we illustrate the consequences of combining $\tilde{\chi}^2$ with these relations. }
 
\subsection{Two-Dimensional Correlated Constraints}
\label{section:res2d}
{As discussed earlier, in order to confront symmetry-driven neutrino mixing schemes or sum rules, comparing theoretical predictions against uncorrelated 1D or 2D experimental intervals leads to an overestimation of the allowed model parameter space. To properly evaluate a given mixing scheme, its analytic predictions must be projected onto the full correlated hypersurface reconstructed via 
Eq.~(\ref{eq:ext2}) and Eq.~(\ref{eq:red2}). Here we constrain the model parameter space by requiring that the constructed test satisfies $\tilde{\chi}^2 \leq 9$. } 
\vspace{.2cm}

\begin{figure}[!h]
    \centering
    \includegraphics[width=0.49\textwidth]{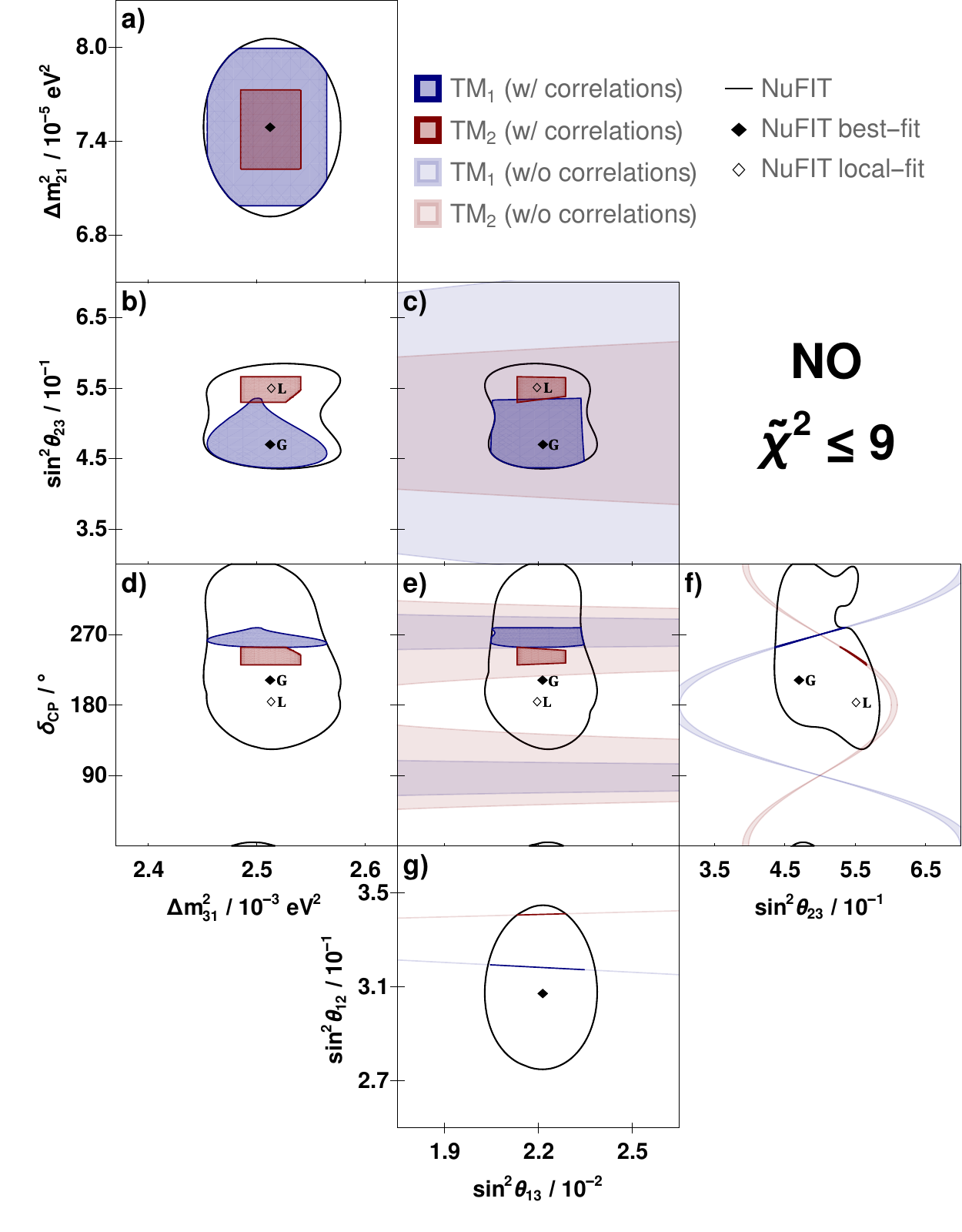}
     \includegraphics[width=0.49\textwidth]{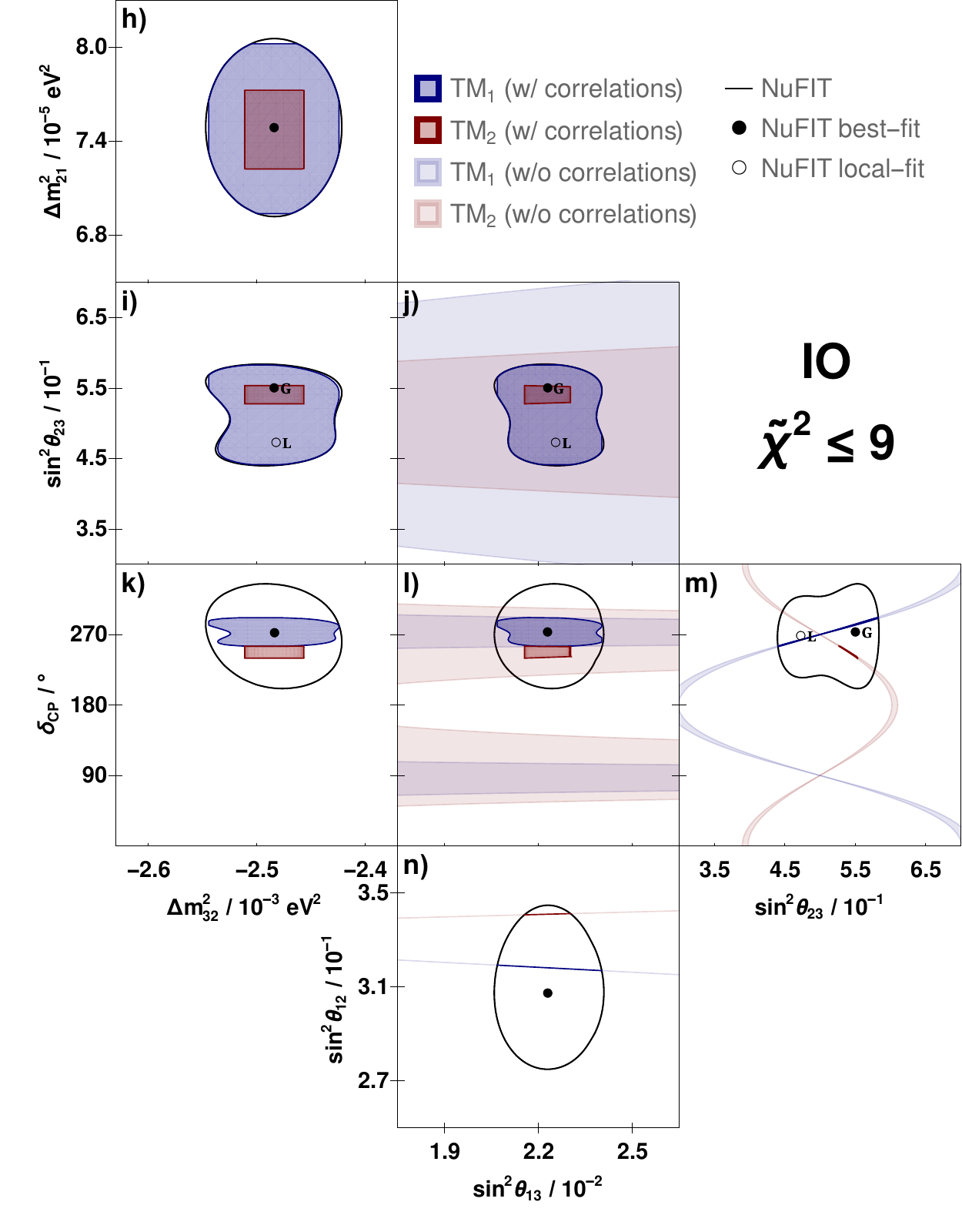}
      \caption{ 
      2D projections of  $\tilde{\chi}^2 \leq 9$ (Eq.~\eqref{eq:tmchi9}) for both mass orderings involving NuFIT correlations and $\rm{TM}_{1(2)}$  
      relations Eqs. \eqref{eq:tms12s13} and \eqref{eq:tmdcps23s13}. Lines and regions description in the text.}
    \label{figure:tm9chi}
\end{figure}

\noindent 
\underline{\bf Trimaximal mixing:} 
For the revamped trimaximal mixing scenarios, the
model-constrained 
2D projections are obtained by combining
$\tilde\chi^2$ with the sum rules given in Eqs. (\ref{eq:tms12s13})–(\ref{eq:tmdcps23s13-2}), which are
implemented directly in the $\tilde\chi^2$ constructed and estimated from the global fit data.
{Thus, for both TM$_1$ and TM$_2$ mixing schemes, we have the following constraints for mass orderings:}
\begin{align}
     \label{eq:tmchi9}
     \nonumber
     \mbox{NO} &:~~~ 
     \tilde\chi^2 (\theta_{13},
     \theta_{12}^{\mbox{TM}_{1(2)}}(\theta_{13}),
     \theta_{23},
     \delta_{\rm CP}^{\mbox{TM}_{1(2)}}(\theta_{13}, \theta_{12}^{\mbox{TM}_{1(2)}}(\theta_{13}),\theta_{23}),
      \Delta m_{21}^2,\Delta m_{31}^2
     ) \leq 9 
     ,\\
     \mbox{IO} &:~~~ 
     \tilde\chi^2 (\theta_{13},
     \theta_{12}^{\mbox{TM}_{1(2)}}(\theta_{13}),
     \theta_{23},
     \delta_{\rm CP}^{\mbox{TM}_{1(2)}}(\theta_{13}, \theta_{12}^{\mbox{TM}_{1(2)}}(\theta_{13}),\theta_{23}),
      \Delta m_{21}^2,\Delta m_{32}^2
     ) \leq 9 . 
\end{align}
In  Fig.~\ref{figure:tm9chi} the results for TM$_1$ and TM$_2$ schemes are given. Where the  panels a)-g) in the left-hand side, show 2D projections of mixing-scheme dependent results for NO in the planes
 $\Delta m_{31(32)}^2$/$\Delta m_{21}^2$, $\Delta m_{31(32)}^2$/$\theta_{23}$,  $\theta_{13}$/$\theta_{23}$, 
$\Delta m_{31(32)}^2$/$\delta_{\rm CP}$, 
$\theta_{13}$/$\delta_{\rm CP}$,
$\theta_{23}$/$\delta_{\rm CP}$, 
$\theta_{13}$/$\theta_{12}$.  
Panels  h)-n) gives corresponding results for the IO of neutrino mass on the right-hand side. 

In this figure, the light-shaded regions first illustrate the algebraic relations following from Eqs.~\eqref{eq:tbm} and~\eqref{u_tm1tm2} for the TM$1$ and TM$2$ mixing schemes, shown in panels c), e), f), j), l), and m), respectively. In this deliberately conservative, overestimated approach, the remaining parameters are varied within their $3\sigma$ ranges (Tab.~\ref{table:nufit60}), without (w/o) accounting for correlations in the oscillation data, as constructed in Eqs.~\eqref{eq:tmdcps23s13} and~\eqref{eq:tmdcps23s13-2}. Following the same approach, panels g) and n) show schematically the relation between $\theta_{12}$ and $\theta_{13}$ given in Eq.~\eqref{eq:tms12s13}, represented by the light-shaded solid line.

The impact of the correlations constructed in Eqs.~\eqref{eq:ext2} and~\eqref{eq:red2} is also shown in Fig.~\ref{figure:tm9chi} for the trimaximal TM$_1$ and TM$_2$ mixing schemes. In all panels dark-shaded regions represent allowed regions with (w/) these correlations. The {correlated} $\rm{TM}_1$ {allowed region} is colored in blue, {correlated} $\rm{TM}_2$ {allowed region} is in red, while the black, solid line contours represent NuFIT border regions at $\chi^2 = 9$. Here, the global minimum (best-fit) is denoted by $\mathbin{\medblackdiamond}$ ($\mathbin{\medblackcircle}$), while local minimum (local-fit) is denoted by $\mathbin{\Diamond}$  ($\mathbin{\mdlgwhtcircle}$) for NO (IO), respectively. If both best-fit (\textit{bf}) and local-fit (\textit{lf}) are present, the symbols are extensively followed by letters \textbf{G} (global NuFIT minimum) and \textbf{L} (local NuFIT minimum). 
We will discuss \textit{bfs}  and \textit{lfs} for $\tilde{\chi}^2$ later on. The choice of colors and markings remains the same throughout the next figures.
In panels c), e), f) g) (j), l), m) n)) for NO (IO), comparing the dark shaded regions with the light shaded regions and black solid line contours {\it we observe that the $\rm{TM}_{1(2)}$ allowed region w/ correlations taken into account is reduced, and narrower than  the individual permitted ranges.}  The strongest reduction of allowed region is observed in the case of $\rm{TM}_{2}$ for both mass orderings, while the weakest reduction of allowed region is observed for $\rm{TM}_{1}$, IO. 
In panels b) and d) (i) and k)) for NO (IO), the shape of the regions is mostly determined by the $\theta_{23}/\Delta m_{31(32)}^2/\delta_{\rm CP}$ 
NuFIT correlations data, involving model-dependent relations, Eqs. \eqref{eq:tms12s13}-\eqref{eq:tmdcps23s13-2}. 

Here, we find that, in some of the planes, the NuFIT \textit{bfs} and \textit{lfs} do not lie within the $\tilde{\chi}^2\leq 9$ region. This is a consequence of the mutual determination of the mixing parameters imposed by the mixing-scheme relations. In the case of $\mathrm{TM}_{1(2)}$, any two of the four mixing parameters $\theta_{12}$, $\theta_{13}$, $\theta_{23}$, and $\delta_{\rm CP}$ determine the remaining two.
In Section \ref{section:res1d} we discuss in more detail how the mixing schemes affect the \textit{bfs};
Figs. \ref{figure:paramno9chi} and \ref{figure:paramio9chi}.
There, in the one-dimensional $\tilde{\chi}^2$ profiles, the \textit{bfs} (\textit{lfs}) correspond to the global (local) minima. 

\begin{figure}[!h]
    \centering
            \includegraphics[width=0.39\textwidth]{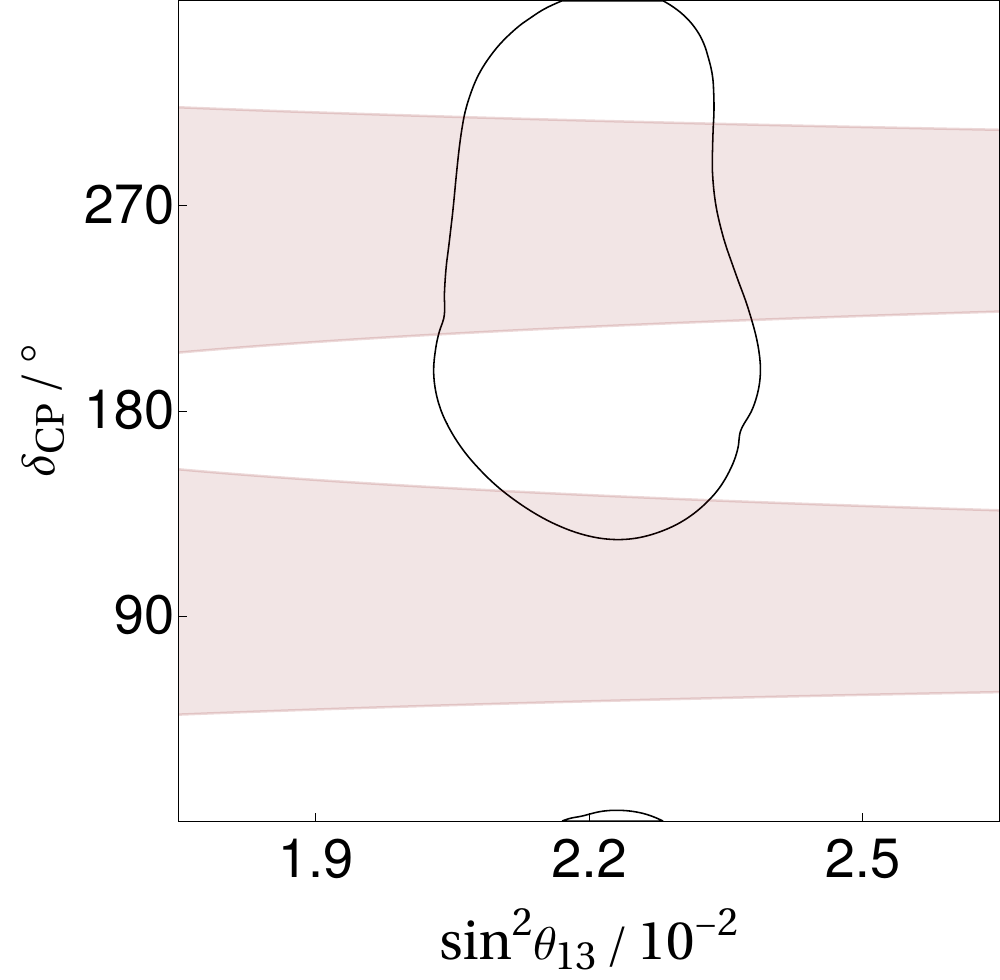}
            \includegraphics[width=0.39\textwidth]{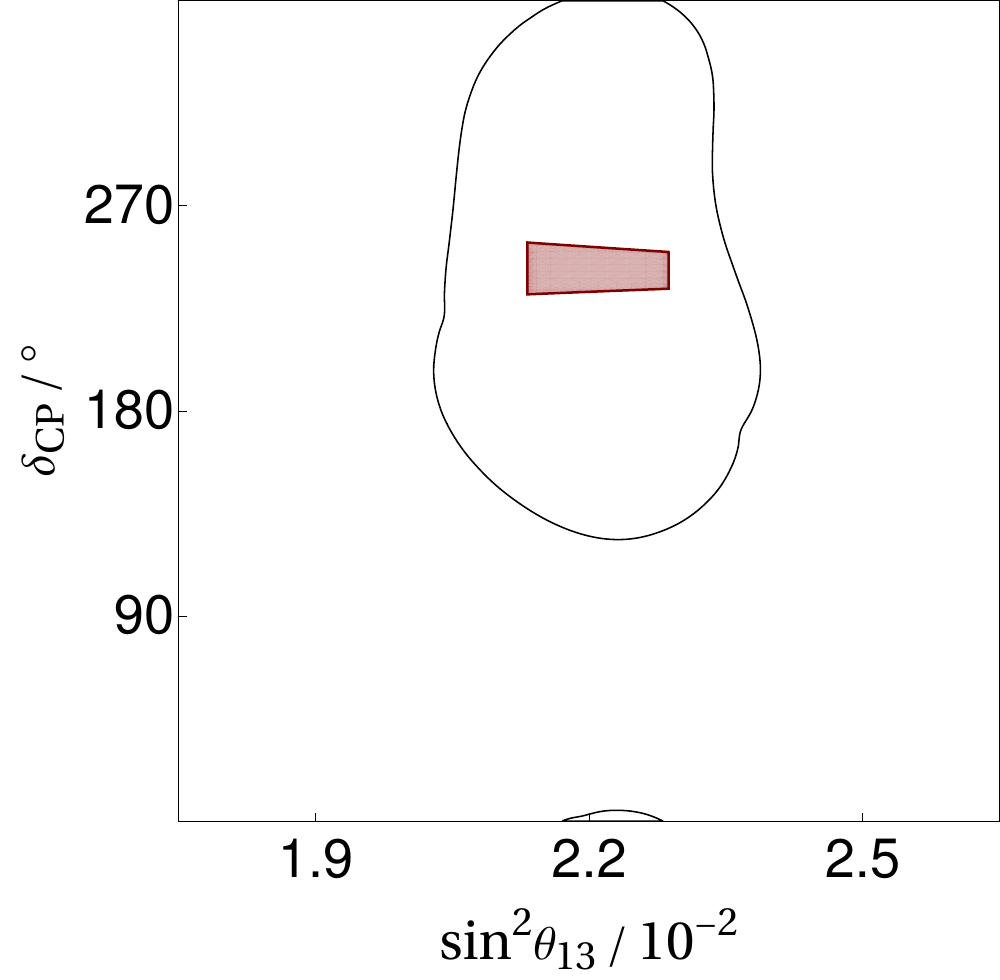}
     \caption{TM$_2$ relations and NuFIT NO data sets (correlations) plotted independently and superimposed on each other (left panel) and when combined using relations in Eq.~(\ref{eq:tmchi9}) (right panel) at 
     $\tilde\chi^2 \leq 9$
     for NO. 
}
    \label{figure:expth}
\end{figure}

Let us emphasize the relevance of the approach and the resulting constraints shown in Fig.~\ref{figure:tm9chi}, 
using TM$_2$ mixing with NO as an example. This is illustrated by comparing the dark-shaded (correlated) regions with the light-shaded (uncorrelated) ones. Incorporating the multi-parameter correlations leads to a substantial contraction of the allowed regions in all two-dimensional projections.
For example, consider Fig.~\ref{figure:expth} plotted in the $\sin^2\theta_{13}-\sin^2\theta_{12}$ plane.
Here, in the left panel, the light red-shaded region depicts the TM$_2$ mixing-scheme relations given in Eq.~\eqref{eq:tms12s13} and Eq.~\eqref{eq:tmdcps23s13-2} 
whereas the black contour shows the NuFIT correlations considered independently and superimposed on these relations.  However, when the combined correlations in Eq.~(\ref{eq:tmchi9}) are imposed,we observe a clear and significant reduction of the allowed area, as indicated in Fig.~\ref{figure:tm9chi}, panel e).
 
\color{black}

\begin{figure}[!h]
    \centering
     \includegraphics[width=0.49\textwidth]{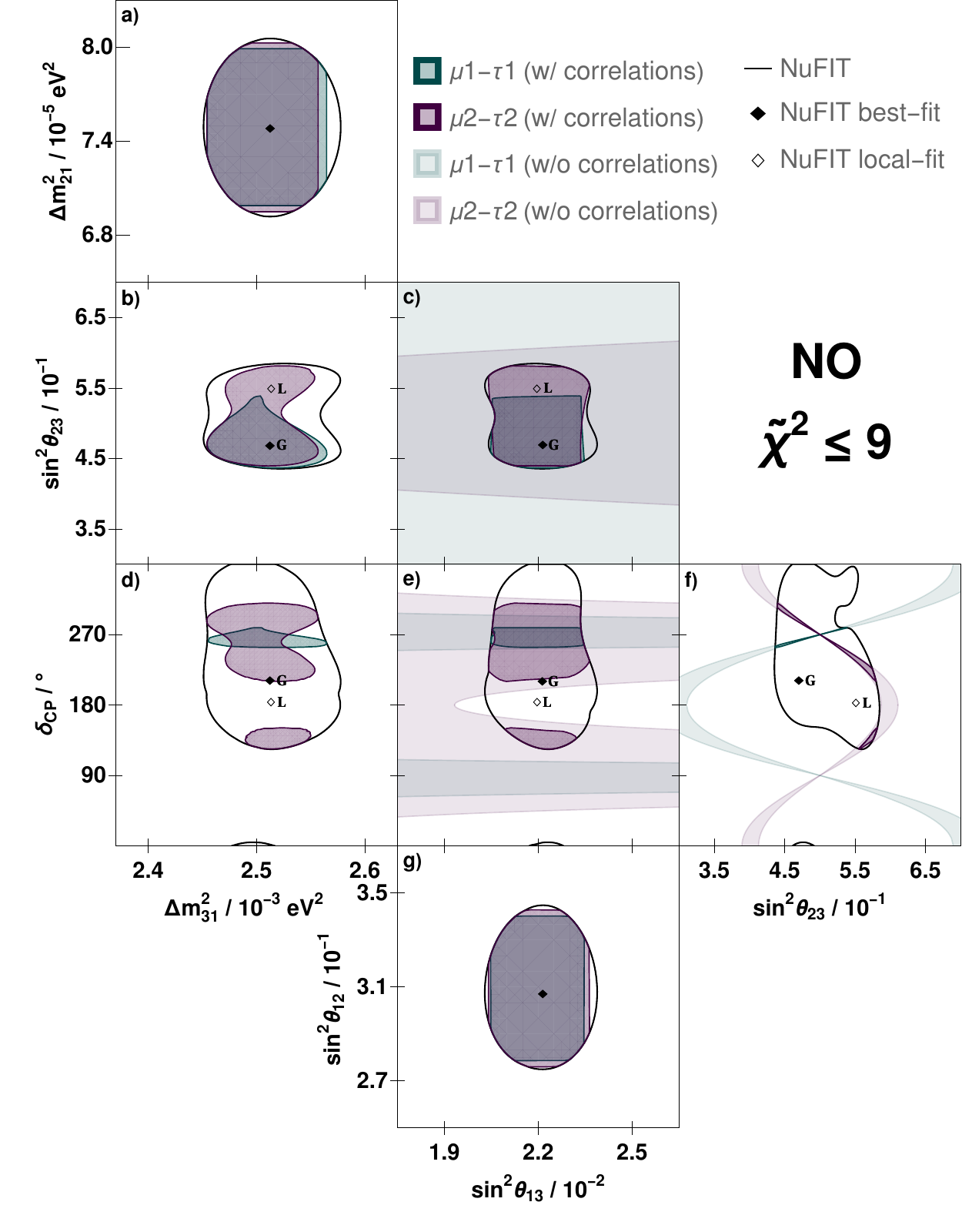}
     \includegraphics[width=0.49\textwidth]{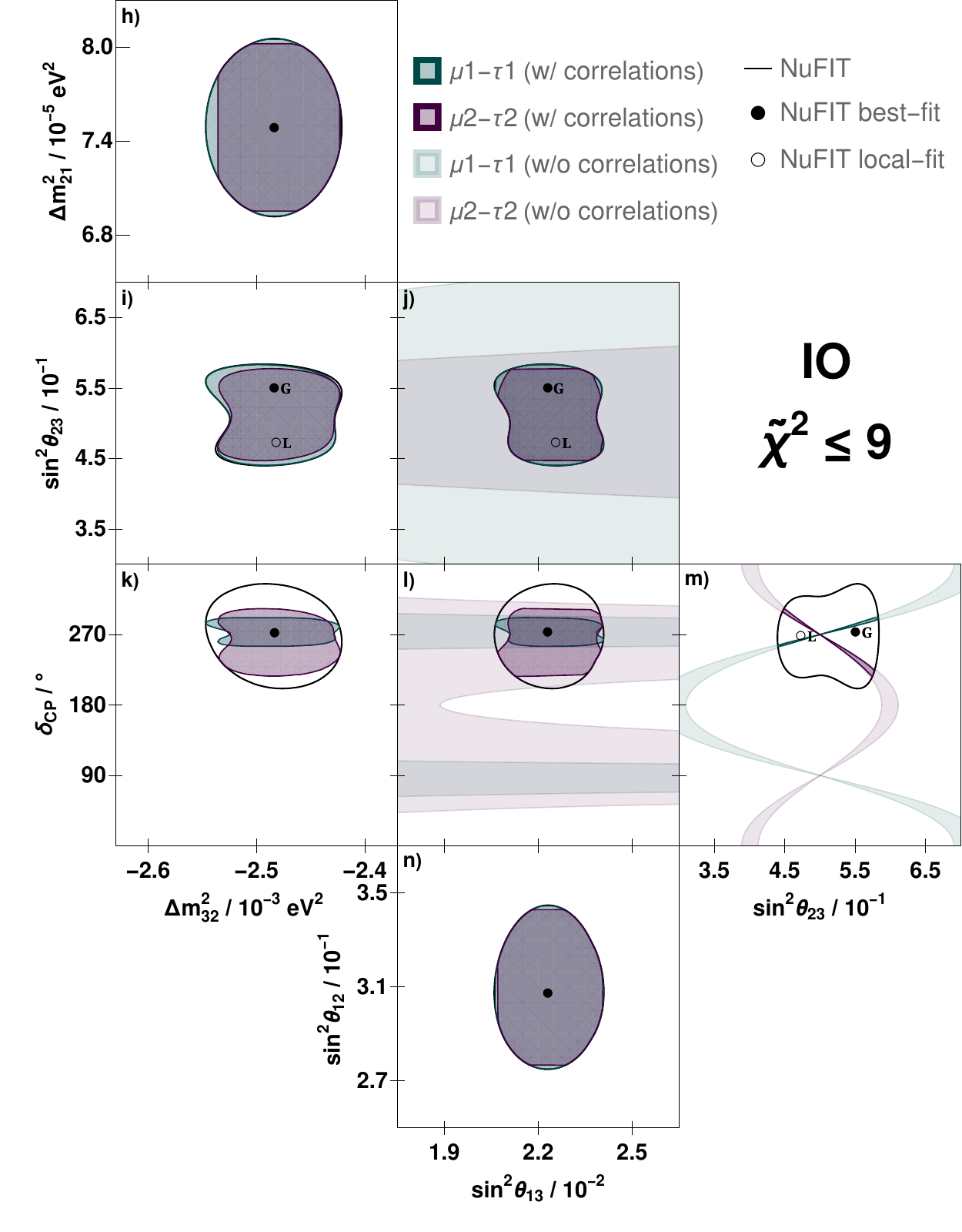}
      \caption{ 
 2D projections of  $\tilde{\chi}^2 \leq 9$ (Eq.~\eqref{eq:mutauchi9}) for both mass orderings involving NuFIT correlations and $\mu1(2)$-$\tau1(2)$
      relations Eq. \eqref{eq:mutaudcps23s13s12}. The region's description in the text.}
    \label{figure:mutau9chi}
\end{figure}

\vspace{.2cm}
\noindent
\underline{\bf Partial $\mu$-$\tau$ reflection symmetry:} {Similarly, for $\mu1$-$\tau1$ and $\mu2$-$\tau2$ mixing schemes where $\delta_{\text{CP}}$ is determined by the three mixing angles through Eqs. (\ref{eq:mutaudcps23s13s12})–(\ref{eq:mutaudcps23s13s12-2}), the corresponding  tested  $\tilde{\chi}^2$ has the following form}
\begin{align}
     \label{eq:mutauchi9}
      \nonumber
     \mbox{NO} &:~~~  \tilde\chi^2 (
 \theta_{13},\theta_{12}, 
 \theta_{23},
 \delta_{\rm CP}^{\mbox{$\mu1(2)$-$\tau1(2)$}}(\theta_{13}, \theta_{12},\theta_{23}), 
 \Delta m_{21}^2,\Delta m_{31}^2
      ) \leq 9 
     ,\\
     \mbox{IO} &:~~~  \tilde\chi^2 (
 \theta_{13},\theta_{12}, 
 \theta_{23},
 \delta_{\rm CP}^{\mbox{$\mu1(2)$-$\tau1(2)$}}(\theta_{13}, \theta_{12},\theta_{23}), 
 \Delta m_{21}^2,\Delta m_{32}^2
      ) \leq 9 .
\end{align}
In  Fig.~\ref{figure:mutau9chi} the constraints for $\mu1$-$\tau1$ and $\mu2$-$\tau2$ mixing  schemes are illustrated in terms of  $\tilde{\chi}^2$ 2D projections.
Just like the previous case, the light-shaded regions (uncorrelated), in panels c), e), f), j), l), m) for NO (IO) represent Eq.~\eqref{eq:mutaudcps23s13s12} and Eq.~\eqref{eq:mutaudcps23s13s12-2} with remaining parameters varying within $3\sigma$ range, w/o correlations. In all panels, a)-n), the dark-shaded region represents allowed region w/ correlations. The $\mu1$-$\tau1$ is colored in green, $\mu2$-$\tau2$ in purple, while the black, solid line contours represent NuFIT correlation border region at $\chi^2 = 9$.  

In panels c), e), f),  j), l), m)  for NO (IO), comparing the dark shaded regions with the light shaded regions and black solid line contours we observe that the $\mu1(2)$-$\tau1(2)$ allowed region w/ correlations taken into account is reduced, and narrower than other compared regions. The strongest reduction of the allowed region is observed in the case of $\mu1$-$\tau1$ for both mass orderings.

$\rm{TM}_{1(2)}$ mixings are constrained descendants of the tribimaximal mixing and are more restrictive schemes. {\it All shaded regions for $\mu1(2)$-$\tau1(2)$ are wider than for $\rm{TM}_{1(2)}$ because $\mu$-$\tau$ reflection symmetry conditions fix only $\delta_{\rm CP}$ leaving $\theta_{12}$ free, while TM$_{1(2)}$ also fixes $\theta_{12}$ through Eq. (\ref{eq:tms12s13}).}
For the same reason as discussed in the TM$_{1(2)}$ case, some of \textit{bfs} and \textit{lfs} are not in the $\tilde{\chi}^2 < 9$ region.
In case of $\mu1(2)$-$\tau1(2)$, we need three mixing parameters to determine simultanously $\theta_{12}, \theta_{13}, \theta_{23},  \delta_{\rm CP}$.  
  
\subsection{ {Implications for $m_{\beta}$ and $m_{\beta\beta}$}}\label{section:mbb}

\begin{figure}[!h]
    \centering
    \includegraphics[width=0.24\textwidth]{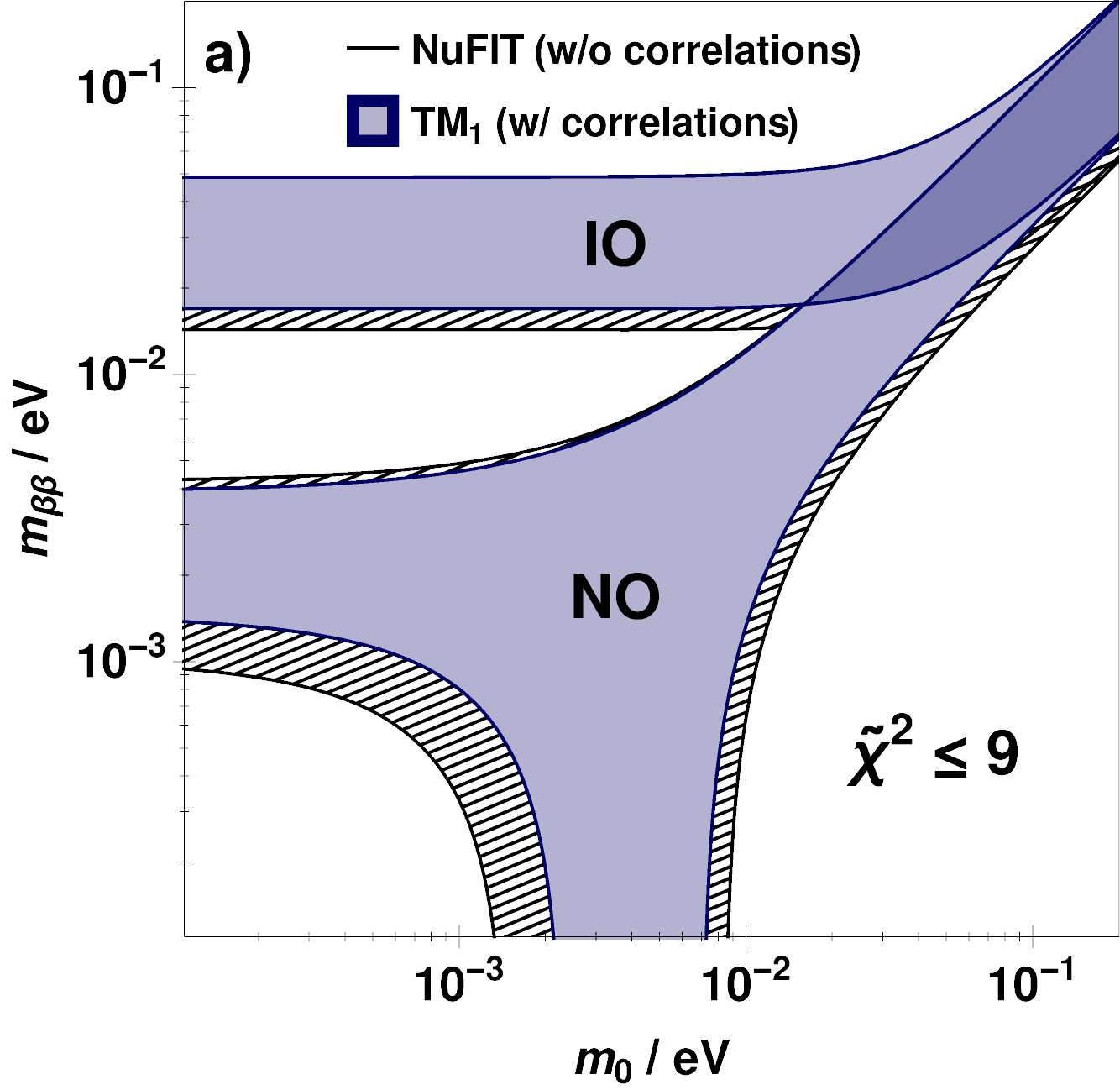}
     \includegraphics[width=0.24\textwidth]{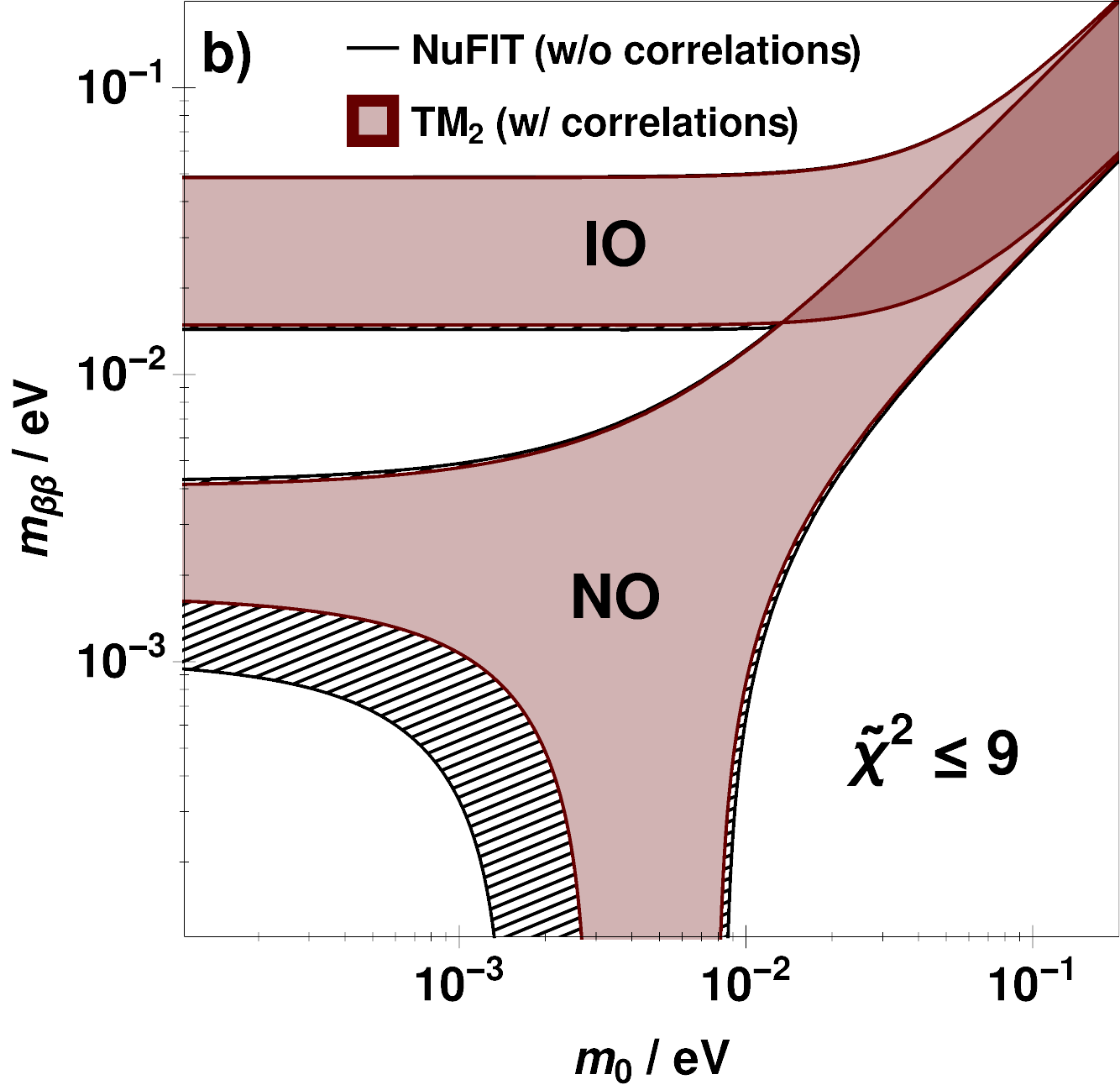}
\includegraphics[width=0.24\textwidth]{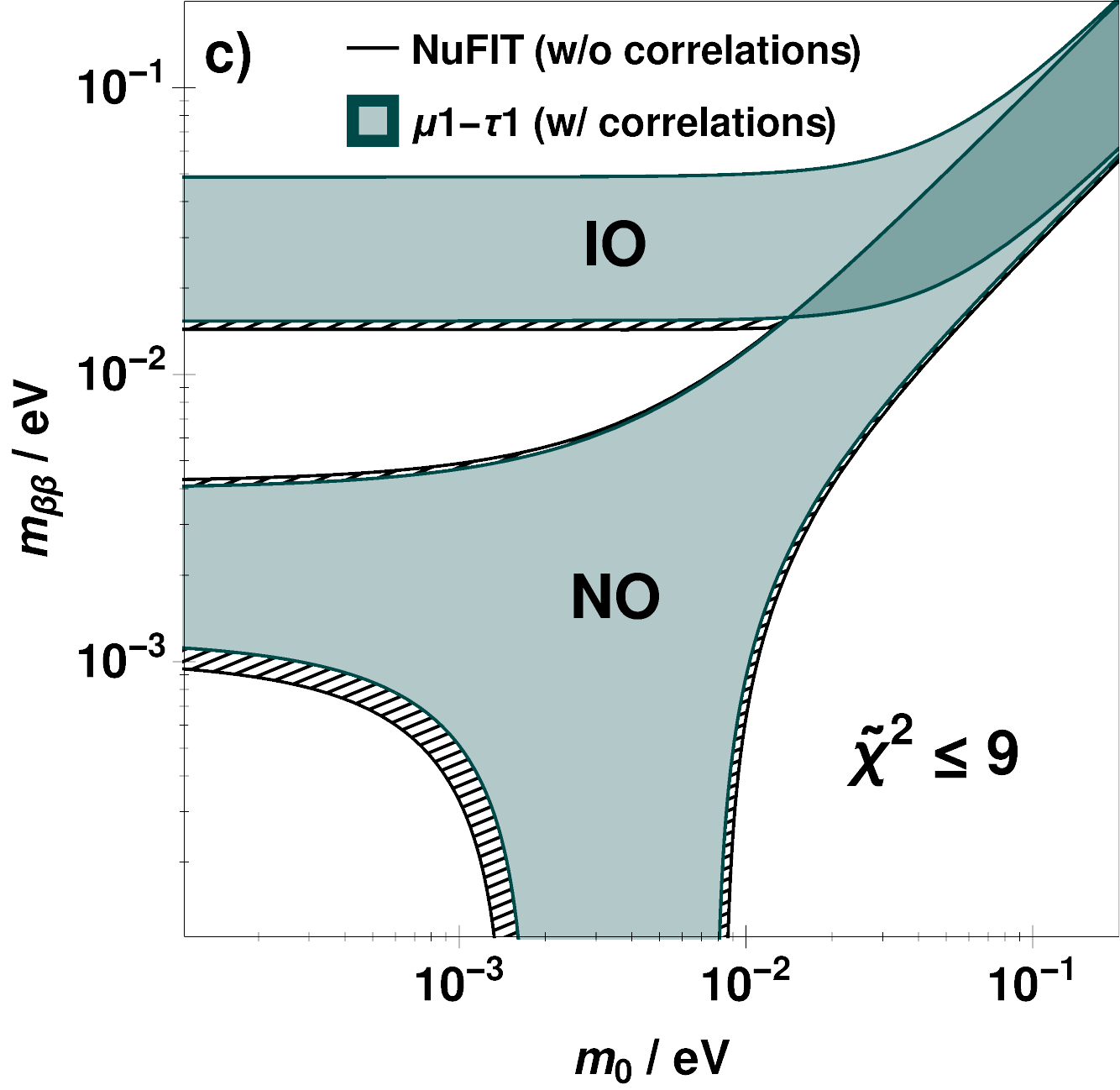}
     \includegraphics[width=0.24\textwidth]{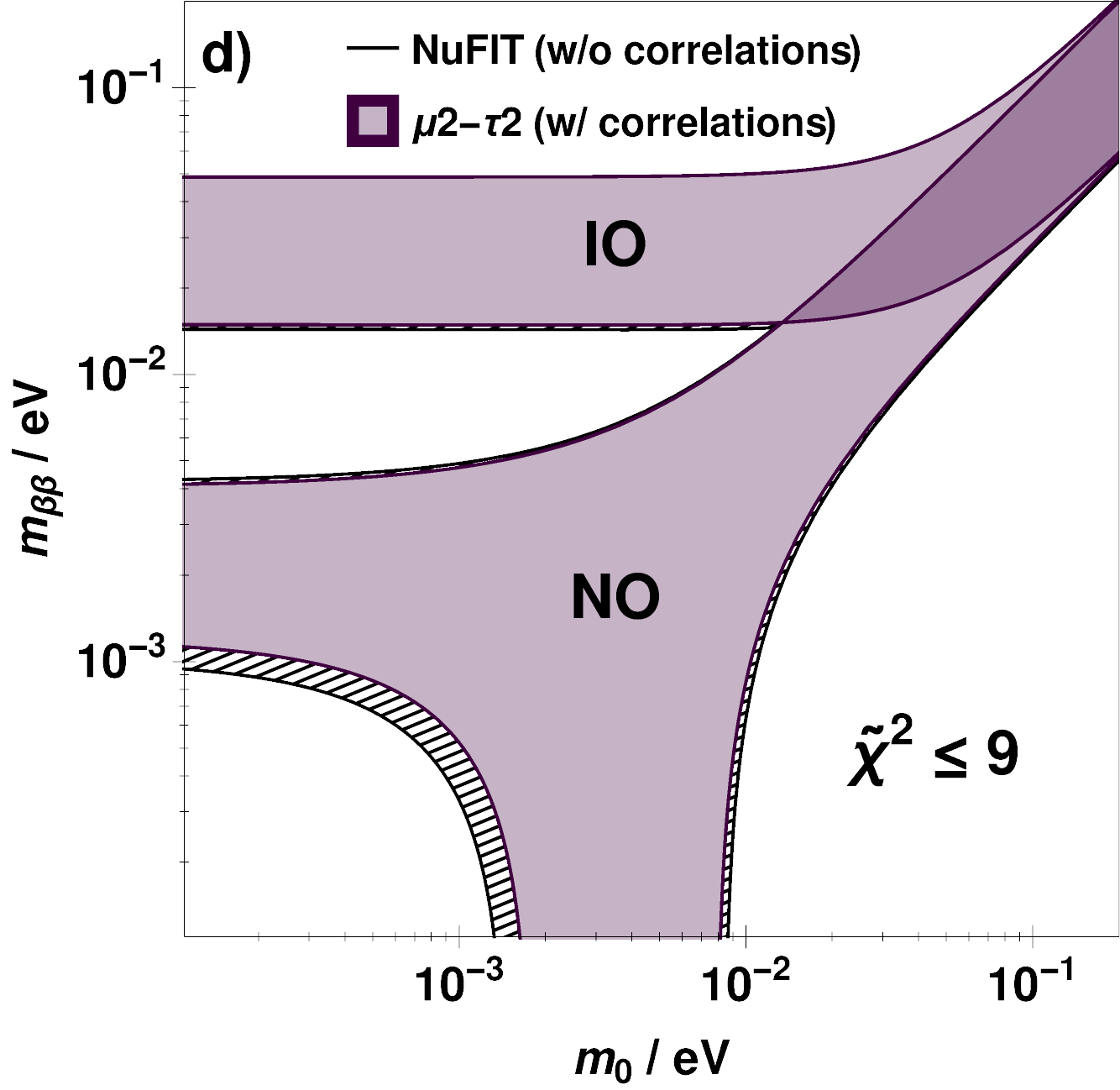}
     \caption{Impact of simultaneous analysis of neutrino mixing parameters on the effective Majorana mass $m_{\beta\beta}$ as a function of the lightest neutrino mass for both mass orderings, incorporating NuFIT correlations and the mixing-parameter relations predicted by the considered mixing schemes.
     }
    \label{figure:mbbtm9chi}
\end{figure}

In this section we show the impact of determined allowed neutrino parameter space with discussed correlations on the effective neutrino masses $m_{\beta}$ and $m_{\beta\beta}$ in the tritium beta and neutrinoless double beta decays, respectively.
The  $m_{\beta}$ and $m_{\beta\beta}$ effective masses are defined, for each mass ordering, as 
\begin{align}
     \label{eq:mbetabetatm}
      \nonumber
     \mbox{NO} &:~
     \begin{cases}
    m_{\beta} = 
     \sqrt{m_{0}^2 + \Delta m_{21}^2 c_{13}^{2} s_{12}^{2}+
     \Delta m_{31}^2 s_{13}^{2}}
     \,, \\
     m_{\beta\beta} = \big|
     m_{0}  c_{13}^{2} c_{12}^{2} e^{i2(\alpha_1-\delta_{\rm CP})}+
     \sqrt{m_{0}^2 + \Delta m_{21}^2}c_{13}^{2} s_{12}^{2} e^{i2(\alpha_2-\delta_{\rm CP})}
     +
      \sqrt{m_{0}^2 + \Delta m_{31}^2}s_{13}^{2} 
     \big|
     \,,
     \end{cases}
     \\
     \mbox{IO} &:~ 
      \begin{cases}
   m_{\beta} = \sqrt{m_{0}^2 - \Delta m_{21}^2 c_{13}^{2} c_{12}^{2}-
     \Delta m_{32}^2 c_{13}^{2}}\\ 
      m_{\beta\beta} = \big|
     \sqrt{m_{0}^2 - \Delta m_{32}^2-\Delta m_{21}^2}c_{13}^{2} c_{12}^{2} e^{i2(\alpha_1-\delta_{\rm CP})}
     +
      \sqrt{m_{0}^2 - \Delta m_{32}^2}c_{13}^{2} s_{12}^{2} e^{i2(\alpha_2-\delta_{\rm CP})}
      +
      m_0 s_{13}^{2} 
     \big|
     \,.
     \end{cases}
\end{align}
{As evident from Eq. (\ref{eq:mbetabetatm}), $m_{\beta}$ is insensitive to any relative phases among the mass eigenstates. It depends on neutrino masses and the mixing angles alone, with no dependence on $\delta_{\rm CP}$ or on the Majorana phases $\alpha_{1,2}$. In contrast,  $m_{\beta\beta}$ retains full sensitivity to $\delta_{\rm CP}$ and to $\alpha_{1,2}$. In both orderings, $m_0$ denotes the mass of the lightest eigenstate, $i.e.$, $m_1$ for NO, $m_3$ for IO, so the entire absolute mass spectrum is fixed by the single free parameter $m_0$ together with the two measured mass-squared differences.}  
Fig.~\ref{figure:mbbtm9chi} shows the corresponding result for   $m_{\beta\beta}$  as a function of the lightest neutrino mass $m_0$. 

\begin{figure}[!h]
    \centering
    \includegraphics[width=0.24\linewidth]{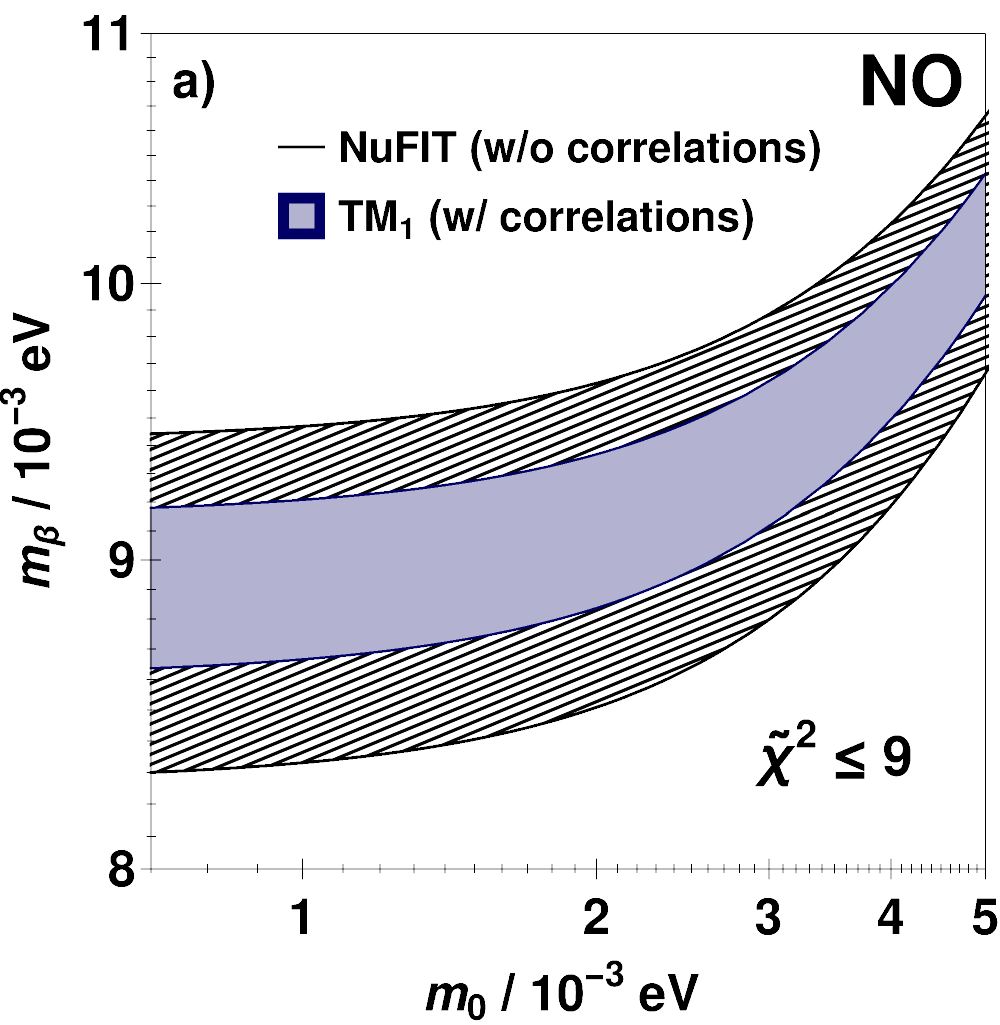}
    \includegraphics[width=0.24\linewidth]{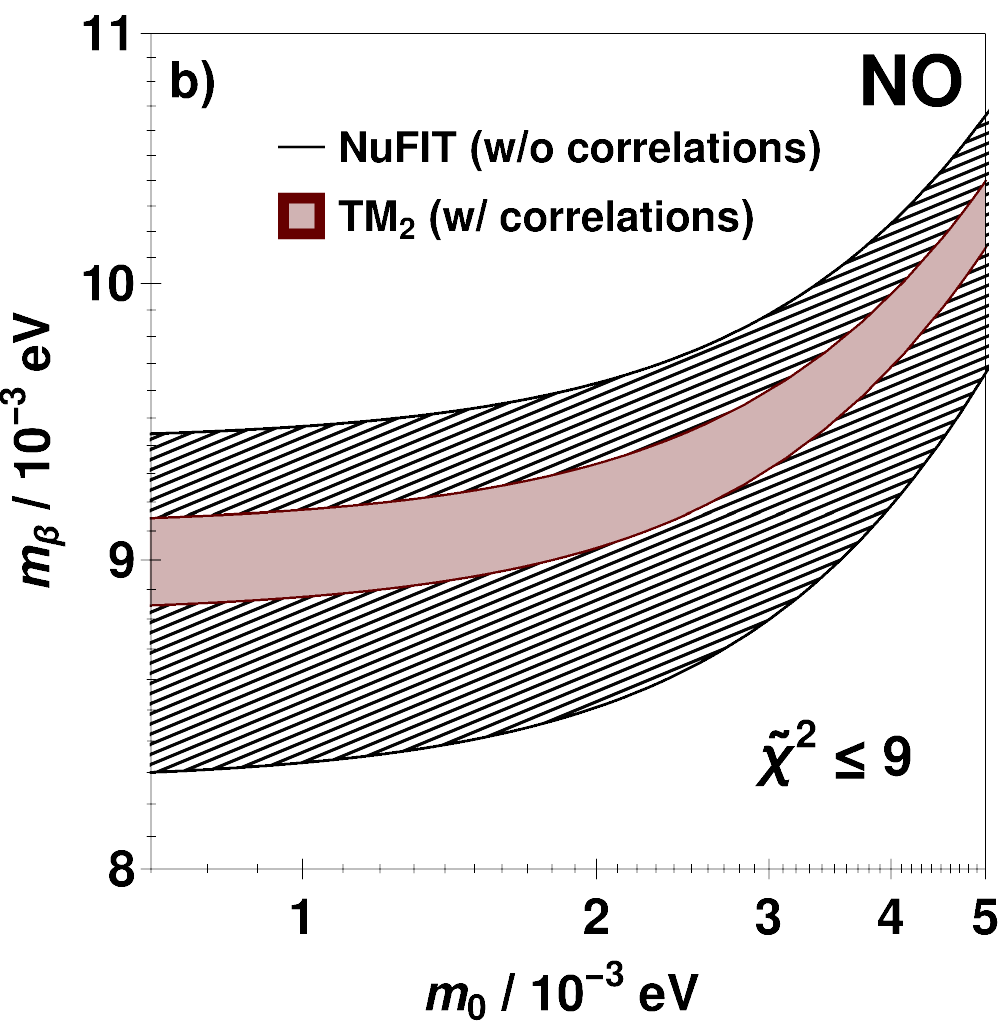}
    \includegraphics[width=0.24\linewidth]{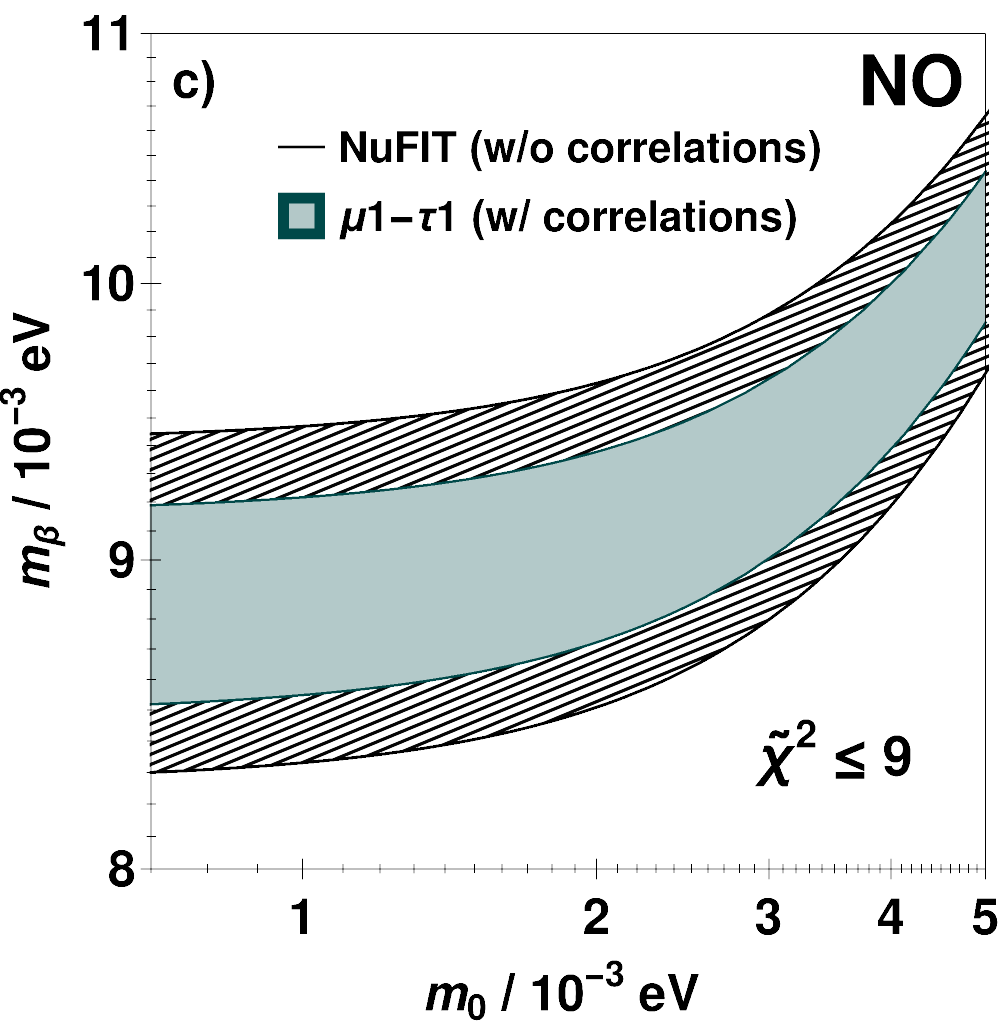}
    \includegraphics[width=0.24\linewidth]{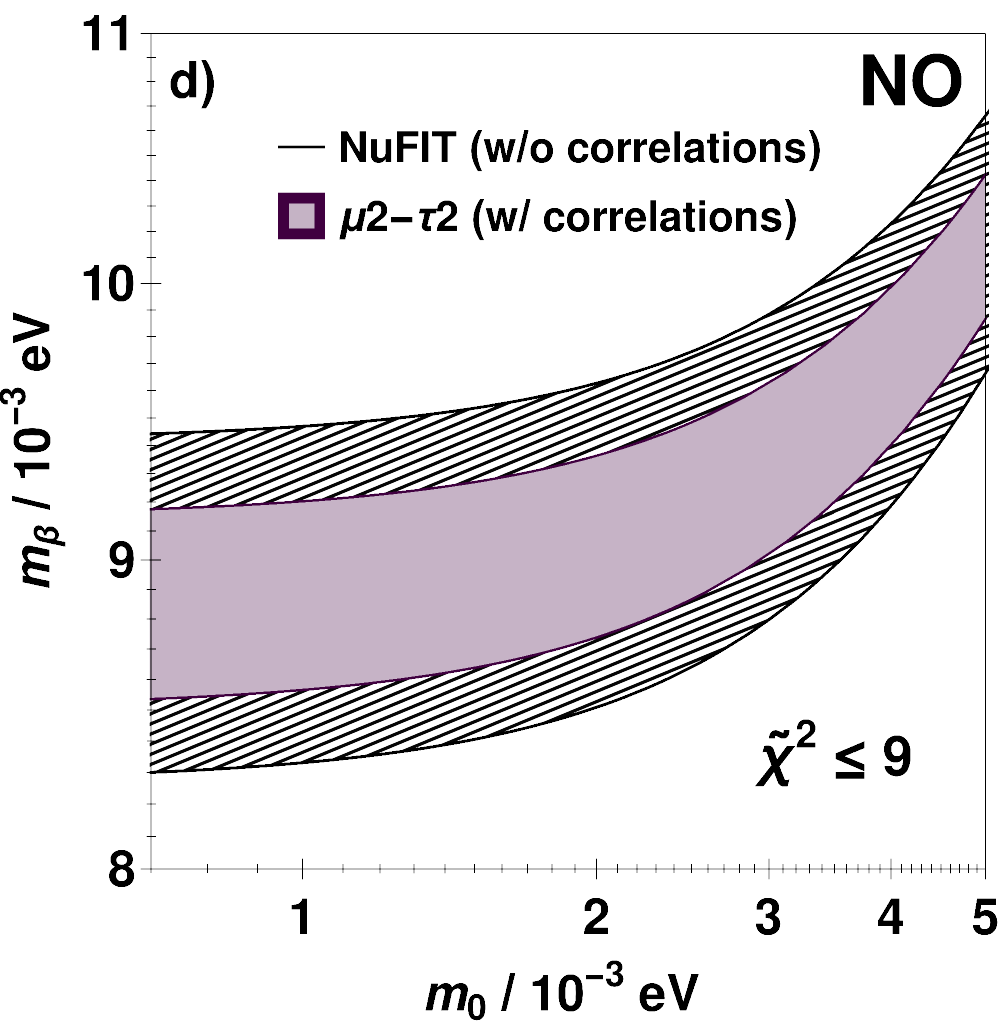}\\
    \includegraphics[width=0.24\linewidth]{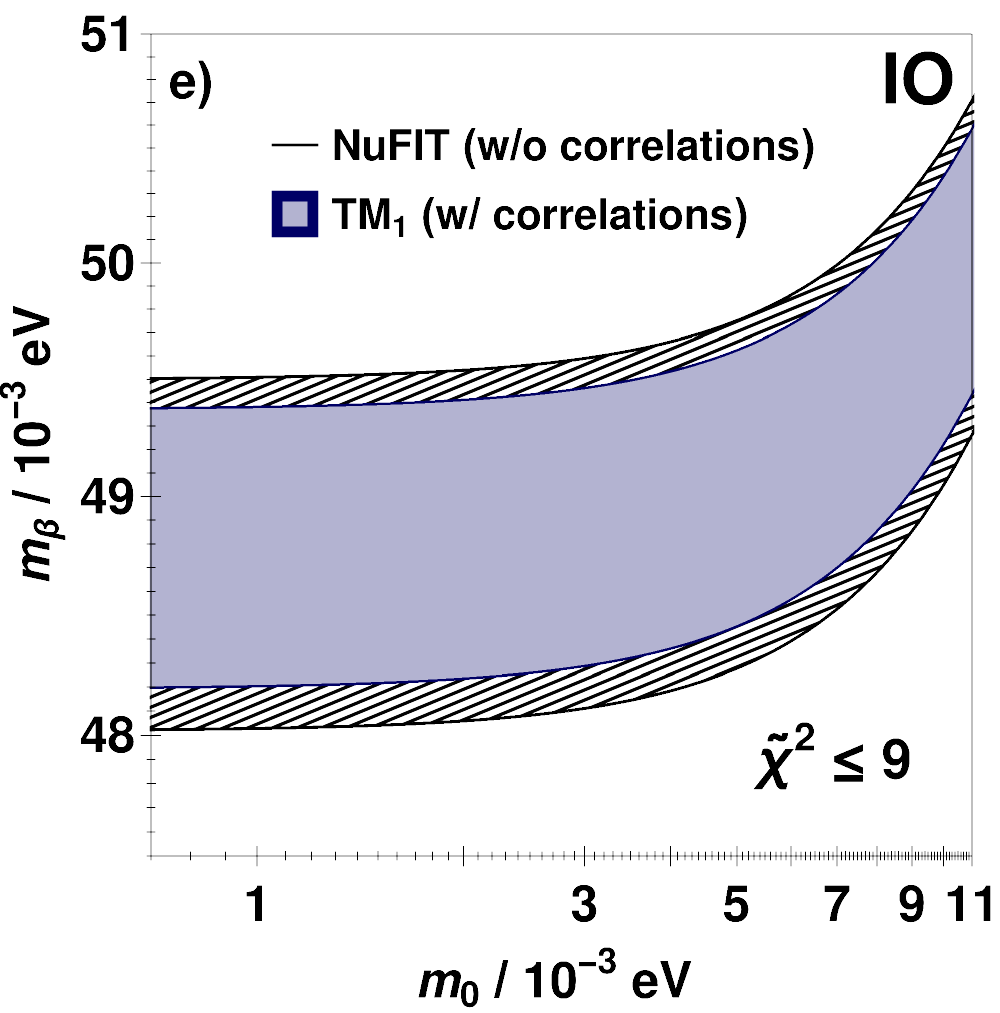}
    \includegraphics[width=0.24\linewidth]{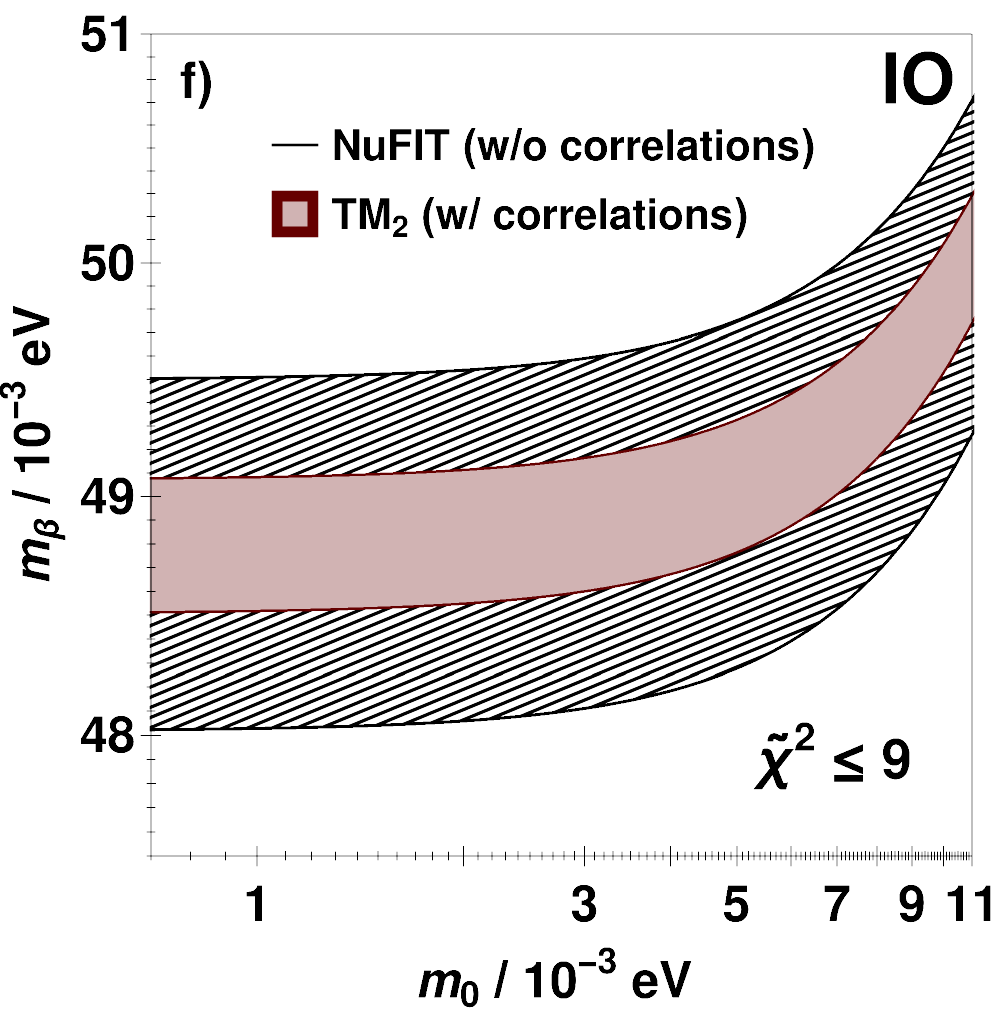}
    \includegraphics[width=0.24\linewidth]{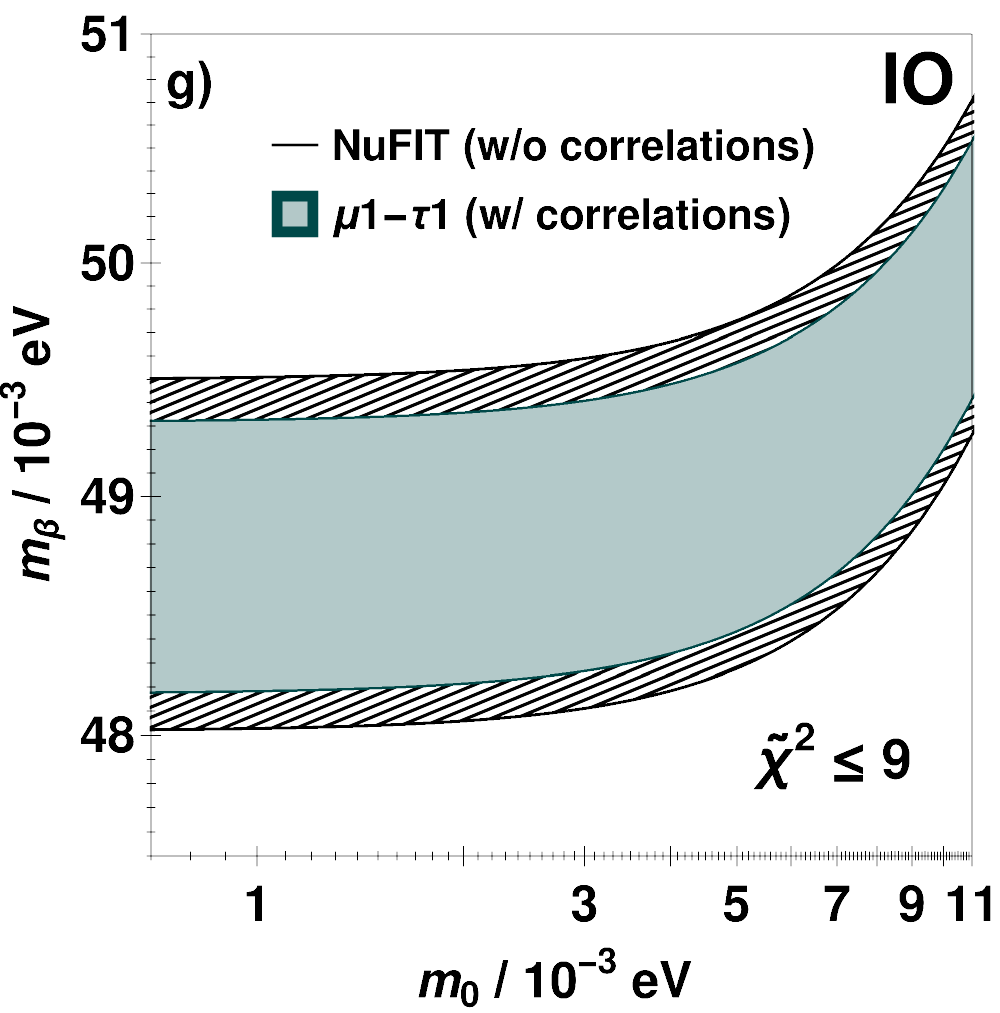}
    \includegraphics[width=0.24\linewidth]{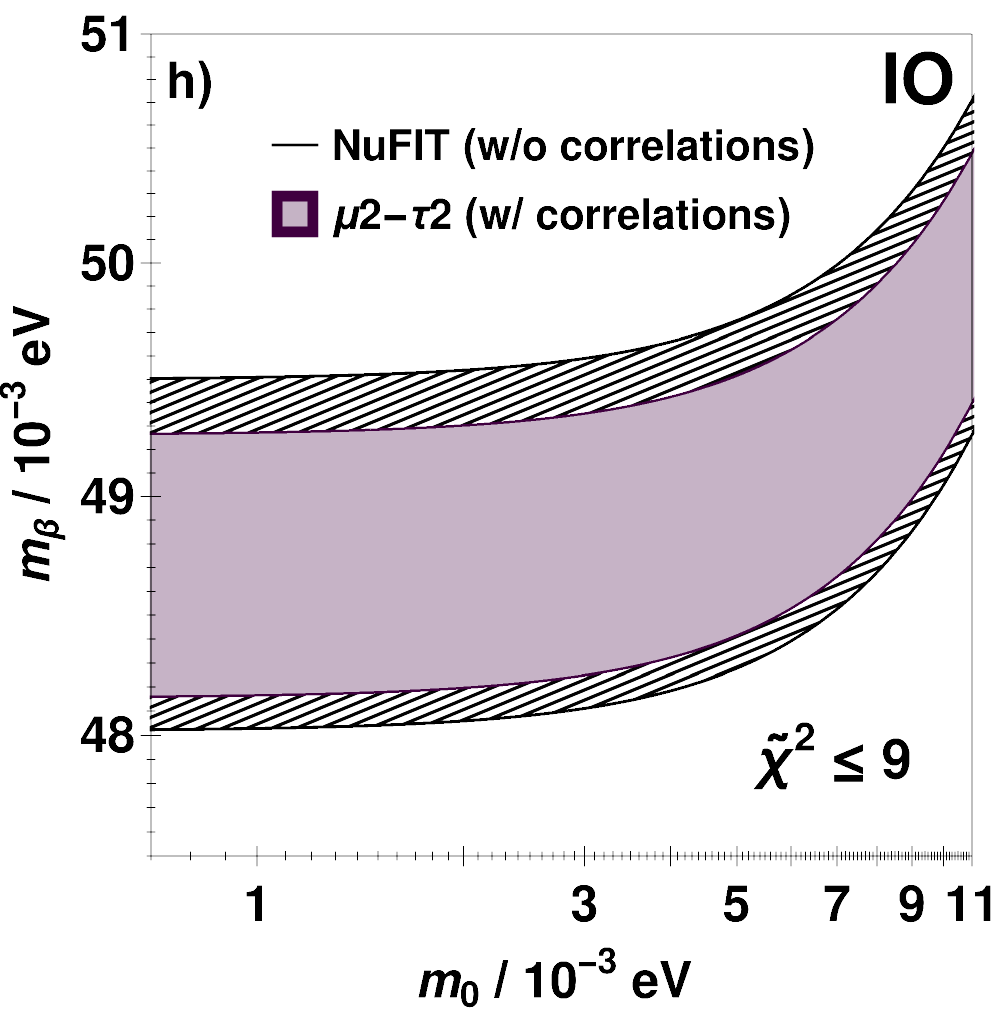}
    \caption{
    Impact of simultaneous analysis of neutrino mixing parameters on the effective Majorana mass $m_{\beta}$ as a function of the lightest neutrino mass for both mass orderings, incorporating NuFIT correlations and the mixing-parameter relations predicted by the considered discrete flavour models.
    }
    \label{fig:mb}
\end{figure}

 The shaded regions denote the allowed parameter space for $m_{\beta\beta}$ obtained by combining the NuFIT correlations with the mixing-scheme-specific relations given in Eqs.~\eqref{eq:tms12s13}-\eqref{eq:mutaudcps23s13s12-2}. The color scheme is described in Section~\ref{section:res2d}. The black solid contour corresponds to the range of $m_{\beta\beta}$ obtained by varying the oscillation parameters independently within their $3\sigma$ intervals (Table~\ref{table:nufit60}), i.e., without imposing either the NuFIT correlations or the mixing-scheme-specific relations. Excluded regions are indicated by hatching.
 
Clearly, accounting for the estimated correlations among all neutrino mixing parameters leads to significantly stronger constraints on $m_{\beta\beta}$. In particular, the allowed regions for the effective Majorana mass are substantially reduced for the $\mathrm{TM}_1$ and $\mathrm{TM}_2$ mixing schemes and for normal mass ordering at low values of the lightest neutrino mass, $m_0$. This narrowing of predictions is valuable for planning future neutrinoless double beta decay experiments such as nEXO~\cite{nEXO:2017nam,nEXO:2021ujk}, LEGEND~\cite{LEGEND:2017cdu,LEGEND:2021bnm}, and CUPID~\cite{CUPID:2019imh}.  For inverted ordering, the impact of correlations on $m_{\beta\beta}$ is more modest, primarily due to the relative insensitivity to mixing parameters at the higher absolute neutrino mass scales characteristic of this ordering.  

On the other hand, Fig.~\ref{fig:mb} shows the corresponding result for   $m_{\beta}$  as a function of the lightest neutrino mass $m_0$.
For normal mass ordering, $m_\beta$ shows the most significant constraints from correlations, particularly in the TM$_{1,2}$ scenarios. This has direct implications for next-generation experiments like KATRIN~\cite{KATRIN:2001ttj, KATRIN:2021dfa} and Project 8~\cite{Project8:2013trt}.  
The correlation-induced constraints are complementary to these direct mass measurements and can help break degeneracies in the interpretation of future observations.

\subsection{1D projections of model dependent neutrino mixing parameters \label{section:res1d}}

\begin{figure}[!h]
    \centering
     \includegraphics[width=0.75\textwidth]{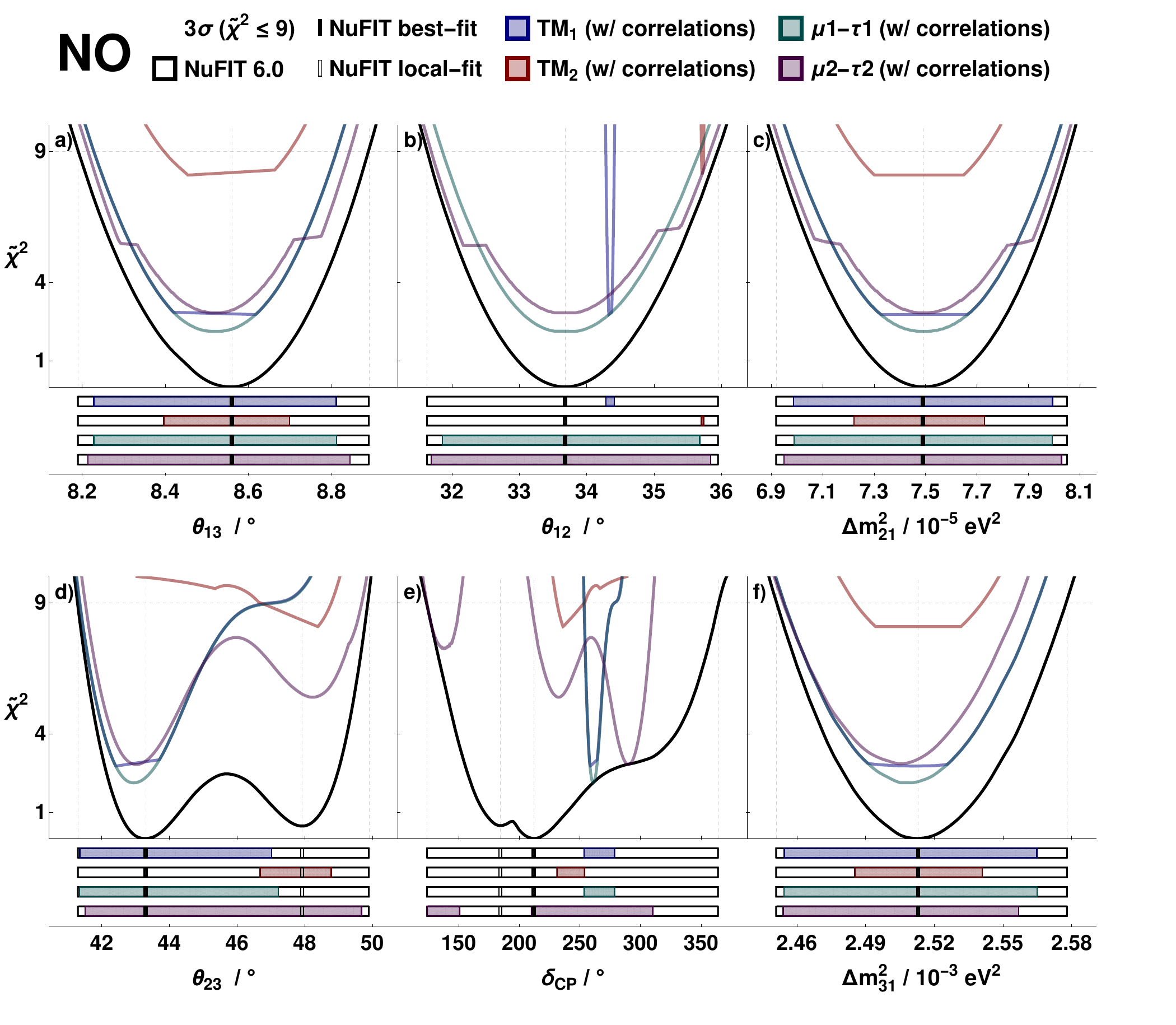}
     \caption{NuFIT (black), $\rm{TM}_{1}$ (blue), $\rm{TM}_{2}$ (red), $\rm{\mu1}$-$\rm{\tau1}$ (green)and $\rm{\mu2}$-$\rm{\tau2}$ (purple) reflection symmetries $\tilde{\chi}^2$ \textit{vs} oscillation parameters plots  and $3 \sigma$ ranges for NO.  
     } 
    \label{figure:paramno9chi}
\end{figure}

\begin{figure}[!h]
    \centering
     \includegraphics[width=0.75\textwidth]{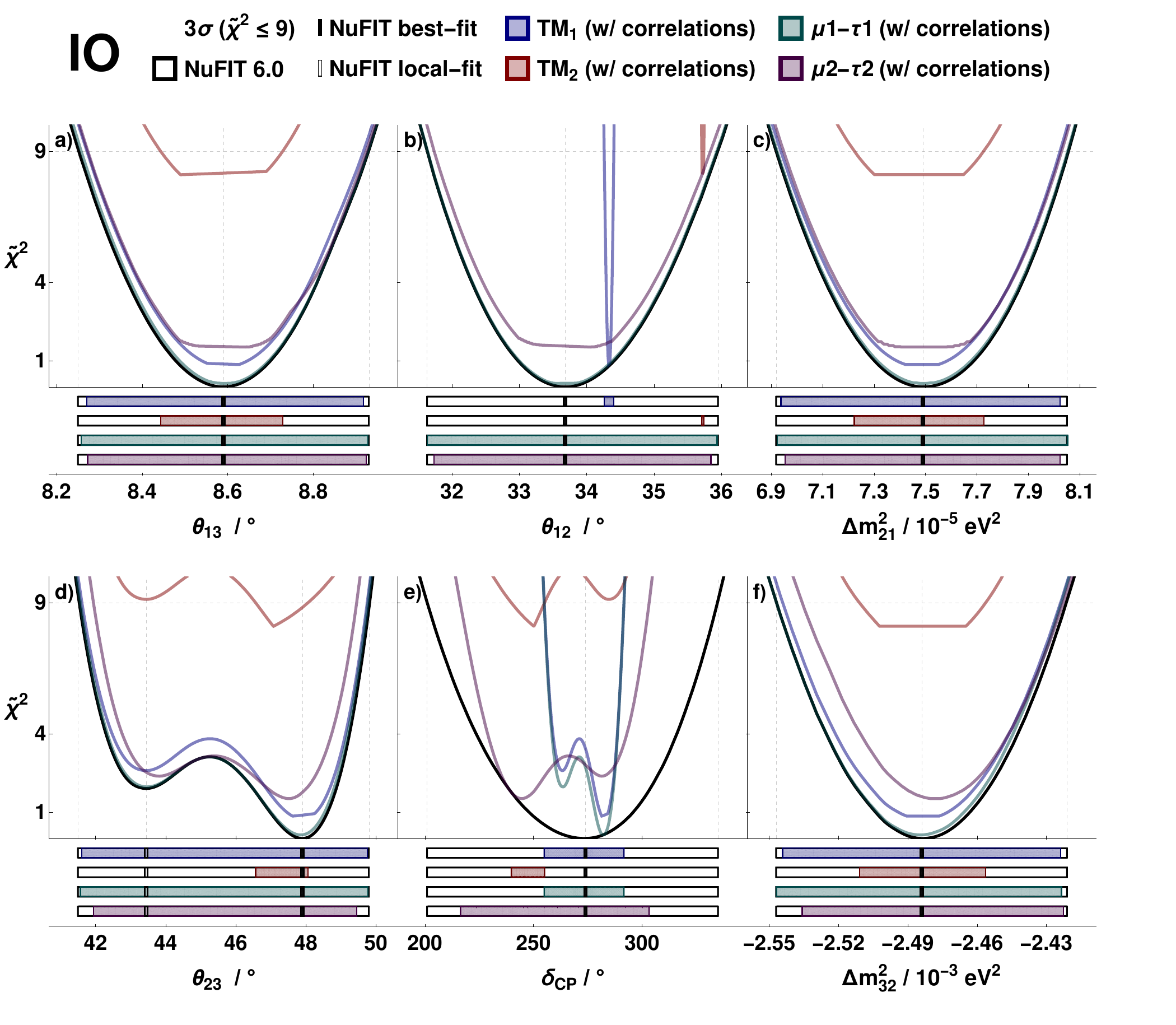}
     \caption{NuFIT (black), $\rm{TM}_{1}$ (blue), $\rm{TM}_{2}$ (red), $\rm{\mu1}$-$\rm{\tau1}$ (green)and $\rm{\mu2}$-$\rm{\tau2}$ (purple) reflection symmetries $\tilde{\chi}^2$ \textit{vs} oscillation parameters plots  and $3 \tilde\sigma$ ranges for IO.   
     } 
    \label{figure:paramio9chi}
\end{figure}

In Figs. \ref{figure:paramno9chi} (NO) and \ref{figure:paramio9chi} (IO) we present $\tilde\chi^2$  \textit{vs} neutrino oscillation parameters plots,  derived from NuFIT correlations and mixing-scheme-specific relations (Eqs. \eqref{eq:tms12s13}, \eqref{eq:tmdcps23s13} and \eqref{eq:mutaudcps23s13s12},\eqref{eq:mutaudcps23s13s12-2}),
for $\rm{TM}_{1}$, $\rm{TM}_{2}$,  $\mu1$-$\tau1$ and $\mu2$-$\tau2$ reflection symmetries, represented respectively by  blue, red, green  and purple lines.   $3 \tilde\sigma$ ranges of the parameters (1D projections of $\tilde\chi^2 \leq 9$ are given  as respectively colored bars.  Here, each curve shows, for every value of the plotted parameter, the lowest $\tilde\chi^2$ the mixing scheme can reach after optimizing over all other parameters.

As expected, the allowed regions for $\mathrm{TM}_{1(2)}$ are fully contained within those of the corresponding $\mu1(2)$--$\tau1(2)$ schemes, reflecting the fact that tri-bimaximal mixing is a special realization of the more general $\mu$--$\tau$ symmetry. Consequently, the $\tilde{\chi}^2(\delta_{\rm CP})$ distributions for the $\mu1$--$\tau1$ and $\mu2$--$\tau2$ schemes are bounded by the NuFIT distributions, while the $\tilde{\chi}^2(\theta_{12})$ distributions for $\mathrm{TM}_{1}$ and $\mathrm{TM}_{2}$ are bounded by those of the corresponding $\mu$--$\tau$ schemes. These contact points determine the mixing-scheme global minimum, denoted by $\tilde{\chi}^2_{\rm min}$.

For a given mixing scheme and mass ordering, $\tilde{\chi}^2_{\rm min}$ is common to all oscillation parameters, whereas the corresponding parameter values define the mixing-scheme best fits. Among the considered models, $\mu1$--$\tau1$ provides the best agreement with the NuFIT data for IO ($\tilde{\chi}^2_{\rm min}=0.1$), while $\mathrm{TM}_{2}$ gives the poorest fit for both NO and IO ($\tilde{\chi}^2_{\rm min}=8.1$).

The horizontal plateaus visible in several $\tilde{\chi}^2$ distributions have two distinct origins. The low-$\tilde{\chi}^2$ plateaus, present in the mass-squared difference distributions for all mixing schemes, arise because $\Delta m_{21}^2$ and $\Delta m_{31(32)}^2$ do not explicitly enter the mixing-scheme relations (Eqs.~\eqref{eq:tms12s13}, \eqref{eq:tmdcps23s13}, and \eqref{eq:mutaudcps23s13s12}, \eqref{eq:mutaudcps23s13s12-2}). Consequently, the corresponding best-fit values are intervals rather than unique points. The intermediate plateaus, clearly visible for example in the $\mu2$--$\tau2$ scheme for NO, originate from the rapid variation of the oscillation parameters around the local minimum of $\tilde{\chi}^2(\delta_{\rm CP})$, itself induced by the combined NuFIT $\delta_{\rm CP}$--$\theta_{23}$ correlations and the mixing-scheme relations.

For NO, the lower-octant preference of $\theta_{23}$ is strengthened in the $\mathrm{TM}_{1}$, $\mu1$--$\tau1$, and $\mu2$--$\tau2$ schemes, whereas only the upper octant remains allowed in $\mathrm{TM}_{2}$, shifting the NuFIT local best fit to the mixing-scheme best fit at $48.39^\circ$. For IO, the higher-octant preference is preserved in all mixing schemes, with only small shifts of the best-fit values.

The $\tilde{\chi}^2(\delta_{\rm CP})$ distributions exhibit a pronounced dependence on $\theta_{23}$ through the combined NuFIT correlations and the mixing-scheme relations. This leads to local minima, and hence mixing-scheme local best fits, for $\mu2$--$\tau2$ in NO and for $\mathrm{TM}_{1}$, $\mu1$--$\tau1$, and $\mu2$--$\tau2$ in IO. As a result, the preferred values of $\delta_{\rm CP}$ shift, the corresponding $3\tilde\sigma$ intervals become significantly narrower, and non-zero values of $\delta_{\rm CP}$ are generally favoured.

Finally, the relation between $\theta_{12}$ and $\theta_{13}$ in Eq.~\eqref{eq:tms12s13} produces noticeable shifts of the best-fit values and substantial reductions of the allowed $3\tilde\sigma$ ranges for $\theta_{13}$ in the   $\mathrm{TM}_{2}$ scheme. Smaller correlation-induced shifts, visible, for example, in the $\theta_{13}$ and $\Delta m_{31(32)}^2$ distributions, manifest as mild preferences for higher or lower parameter values.
In general, TM$_2$ is less preferred compared to TM$_1$. 
 
\begin{table}[h!]
 \centering
    \scriptsize
    \setlength{\tabcolsep}{3pt}   
    \begin{tabularx}{\textwidth}{c c c c c c c c}
    \hline \hline
             Parameter &  
             & &
         NuFIT 6.0
         &  
        $\rm{TM}_1$ 
        &  
        $\rm{TM}_2$ 
        &
        $\rm{\mu1}$-$\rm{\tau1}$
        &
        $\rm{\mu2}$-$\rm{\tau2}$
        \\
         \hline 
          $\theta_{13}/^{\circ}$
        & NO
        & $3\tilde\sigma$
        &$8.19 \xrightarrow{} 8.89$
        &$8.22 \xrightarrow{} 8.82$
        &${\bf 8.39 \xrightarrow{} 8.70}$
        &$8.22 \xrightarrow{} 8.82$
        &$8.21 \xrightarrow{} 8.85$
        \\
        & 
        & \textit{bf}
         &8.56
        &8.62   &{\bf 8.45}   &8.51   &8.53
        \\
        & IO
        & $3\tilde\sigma$
        &$8.25 \xrightarrow{} 8.93$
        &$8.27 \xrightarrow{} 8.92$
        &${\bf 8.44 \xrightarrow{} 8.73}$
        &$8.25 \xrightarrow{} 8.93$
        &$8.27 \xrightarrow{} 8.93$
        \\
         & 
         & \textit{bf}
          &8.59
        &8.63   &{\bf 8.49}     &8.60   &8.64
        \\
        \hline
        $\theta_{12}/^{\circ}$
        & NO
        & $3\tilde\sigma$
        &$31.63 \xrightarrow{} 35.95$
        &${\bf 34.28 \xrightarrow{} 34.42}$
        &${\bf 35.70 \xrightarrow{} 35.74}$
        &$31.85 \xrightarrow{} 35.68$
        &$31.69 \xrightarrow{} 35.84$
        \\
        & 
        & \textit{bf}
          &33.68
        &{\bf 34.33}   &{\bf 35.71}   &33.84   &33.84
        \\
        & IO
        & $3\tilde\sigma$
        &$31.63 \xrightarrow{} 35.95$
        &${\bf 34.26 \xrightarrow{} 34.41}$
        &${\bf 35.71 \xrightarrow{} 35.74}$
        &$31.63 \xrightarrow{} 35.94$
        &$31.73 \xrightarrow{} 35.85$
        \\
        & 
        & \textit{bf}
        &33.68
        &{\bf 34.33}   &{\bf 35.72}   &33.53   &34.06
        \\
        \hline
        $\theta_{23}/^{\circ}$
        & NO
        & $3\tilde\sigma$
        &$41.3 \xrightarrow{} 49.9$
        &$41.3 \xrightarrow{} 47.1$
        &${\bf 46.6 \xrightarrow{} 48.8}$
        &$41.3 \xrightarrow{} 47.3$
        &$41.5 \xrightarrow{} 49.7$
        \\
        & 
        & \textit{bf}
        &43.3
        &42.43   &{\bf 48.39}    &42.94   &43.00 
        \\
        & IO
        & $3\tilde\sigma$
        &$41.5 \xrightarrow{} 49.8$
        &$41.6 \xrightarrow{} 49.8$
        &${\bf 46.5 \xrightarrow{} 48.1}$
        &$41.5 \xrightarrow{} 49.8$
        &$41.9 \xrightarrow{} 49.5$
        \\
        & 
        & \textit{bf}
        &47.9
        &47.63   &{\bf 47.08}    &47.85   &47.53
        \\
        \hline
        $\delta_{\rm CP}/^{\circ}$
        & NO
        & $3\tilde\sigma$
        &$124 \xrightarrow{} 364$
        &${\bf 253.5 \xrightarrow{} 278.7}$
        &${\bf 231.1 \xrightarrow{} 254.0}$
        &${\bf 253.5 \xrightarrow{} 278.8 }$
        &${\bf 123.7 \xrightarrow{} 151.1~~ \&}$
        \\  
        &  
        &  
        & 
        &
        & 
        & 
        &${\bf   210.1 \xrightarrow{} 310.2}$
        \\   & 
        & \textit{bf}
        &212
        &{\bf 259}   &{\bf 236}   &{\bf 261}   &{\bf 290}   
        \\
        & IO
        & $3\tilde\sigma$
        &$201 \xrightarrow{} 335$
        &${\bf 254.9 \xrightarrow{} 291.7}$
        &${\bf 239.8 \xrightarrow{} 255.2}$
        &${\bf 254.9 \xrightarrow{} 291.8}$
        &${\bf 216.1 \xrightarrow{} 303.3}$
        \\
        & 
        & \textit{bf}
        &274
        &{\bf 281}   &{\bf 250}      &{\bf 282}   &{\bf 244}
        \\
        \hline
        $\Delta m^2_{21}/{10^{-5} \rm{eV}^2}$
        & NO
        & $3\tilde\sigma$
        &$6.92 \xrightarrow{} 8.05$
        &$6.98 \xrightarrow{} 8.00$
        &${\bf 7.22 \xrightarrow{} 7.73}$
        &$6.98 \xrightarrow{} 8.00$
        &$6.94 \xrightarrow{} 8.04$
        \\
        & 
        & \textit{bf}
        &$7.49$
        &$7.33 \xrightarrow{} 7.66$
        &${\bf 7.30 \xrightarrow{} 7.65}$
        &$7.48 \xrightarrow{} 7.51$
        &$7.47 \xrightarrow{} 7.51$
        \\
        & IO
        & $3\tilde\sigma$
        &$6.92 \xrightarrow{} 8.05$
        &$6.93 \xrightarrow{} 8.03$
        &${\bf 7.22 \xrightarrow{} 7.73}$
        &$6.92 \xrightarrow{} 8.05$
        &$6.94 \xrightarrow{} 8.03$
        \\
        & 
        & \textit{bf}
        &$7.49$
        &$7.42 \xrightarrow{} 7.55$
        &${\bf 7.30 \xrightarrow{} 7.65}$
        &$7.48 \xrightarrow{} 7.50$
        &$7.40 \xrightarrow{} 7.58$
        \\
        \hline
        $\Delta m^2_{31}/{10^{-3} \rm{eV}^2}$
        & NO
        & $3\tilde\sigma$
        &$2.451 \xrightarrow{} 2.578$
        &$2.454 \xrightarrow{} 2.565$
        &$2.485 \xrightarrow{} 2.541$
        &$2.454 \xrightarrow{} 2.565$
        &$2.454 \xrightarrow{} 2.557$
        \\
        &
        & \textit{bf}
        &$2.513$
        &$2.510 \xrightarrow{} 2.513$
        &$2.494 \xrightarrow{} 2.532$
        &$2.506 \xrightarrow{} 2.510$
        &$2.505$
        \\
        $\Delta m^2_{32}/{10^{-3} \rm{eV}^2}$
        & IO
        & $3\tilde\sigma$
        &$-2.547 \xrightarrow{} -2.421$
        &$-2.545 \xrightarrow{} -2.423$
        &$-2.511 \xrightarrow{} -2.456$
        &$-2.547 \xrightarrow{} -2.423$
        &$-2.536 \xrightarrow{} -2.422$
        \\
        & 
        & \textit{bf}
        &$-2.484$
        &$-2.491 \xrightarrow{} -2.476$
        &$-2.502 \xrightarrow{} -2.465$
        &$-2.485 \xrightarrow{} -2.484$
        &$-2.480 \xrightarrow{} -2.476$
        \\
        \hline
        $\tilde{\chi}^2_{min}$
        & NO
        &
        &0
        &$2.8$
        &$\color{red}\mathbf{8.1}$
        &$2.1$
        &$2.8$
        \\
        & IO
        &
        &0
        &$0.9$
        &$\color{red}\mathbf{8.1}$
        &$\color{blue}\mathbf{0.1}$
        &$1.5$
        \\
        \hline
        $\tilde\sigma_{min}$
        & NO
        &
        &$-$
        &$1.7$
        &$\color{red}\mathbf{2.8}$
        &$1.4$
        &$1.7$
        \\
        & IO
        &
        &$-$
        &$0.9$
        &$\color{red}\mathbf{2.8}$
        &$\color{blue}\mathbf{0.3}$
        &$1.2$
        \\
        \hline
        $\tilde{p}$
        & NO
        &
        &$-$
        & 0.10 
        & $\color{red}\mathbf{0.004}$  
        & 0.15 
        & 0.09  
        \\
        & IO
        &
        &$-$
        & 0.35  
        & $\color{red}\mathbf{0.004}$  
        & $\color{blue}\mathbf{0.71}$  
        & 0.22 
        \\
        \hline \hline
    \end{tabularx}
    \caption{
    NuFIT v6.0, $\mathrm{TM}_{1(2)}$ and $\mu{1(2)}$--$\tau{1(2)}$ reflection symmetry $3\tilde{\sigma}$ ranges, and \textit{bfs}. The boldfaced and colored entries highlight the most relevant results; see the discussion in the text. For the neutrino masses, the \textit{bfs} are given as ranges corresponding to the same $\tilde{\chi}^2_{\min}$, i.e., all mass-parameter values within a given range provide an equally good fit.}  
\label{tab:param_data}
\end{table}

In Table~\ref{tab:param_data}, we summarize the derived constraints on the neutrino mixing parameters. As can be seen, by taking into account relations among the model parameters and exploiting the complete 6D global-fit information, we obtain significantly more stringent constraints than those derived from the NuFIT~6.0 data alone. The boldface entries indicate the most stringent constraints discussed above.
For the purpose of evaluating the consistency between the models and the experimental data, the key discriminator is the $\tilde\chi^2_{min}$, which is closely connected to the goodness-of-fit.

From a physics standpoint, the key quantities are the values of 
$\tilde{\chi}^2_{\min}$, which indicates how well each model reproduces 
the global-fit data. In Table~\ref{tab:param_data}, the best-fit result 
is highlighted in blue for the $\mu1$--$\tau1$ model in the IO case, 
while the weakest fit is shown in red for TM$_2$. The same color-coding 
convention is adopted in the next section for Table~\ref{tab:chimin}, 
where we present the results based on the NuFIT~6.1 data.
The $3\tilde{\sigma}$ ranges are provided to illustrate how the 
different model assumptions affect the allowed ranges of the neutrino 
oscillation parameters. More importantly, they demonstrate the stronger 
determination of $\delta_{\rm CP}$ and, for the TM$_1$ and TM$_2$ cases, 
of $\theta_{12}$, as well as an enhanced preference for a particular 
$\theta_{23}$ octant (see Figs.~\ref{figure:paramno9chi} and 
\ref{figure:paramio9chi}).
In addition, we provide two auxiliary quantities in Table~\ref{tab:chimin}: the goodness-of-fit 
probability, $\tilde{p}$, and the corresponding significance, 
$\tilde{\sigma}_{\min}$, expressed in units of standard deviations and 
obtained assuming a $\chi^2$ distribution with one degree of freedom. 
The goodness-of-fit probability is defined as
$\tilde{p}=P\left(\chi^2 \geq \tilde{\chi}^2_{\min}\right)$.
Thus, $\tilde{p}$ represents the probability of obtaining a $\chi^2$ 
value at least as large as the observed minimum, $\tilde{\chi}^2_{\min}$. 
We emphasize that $\tilde{p}$ should not be interpreted as the 
probability that the model is correct. Rather, it is an auxiliary 
statistical measure of the goodness of fit, allowing us to quantify and 
compare how well the different models reproduce the global-fit data.
The quantity $\tilde{\sigma}_{min}$ is evaluated under the same assumptions and measures the separation, in units of standard deviations, between the oscillation-parameter values corresponding to $\tilde{\chi}^2_{min}$ (under the constraints of the considered model) and the global-fit best-fit point in the oscillation-parameter space.

\section{Effects of NuFIT 6.1 with JUNO} 
\label{section:nufit61}

In the previous sections, we presented 2- and 1-D projections of the results obtained for combined mixing-scheme relations and estimated correlations derived from global fits. The presented regions and intervals change with each successive global-fit dataset, and the corresponding calculations are computationally intensive, see \cite{GitHubCorrelations}.  In Tab.~\ref{tab:chimin}, we report $\tilde\chi^2_{min}$ for the NuFIT v6.0 and v6.1 data combined with models under consideration. 

\begin{table}[!h]
\begin{tabular}{llcc|cc|cc|cc}
\hline
 & & \multicolumn{2}{c|}{TM$_1$} &
\multicolumn{2}{c|}{TM$_2$} &
\multicolumn{2}{c|}{$\mu1$-$\tau1$} &
\multicolumn{2}{c}{$\mu2$-$\tau2$} \\
 NuFIT & & v6.0 & v6.1 & v6.0 & v6.1 & v6.0 & v6.1 & v6.0 & v6.1 \\
\hline
\multirow{2}{*}{$\tilde{\chi}^2_{min}$}
& NO & 2.77 & 3.88 & 8.09 & $\color{red}\mathbf{21.42}$ & 2.12 & 2.35 & 2.83 & 3.16 \\
& IO & 0.86 & 1.86 & 8.10 & $\color{red}\mathbf{21.43}$ & 0.13 & $\color{blue}\mathbf{0.12}$ & 1.52 & 1.59 \\
\hline
\multirow{2}{*}{$\tilde{p}$}
& NO & 0.10 & 0.05 & 0.004 & $\color{red}\mathbf{0.0001}$ & 0.15 & 0.12 & 0.09 & 0.08 \\
& IO & 0.35 & 0.17 & 0.004 & $\color{red}\mathbf{0.0001}$ & 0.71 & $\color{blue}\mathbf{0.73}$ & 0.22 & 0.21 \\
\hline
\multirow{2}{*}{$\tilde{\sigma}_{min}$}
& NO & 1.66 & 1.97 & 2.84 & $\color{red}\mathbf{4.63}$ & 1.46 & 1.53 & 1.68 & 1.78 \\
& IO & 0.93 & 1.36 & 2.85 & $\color{red}\mathbf{4.63}$ & 0.37 & $\color{blue}\mathbf{0.35}$ & 1.23 & 1.26 \\
\hline
\end{tabular}
\caption{Smallest $\tilde{\chi}^2$ values attained in multidimensional osciallation-parameter space and the goodness-of-fit. Similarly as in the previouse table, the best-fit result is
highlighted in blue while the weakest fit is shown in
red.} \label{tab:chimin} 
\end{table}

\begin{figure}[!h]
    \centering
      \includegraphics[width=0.35\textwidth]{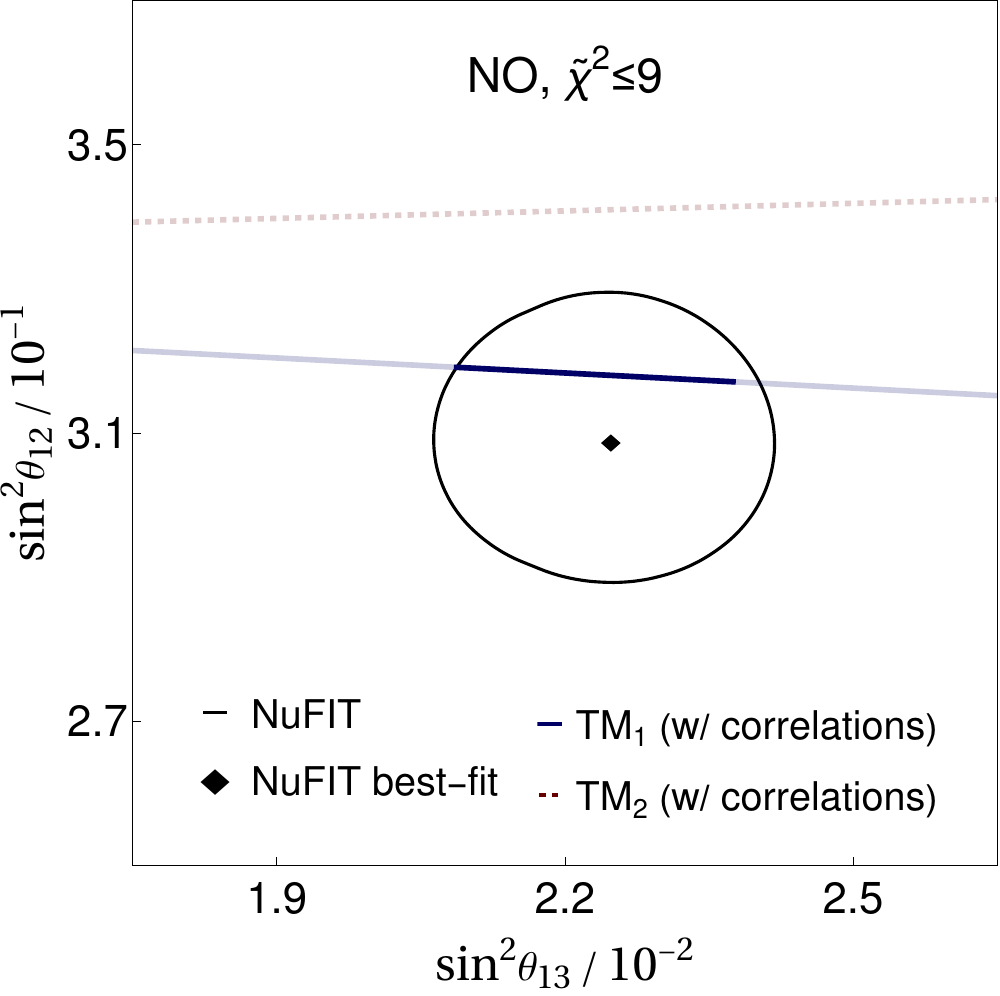}
      \includegraphics[width=0.35\textwidth]{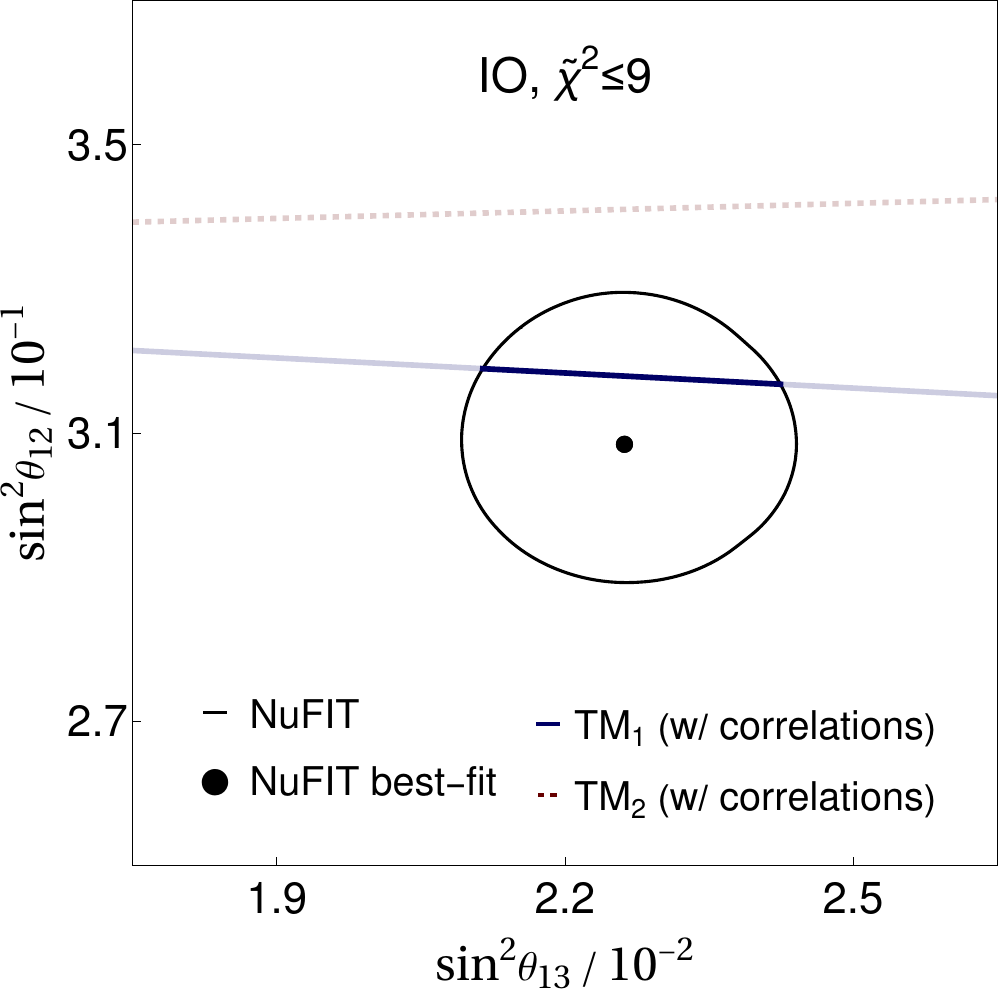}
       \caption{
      In the $\theta_{13}$--$\theta_{12}$ plane, for the NuFIT 6.1 data with $\tilde\chi^2 \leq 9$, the correlated TM$_1$ result is shown by the solid blue line, while the light-blue shaded line represents the theoretical TM$_1$ prediction. The theoretical TM$_2$ prediction is shown by the red dashed line. The plot clearly shows that the theoretical TM$_2$ prediction lies outside the region enclosed by the black contour representing the NuFIT data. No correlated TM$_2$ solutions satisfy the $\tilde\chi^2 \leq 9$ criterion; therefore, this result is not shown. Only the uncorrelated theoretical TM$_2$ prediction is displayed. 
      } 
    \label{figure:t12t13v61}
\end{figure}

\begin{figure}[!h]
    \centering
     \includegraphics[width=0.35\textwidth]{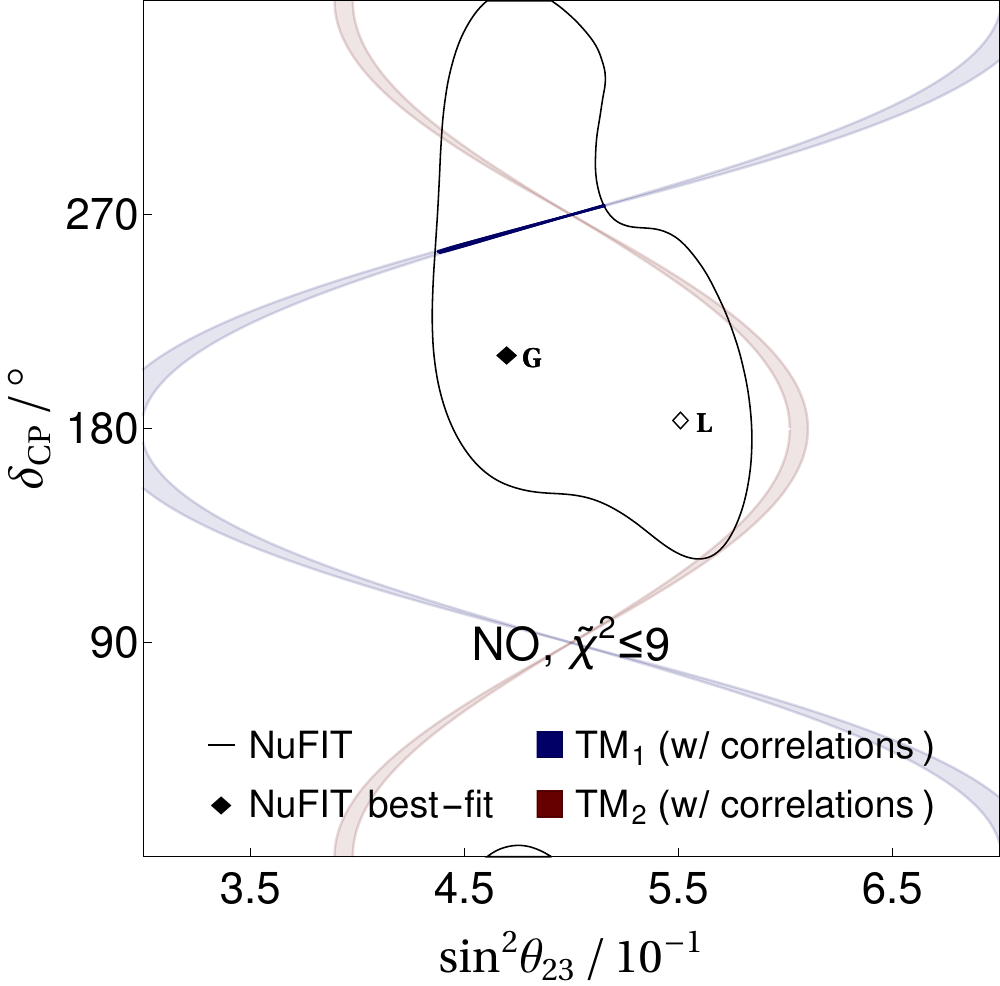}
     \includegraphics[width=0.35\textwidth]{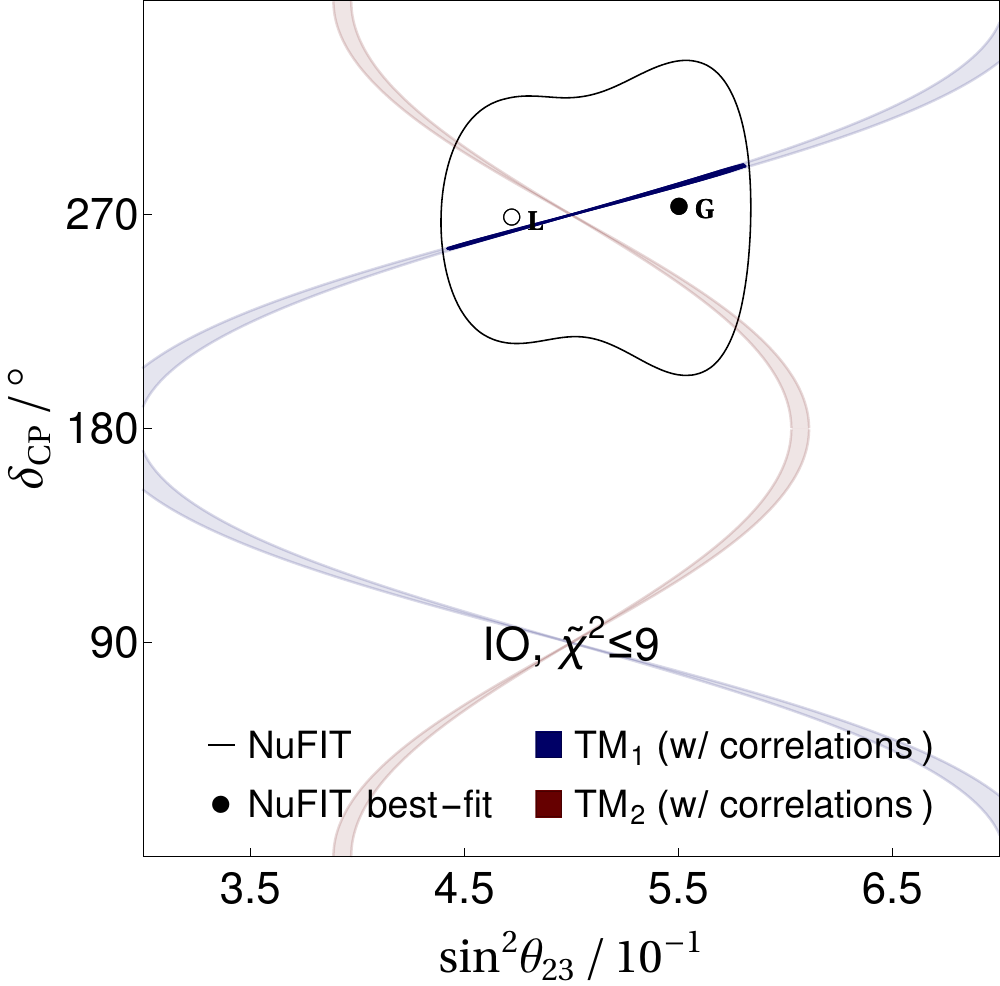}
     \caption{
     For the NuFIT 6.1 data in $\theta_{23}$--$\delta_{\rm CP}$ plane, TM$_1$ and TM$_2$ are shown by the light-blue and light-red shaded regions, respectively. The correlated TM$_1$ result is shown by the dark-blue shaded region. As discussed above, no correlated TM$_2$ solutions satisfy the $\tilde\chi^2 \leq 9$ criterion; hence, the corresponding dark-red region is absent.
     }
    \label{figure:tmdcpt23v61}
\end{figure}

\begin{figure}[!h]
    \centering
      \includegraphics[width=0.35\textwidth]{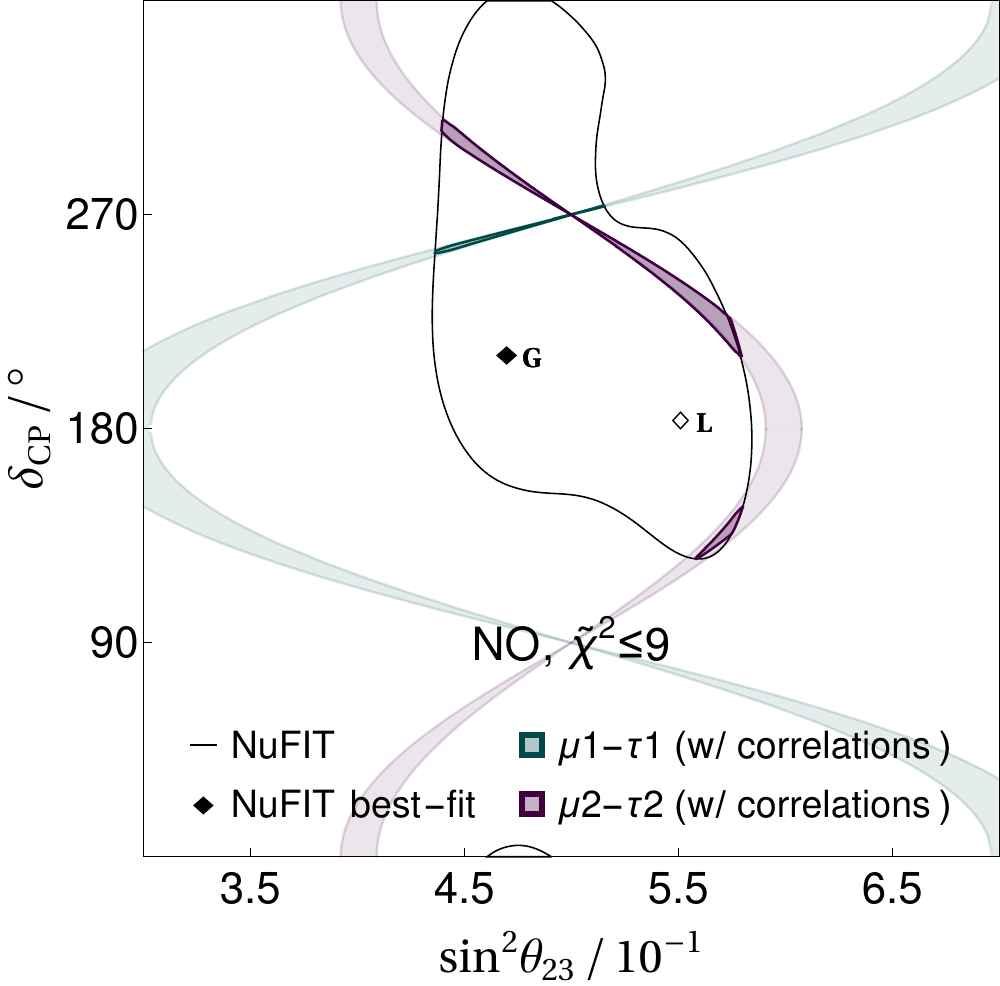}
      \includegraphics[width=0.35\textwidth]{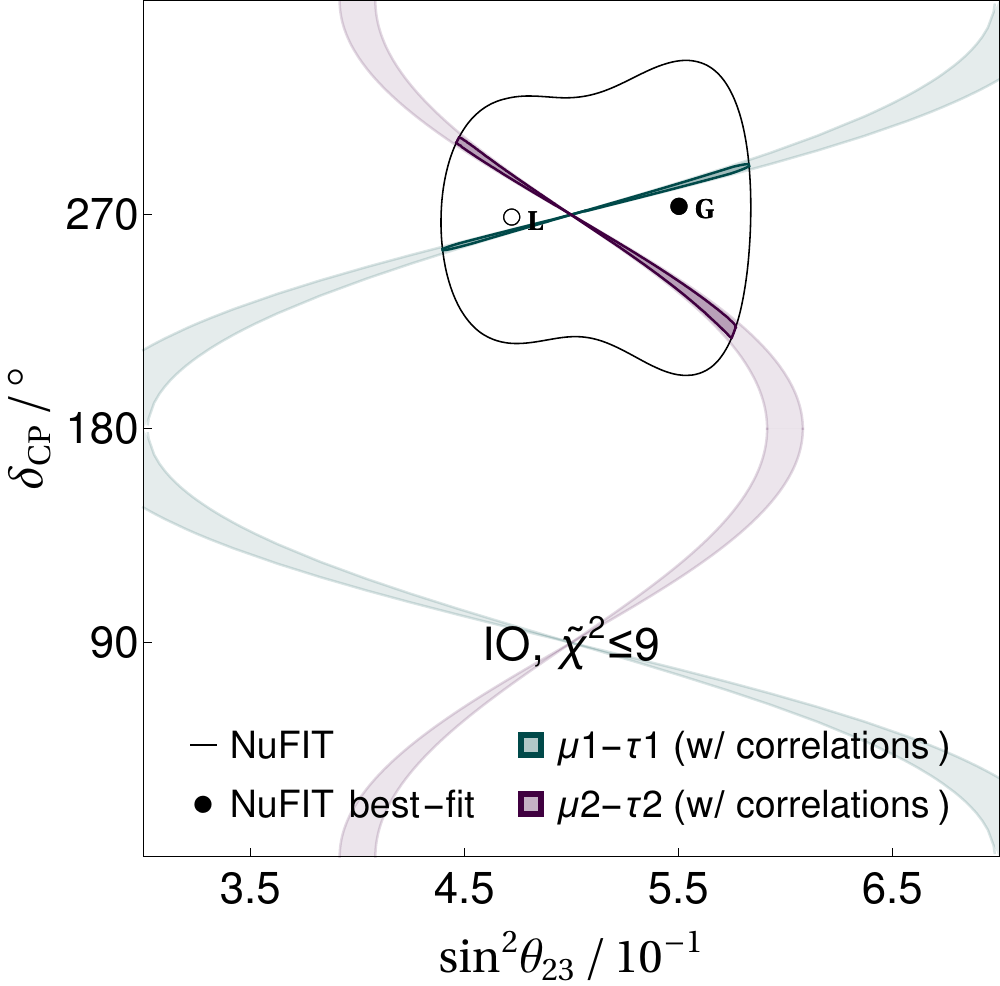}
        \caption{
        For the NuFIT 6.1 data in $\theta_{23}$--$\delta_{\rm CP}$ plane, the $\mu1$--$\tau1$ and $\mu2$--$\tau2$ results are shown in green and purple, respectively. Both results are present. When correlations are included, the darker shaded regions largely overlap with the corresponding lighter theoretical regions, indicating good agreement between the model predictions and the experimental correlated data. 
        }
    \label{figure:mutaudcpt23v61}
\end{figure}

The results show an overall strengthening of the preference patterns discussed earlier. In particular, TM$_2$ now lies outside the criterion $\tilde\chi^2 \leq 9$. Consequently, the red colored, dark-shaded  regions are absent in Figs.~\ref{figure:t12t13v61} and \ref{figure:tmdcpt23v61} which have been obtained for NuFIT 6.1. Interestingly, for the most recent dataset, TM$_2$ is strongly disfavored for both mass orderings and all the considered models (except $\mu1$-$\tau1$) also experience an overall increase in disfavoring. The notable exception is the $\mu1$-$\tau1$ scenario (for IO), where  $\tilde\chi^2_{min}$ is lower than for the previous datasets, pointing to an improved fit ($\tilde\chi^2_{min}$ reduced $0.13 \rightarrow 0.12$).
Fig. \ref{figure:mutaudcpt23v61} reveals a reduction of the dark-shaded regions compared with the earlier global-fit results for both mass orderings. The reduction is consistent with improved precision of the global-fit data. The stronger model rejection is rather corroborated by the corresponding changes in $\tilde\chi^2_{min}$ (related to parameter combinations, see Table \ref{tab:chimin}). The $\tilde\chi^2_{min}$ changes disfavor more decisively all models except  $\mu1$-$\tau1$ (IO).

\section{Summary and Conclusions}
\label{section:conc}

In this work, we confront neutrino mixing-schemes which originates from theoretical flavor models with experimental neutrino oscillation data, going beyond the standard 1D parameter ranges from global fits. We present multidimensional analyses that incorporate full correlations using the $\tilde\chi^2$ function constructed in Section~\ref{section:method}, thereby improving the accuracy of model viability assessments. 

Our approach offers methodological advantages where (i) we construct a conservative estimate of the full multi-parameter $\tilde\chi^2$ surface that preserves the complete correlations based on the available global fits; (ii) test individual points in the neutrino parameter space where each point can be evaluated independently for both model compliance (e.g., predicted relations in TM$_{1,2}$ mixing-schemes) and compatibility with global experimental data; 
(iii) for any point satisfying a given mixing-schemes's predictive relations, we calculate $\tilde\chi^2$ where the minimum value, $\tilde\chi^2_{\text{min}}$, determines the best achievable agreement for that model, naturally defining the allowed parameter regions at any desired confidence level; (iv) the methodology is model-independent and applies to any theoretical mixing scheme or sum rule with predictive parameter relations. To facilitate community use, we provide a Mathematica package containing these tools \cite{GitHubCorrelations}. 
\\
 
To illustrate the advantages of the proposed methodology, as examples,  we apply this framework to tri-bimaximal mixing variations ($\text{TM}_1$, $\text{TM}_2$), and to $\mu$--$\tau$ reflection symmetries ($\mu1$--$\tau1$, $\mu2$--$\tau2$),  yielding several noteworthy phenomenological insights:
\begin{itemize}
    \item[(i)] {Model Preference ($\text{TM}_1$ vs.\ $\text{TM}_2$ vs.\ $\mu1(2)$--$\tau1(2)$):} The updated determination of $\theta_{12}$ creates severe tension with the $\text{TM}_2$ mixing scheme, whose predicted relation between $\theta_{12}$ and $\theta_{13}$ results in a significantly higher $\tilde\chi^2_{\text{min}}$. Consequently, $\text{TM}_2$ is tightly constrained by NuFIT~v6.0 data (Table~\ref{tab:param_data}) and strongly disfavored by NuFIT~v6.1 (Table~\ref{tab:chimin}), whereas $\text{TM}_1$ remains fully compatible with NuFIT correlations, yielding $\tilde\chi^2_{\text{min}}$ values close to the global minimum. Additionally, $\mu$--$\tau$ reflection symmetries remain valuable models, with a slight preference for $\mu1$--$\tau1$ reflection symmetry.

The extent to which the constructed $\tilde\chi^2$ restricts neutrino parameters within specific models can already be seen in the 2D projections in Fig.~\ref{figure:tm9chi}. Incorporating full correlations cleanly leads to much narrower parameter ranges across multiple projections. In particular, Fig.~\ref{figure:expth} illustrates how mixing-scheme constraints restrict global-fit parameters (left panel), while fully incorporating correlations leads to a significantly tighter range of oscillation parameters (right panel). Furthermore, implementing NuFIT~v6.1 with JUNO data makes the impact of correlations even stronger, as demonstrated in Figs.~\ref{figure:tmdcpt23v61} and \ref{figure:mutaudcpt23v61}.

    \item[(ii)] {Narrowing of the allowed oscillation-parameter regions:} Unlike the standard procedure, which relies on the overlap of independent regions inferred from mixing-scheme relations and global-fit constraints, the correlation-based method yields considerably narrower allowed regions. This improvement is particularly evident in Fig.~\ref{figure:expth}, as well as in panels (c), (e), (f), (j), (l), and (m) of Figs.~\ref{figure:tm9chi}  and \ref{figure:mutau9chi}.

    \item[(iii)] {Impact on Mass Observables ($m_\beta$ and $m_{\beta\beta}$):} Because correlated mixing-angle and mass-splitting data feed directly into neutrino mass sum rules, our correlation-respecting $\tilde\chi^2$ function restricts $m_\beta$ and $m_{\beta\beta}$ for neutrinoless double beta decay ($0\nu\beta\beta$). We observe that $m_\beta$ is most strongly affected under normal ordering (NO), and substantial reductions in $m_{\beta\beta}$ occur for both $\text{TM}_1$ and $\text{TM}_2$ at the lowest neutrino mass ($m_0$).

    \item[(iv)] {Constraints on Mass-Squared Differences:} Imposing NuFIT correlations inside $\tilde\chi^2$ yields the unique feature of constraining allowed regions in the solar and atmospheric mass-splitting planes ($\Delta m_{21}^2$, $\Delta m_{31(32)}^2$), even though these parameters are not explicitly part of the mixing-scheme symmetry relations (Eqs.~\eqref{eq:tms12s13}-\eqref{eq:mutaudcps23s13s12-2}). This effect is explicitly visible in Figs.~\ref{figure:tm9chi} and \ref{figure:mutau9chi}.

    \item[(v)] {Octant and Phase Isolation:} The method cleanly isolates the impact of flavor symmetries on the $\theta_{23}$ octant preference and the $\delta_{\text{CP}}$ parameter space.
\end{itemize}

Once built, the $\tilde\chi^2$ methodology serves as a ready-made, correlation-respecting module for any mixing scheme or sum rule within broader global fit packages. As neutrino physics enters the precision era, accounting for experimental correlations will become increasingly pivotal. Smaller uncertainties from next-generation experiments, such as DUNE, T2HK, and JUNO, will amplify these correlations, potentially leading to separation of models whose allowed regions currently overlap. As demonstrated here for the $\text{TM}_2$ mixing scheme, incorporating full correlation structures can shift a model from marginal compatibility to exclusion at $>3\sigma$, making robust statistical frameworks essential for unravelling the underlying flavor dynamics.

\section*{Acknowledgements}

We thank Monojit Ghosh, and Federico S\'anchez for useful remarks. 
This work has been supported in part by the Polish National Science Center (NCN) under grant 2020/37/B/ ST2/02371, the Research Excellence Initiative of the University of
Silesia in Katowice and the Swiss National Science Foundation (SNSF) under grant MAPS
IZ11Z0 230193.

\appendix

\bibliographystyle{elsarticle-num-ID}
\bibliography{bibliography}

@article{Duda:2025vmc,
    author = "Duda, Jaros{\l}aw and Gluza, Janusz and Karmakar, Biswajit",
    title = "{CP-conserving SO(3) parameterization of the neutrino mixing matrix}",
    eprint = "2509.25328",
    archivePrefix = "arXiv",
    primaryClass = "hep-ph",
    doi = "10.1016/j.physletb.2026.140219",
    journal = "Phys. Lett. B",
    volume = "874",
    pages = "140219",
    year = "2026"
}

@article{Agarwalla:2013ju,
    author = "Agarwalla, Sanjib Kumar and Prakash, Suprabh and Sankar, S. Uma",
    title = "{Resolving the octant of theta23 with T2K and NOvA}",
    eprint = "1301.2574",
    archivePrefix = "arXiv",
    primaryClass = "hep-ph",
    reportNumber = "IFIC-12-85",
    doi = "10.1007/JHEP07(2013)131",
    journal = "JHEP",
    volume = "07",
    pages = "131",
    year = "2013"
}

@article{T2K:2025wet,
    author = "Abubakar, S. and others",
    collaboration = "T2K, NOvA",
    title = "{Joint neutrino oscillation analysis from the T2K and NOvA experiments}",
    eprint = "2510.19888",
    archivePrefix = "arXiv",
    primaryClass = "hep-ex",
    reportNumber = "FERMILAB-PUB-25-0132-PPD",
    doi = "10.1038/s41586-025-09599-3",
    journal = "Nature",
    volume = "646",
    number = "8086",
    pages = "818--824",
    year = "2025"
}

@article{Barger:2001yr,
    author = "Barger, V. and Marfatia, D. and Whisnant, K.",
    title = "{Breaking eight fold degeneracies in neutrino CP violation, mixing, and mass hierarchy}",
    eprint = "hep-ph/0112119",
    archivePrefix = "arXiv",
    reportNumber = "AMES-HET-01-12, BUHEP-01-32, MADPH-01-1250",
    doi = "10.1103/PhysRevD.65.073023",
    journal = "Phys. Rev. D",
    volume = "65",
    pages = "073023",
    year = "2002"
}

@article{Ge:2011qn,
    author = "Ge, Shao-Feng and Dicus, Duane A. and Repko, Wayne W.",
    title = "{Residual Symmetries for Neutrino Mixing with a Large $\theta_{13}$ and Nearly Maximal $\delta_D$}",
    eprint = "1108.0964",
    archivePrefix = "arXiv",
    primaryClass = "hep-ph",
    doi = "10.1103/PhysRevLett.108.041801",
    journal = "Phys. Rev. Lett.",
    volume = "108",
    pages = "041801",
    year = "2012"
}

@article{Ganguly:2023jml,
    author = "Ganguly, Joy and Gluza, Janusz and Karmakar, Biswajit and Mahapatra, Satyabrata",
    title = "{Phenomenology of the flavor symmetric scoto-seesaw model with dark matter and TM1 mixing}",
    eprint = "2311.15997",
    archivePrefix = "arXiv",
    primaryClass = "hep-ph",
    doi = "10.1103/PhysRevD.110.035012",
    journal = "Phys. Rev. D",
    volume = "110",
    number = "3",
    pages = "035012",
    year = "2024"
}

@article{Chulia:2025nsa,
    author = "Chuli{\'a}, Salvador Centelles and Kumar, Ranjeet",
    title = "{Minimal \(A_{4}\) type-II seesaw realization of testable neutrino mass sum rules}",
    eprint = "2512.22343",
    archivePrefix = "arXiv",
    primaryClass = "hep-ph",
    doi = "10.1103/ndvc-1vpl",
    journal = "Phys. Rev. D",
    volume = "113",
    number = "5",
    pages = "055023",
    year = "2026"
}

@article{CentellesChulia:2023osj,
    author = "Centelles Chuli{\'a}, Salvador and Kumar, Ranjeet and Popov, Oleg and Srivastava, Rahul",
    title = "{Neutrino mass sum rules from modular A4 symmetry}",
    eprint = "2308.08981",
    archivePrefix = "arXiv",
    primaryClass = "hep-ph",
    doi = "10.1103/PhysRevD.109.035016",
    journal = "Phys. Rev. D",
    volume = "109",
    number = "3",
    pages = "035016",
    year = "2024"
}

@article{Sugama:2023sjf,
    author = "Sugama, Sho and Yasuda, Osamu",
    title = "{Octant degeneracy and plots of parameter degeneracy in neutrino oscillations revisited}",
    eprint = "2308.15071",
    archivePrefix = "arXiv",
    primaryClass = "hep-ph",
    doi = "10.1103/PhysRevD.109.035007",
    journal = "Phys. Rev. D",
    volume = "109",
    number = "3",
    pages = "035007",
    year = "2024"
}

@article{JUNO:2025gmd,
    author = "Abusleme, Angel and others",
    collaboration = "JUNO",
    title = "{Measurement of reactor neutrino oscillation with the first JUNO data}",
    eprint = "2511.14593",
    archivePrefix = "arXiv",
    primaryClass = "hep-ex",
    doi = "10.1038/s41586-026-10538-z",
    journal = "Nature",
    volume = "654",
    number = "8118",
    pages = "343--348",
    year = "2026"
}

@article{Gehrlein:2020jnr,
    author = "Gehrlein, J. and Spinrath, M.",
    title = "{Leptonic Sum Rules from Flavour Models with Modular Symmetries}",
    eprint = "2012.04131",
    archivePrefix = "arXiv",
    primaryClass = "hep-ph",
    doi = "10.1007/JHEP03(2021)177",
    journal = "JHEP",
    volume = "03",
    pages = "177",
    year = "2021"
}

@article{Antusch:2007rk,
    author = "Antusch, S. and Huber, P. and King, S. F. and Schwetz, T.",
    title = "{Neutrino mixing sum rules and oscillation experiments}",
    eprint = "hep-ph/0702286",
    archivePrefix = "arXiv",
    reportNumber = "CERN-PH-TH-2007-036, FTU-AM-07-03, IFT-UAM-CSIC-07-08, SHEP-07-05",
    doi = "10.1088/1126-6708/2007/04/060",
    journal = "JHEP",
    volume = "04",
    pages = "060",
    year = "2007"
}

@article{Kajiyama:2007gx,
    author = "Kajiyama, Yuji and Raidal, Martti and Strumia, Alessandro",
    title = "{The Golden ratio prediction for the solar neutrino mixing}",
    eprint = "0705.4559",
    archivePrefix = "arXiv",
    primaryClass = "hep-ph",
    reportNumber = "IFUP-TH-11-2007",
    doi = "10.1103/PhysRevD.76.117301",
    journal = "Phys. Rev. D",
    volume = "76",
    pages = "117301",
    year = "2007"
}

@article{Altarelli:2010gt,
    author = "Altarelli, Guido and Feruglio, Ferruccio",
    title = "{Discrete Flavor Symmetries and Models of Neutrino Mixing}",
    eprint = "1002.0211",
    archivePrefix = "arXiv",
    primaryClass = "hep-ph",
    reportNumber = "RM3-TH-10-01, CERN-PH-TH-2010-016, DFPD-10-TH-02",
    doi = "10.1103/RevModPhys.82.2701",
    journal = "Rev. Mod. Phys.",
    volume = "82",
    pages = "2701--2729",
    year = "2010"
}

@article{Fukugita:1998vn,
    author = "Fukugita, M. and Tanimoto, M. and Yanagida, T.",
    title = "{Atmospheric neutrino oscillation and a phenomenological lepton mass matrix}",
    eprint = "hep-ph/9709388",
    archivePrefix = "arXiv",
    reportNumber = "UT-788",
    doi = "10.1103/PhysRevD.57.4429",
    journal = "Phys. Rev. D",
    volume = "57",
    pages = "4429--4432",
    year = "1998"
}

@article{Ge:2011ih,
    author = "Ge, Shao-Feng and Dicus, Duane A. and Repko, Wayne W.",
    title = "{$\mathbb{Z}_2$ Symmetry Prediction for the Leptonic Dirac CP Phase}",
    eprint = "1104.0602",
    archivePrefix = "arXiv",
    primaryClass = "hep-ph",
    doi = "10.1016/j.physletb.2011.06.096",
    journal = "Phys. Lett. B",
    volume = "702",
    pages = "220--223",
    year = "2011"
}

@article{Ding:2011cm,
    author = "Ding, Gui-Jun and Everett, Lisa L. and Stuart, Alexander J.",
    title = "{Golden Ratio Neutrino Mixing and $A_5$ Flavor Symmetry}",
    eprint = "1110.1688",
    archivePrefix = "arXiv",
    primaryClass = "hep-ph",
    reportNumber = "MADPH-11-1576",
    doi = "10.1016/j.nuclphysb.2011.12.004",
    journal = "Nucl. Phys. B",
    volume = "857",
    pages = "219--253",
    year = "2012"
}

@article{Kobayashi:1973fv,
    author = "Kobayashi, Makoto and Maskawa, Toshihide",
    title = "{CP Violation in the Renormalizable Theory of Weak Interaction}",
    reportNumber = "KUNS-242",
    doi = "10.1143/PTP.49.652",
    journal = "Prog. Theor. Phys.",
    volume = "49",
    pages = "652--657",
    year = "1973"
}

@article{Chakraborty:2018dew,
    author = "Chakraborty, Kaustav and Deepthi, K. N. and Goswami, Srubabati and Joshipura, Anjan S. and Nath, Newton",
    title = "{Exploring partial $\mu$-$\tau$ reflection symmetry at DUNE and Hyper-Kamiokande}",
    eprint = "1804.02022",
    archivePrefix = "arXiv",
    primaryClass = "hep-ph",
    doi = "10.1103/PhysRevD.98.075031",
    journal = "Phys. Rev. D",
    volume = "98",
    number = "7",
    pages = "075031",
    year = "2018"
}

@article{King:2013eh,
    author = "King, Stephen F. and Luhn, Christoph",
    title = "{Neutrino Mass and Mixing with Discrete Symmetry}",
    eprint = "1301.1340",
    archivePrefix = "arXiv",
    primaryClass = "hep-ph",
    reportNumber = "IPPP-12-100, DCPT-12-200",
    doi = "10.1088/0034-4885/76/5/056201",
    journal = "Rept. Prog. Phys.",
    volume = "76",
    pages = "056201",
    year = "2013"
}

@article{Albright:2010ap,
    author = "Albright, Carl H. and Dueck, Alexander and Rodejohann, Werner",
    title = "{Possible Alternatives to Tri-bimaximal Mixing}",
    eprint = "1004.2798",
    archivePrefix = "arXiv",
    primaryClass = "hep-ph",
    reportNumber = "FERMILAB-PUB-10-079-T",
    doi = "10.1140/epjc/s10052-010-1492-2",
    journal = "Eur. Phys. J. C",
    volume = "70",
    pages = "1099--1110",
    year = "2010"
}

@article{NOvA:2019cyt,
    author = "Acero, M. A. and others",
    collaboration = "NOvA",
    title = "{First Measurement of Neutrino Oscillation Parameters using Neutrinos and Antineutrinos by NOvA}",
    eprint = "1906.04907",
    archivePrefix = "arXiv",
    primaryClass = "hep-ex",
    reportNumber = "FERMILAB-PUB-19-272-ND",
    doi = "10.1103/PhysRevLett.123.151803",
    journal = "Phys. Rev. Lett.",
    volume = "123",
    number = "15",
    pages = "151803",
    year = "2019"
}

@article{deMedeirosVarzielas:2012apl,
    author = "de Medeiros Varzielas, Ivo and Lavoura, Lu\'\i{}s",
    title = "{Flavour models for $TM_{1}$ lepton mixing}",
    eprint = "1212.3247",
    archivePrefix = "arXiv",
    primaryClass = "hep-ph",
    reportNumber = "DO-TH-12-40, CFTP-12-018",
    doi = "10.1088/0954-3899/40/8/085002",
    journal = "J. Phys. G",
    volume = "40",
    pages = "085002",
    year = "2013"
}

@article{Borah:2017dmk,
    author = "Borah, Debasish and Karmakar, Biswajit",
    title = "{$A_4$ flavour model for Dirac neutrinos: Type I and inverse seesaw}",
    eprint = "1712.06407",
    archivePrefix = "arXiv",
    primaryClass = "hep-ph",
    doi = "10.1016/j.physletb.2018.03.047",
    journal = "Phys. Lett. B",
    volume = "780",
    pages = "461--470",
    year = "2018"
}

@article{Karmakar:2015jza,
    author = "Karmakar, Biswajit and Sil, Arunansu",
    title = "{Spontaneous CP violation in lepton-sector: A common origin for $\theta_{13}$, the Dirac CP phase, and leptogenesis}",
    eprint = "1509.07090",
    archivePrefix = "arXiv",
    primaryClass = "hep-ph",
    doi = "10.1103/PhysRevD.93.013006",
    journal = "Phys. Rev. D",
    volume = "93",
    number = "1",
    pages = "013006",
    year = "2016"
}

@article{Pontecorvo:1957qd,
    author = "Pontecorvo, B.",
    title = "{Inverse beta processes and nonconservation of lepton charge}",
    journal = "Zh. Eksp. Teor. Fiz.",
    volume = "34",
    pages = "247",
    year = "1957"
}

@article{Rodejohann:2008ir,
    author = "Rodejohann, Werner",
    title = "{Unified Parametrization for Quark and Lepton Mixing Angles}",
    eprint = "0810.5239",
    archivePrefix = "arXiv",
    primaryClass = "hep-ph",
    doi = "10.1016/j.physletb.2008.12.010",
    journal = "Phys. Lett. B",
    volume = "671",
    pages = "267--271",
    year = "2009"
}

@article{nEXO:2021ujk,
    author = "Adhikari, G. and others",
    collaboration = "nEXO",
    title = "{nEXO: neutrinoless double beta decay search beyond 10$^{28}$ year half-life sensitivity}",
    eprint = "2106.16243",
    archivePrefix = "arXiv",
    primaryClass = "nucl-ex",
    doi = "10.1088/1361-6471/ac3631",
    journal = "J. Phys. G",
    volume = "49",
    number = "1",
    pages = "015104",
    year = "2022"
}

@article{Ishimori:2010au,
    author = "Ishimori, Hajime and Kobayashi, Tatsuo and Ohki, Hiroshi and Shimizu, Yusuke and Okada, Hiroshi and Tanimoto, Morimitsu",
    title = "{Non-Abelian Discrete Symmetries in Particle Physics}",
    eprint = "1003.3552",
    archivePrefix = "arXiv",
    primaryClass = "hep-th",
    reportNumber = "KUNS-2260",
    doi = "10.1143/PTPS.183.1",
    journal = "Prog. Theor. Phys. Suppl.",
    volume = "183",
    pages = "1--163",
    year = "2010"
}

@article{Gehrlein:2016wlc,
    author = "Gehrlein, Julia and Merle, Alexander and Spinrath, Martin",
    title = "{Predictivity of Neutrino Mass Sum Rules}",
    eprint = "1606.04965",
    archivePrefix = "arXiv",
    primaryClass = "hep-ph",
    reportNumber = "MPP-2016-127, TTP16-022",
    doi = "10.1103/PhysRevD.94.093003",
    journal = "Phys. Rev. D",
    volume = "94",
    number = "9",
    pages = "093003",
    year = "2016"
}

@article{Karmakar:2014dva,
    author = "Karmakar, Biswajit and Sil, Arunansu",
    title = "{Nonzero $\theta_{13}$ and leptogenesis in a type-I seesaw model with $A_4$ symmetry}",
    eprint = "1407.5826",
    archivePrefix = "arXiv",
    primaryClass = "hep-ph",
    doi = "10.1103/PhysRevD.91.013004",
    journal = "Phys. Rev. D",
    volume = "91",
    pages = "013004",
    year = "2015"
}

@article{Gehrlein:2017ryu,
    author = "Gehrlein, Julia and Spinrath, Martin",
    title = "{Neutrino Mass Sum Rules and Symmetries of the Mass Matrix}",
    eprint = "1704.02371",
    archivePrefix = "arXiv",
    primaryClass = "hep-ph",
    reportNumber = "IFT-UAM-CSIC-17-021, FTUAM-17-5, NCTS-PH1701, TTP16-030",
    doi = "10.1140/epjc/s10052-017-4817-6",
    journal = "Eur. Phys. J. C",
    volume = "77",
    number = "5",
    pages = "281",
    year = "2017"
}

@article{Adulpravitchai:2009bg,
    author = "Adulpravitchai, Adisorn and Blum, Alexander and Rodejohann, Werner",
    title = "{Golden Ratio Prediction for Solar Neutrino Mixing}",
    eprint = "0903.0531",
    archivePrefix = "arXiv",
    primaryClass = "hep-ph",
    doi = "10.1088/1367-2630/11/6/063026",
    journal = "New J. Phys.",
    volume = "11",
    pages = "063026",
    year = "2009"
}

@article{Harrison:2002kp,
    author = "Harrison, P. F. and Scott, W. G.",
    title = "{Symmetries and generalizations of tri - bimaximal neutrino mixing}",
    eprint = "hep-ph/0203209",
    archivePrefix = "arXiv",
    reportNumber = "RAL-TR-2002-008",
    doi = "10.1016/S0370-2693(02)01753-7",
    journal = "Phys. Lett. B",
    volume = "535",
    pages = "163--169",
    year = "2002"
}

@article{Petcov:2018snn,
    author = "Petcov, S. T. and Titov, A. V.",
    title = "{Assessing the Viability of $A_4$, $S_4$ and $A_5$ Flavour Symmetries for Description of Neutrino Mixing}",
    eprint = "1804.00182",
    archivePrefix = "arXiv",
    primaryClass = "hep-ph",
    reportNumber = "SISSA 15/2018/FISI, IPMU18-0058, IPPP/18/22, SISSA-15-2018-FISI, IPPP-18-22",
    doi = "10.1103/PhysRevD.97.115045",
    journal = "Phys. Rev. D",
    volume = "97",
    number = "11",
    pages = "115045",
    year = "2018"
}

@article{Maki:1962mu,
    author = "Maki, Ziro and Nakagawa, Masami and Sakata, Shoichi",
    title = "{Remarks on the unified model of elementary particles}",
    doi = "10.1143/PTP.28.870",
    journal = "Prog. Theor. Phys.",
    volume = "28",
    pages = "870--880",
    year = "1962"
}

@article{Harrison:2002et,
    author = "Harrison, P. F. and Scott, W. G.",
    title = "{mu - tau reflection symmetry in lepton mixing and neutrino oscillations}",
    eprint = "hep-ph/0210197",
    archivePrefix = "arXiv",
    reportNumber = "RAL-TR-2002-023",
    doi = "10.1016/S0370-2693(02)02772-7",
    journal = "Phys. Lett. B",
    volume = "547",
    pages = "219--228",
    year = "2002"
}

@article{Datta:2021zzf,
    author = "Datta, Arghyajit and Karmakar, Biswajit and Sil, Arunansu",
    title = "{Flavored leptogenesis and neutrino mass with A$_{4}$ symmetry}",
    eprint = "2106.06773",
    archivePrefix = "arXiv",
    primaryClass = "hep-ph",
    doi = "10.1007/JHEP12(2021)051",
    journal = "JHEP",
    volume = "12",
    pages = "051",
    year = "2021"
}

@article{Barger:1998ta,
    author = "Barger, Vernon D. and Pakvasa, S. and Weiler, Thomas J. and Whisnant, K.",
    title = "{Bimaximal mixing of three neutrinos}",
    eprint = "hep-ph/9806387",
    archivePrefix = "arXiv",
    reportNumber = "MADPH-98-1063, HAWAII-511-905-98, VAND-TH-98-08, AMES-HET-98-09",
    doi = "10.1016/S0370-2693(98)00880-6",
    journal = "Phys. Lett. B",
    volume = "437",
    pages = "107--116",
    year = "1998"
}

@article{Harrison:2002er,
    author = "Harrison, P. F. and Perkins, D. H. and Scott, W. G.",
    title = "{Tri-bimaximal mixing and the neutrino oscillation data}",
    eprint = "hep-ph/0202074",
    archivePrefix = "arXiv",
    reportNumber = "RAL-TR-2002-002",
    doi = "10.1016/S0370-2693(02)01336-9",
    journal = "Phys. Lett. B",
    volume = "530",
    pages = "167",
    year = "2002"
}

@article{Everett:2008et,
    author = "Everett, Lisa L. and Stuart, Alexander J.",
    title = "{Icosahedral (A(5)) Family Symmetry and the Golden Ratio Prediction for Solar Neutrino Mixing}",
    eprint = "0812.1057",
    archivePrefix = "arXiv",
    primaryClass = "hep-ph",
    reportNumber = "MADPH-08-1525",
    doi = "10.1103/PhysRevD.79.085005",
    journal = "Phys. Rev. D",
    volume = "79",
    pages = "085005",
    year = "2009"
}

@article{Baltz:1998ey,
    author = "Baltz, Anthony J. and Goldhaber, Alfred Scharff and Goldhaber, Maurice",
    title = "{The Solar neutrino puzzle: An Oscillation solution with maximal neutrino mixing}",
    eprint = "hep-ph/9806540",
    archivePrefix = "arXiv",
    doi = "10.1103/PhysRevLett.81.5730",
    journal = "Phys. Rev. Lett.",
    volume = "81",
    pages = "5730--5733",
    year = "1998"
}

@article{Ahn:2012nd,
    author = "Ahn, J. K. and others",
    collaboration = "RENO",
    title = "{Observation of Reactor Electron Antineutrino Disappearance in the RENO Experiment}",
    eprint = "1204.0626",
    archivePrefix = "arXiv",
    primaryClass = "hep-ex",
    doi = "10.1103/PhysRevLett.108.191802",
    journal = "Phys. Rev. Lett.",
    volume = "108",
    pages = "191802",
    year = "2012"
}

@article{LEGEND:2021bnm,
    author = "Abgrall, N. and others",
    collaboration = "LEGEND",
    title = "{The Large Enriched Germanium Experiment for Neutrinoless $\beta\beta$ Decay}: {LEGEND-1000 Preconceptual Design Report}",
    eprint = "2107.11462",
    archivePrefix = "arXiv",
    primaryClass = "physics.ins-det",
    month = "7",
    year = "2021"
}

@article{Vissani:1997pa,
    author = "Vissani, Francesco",
    title = "{A Study of the scenario with nearly degenerate Majorana neutrinos}",
    eprint = "hep-ph/9708483",
    archivePrefix = "arXiv",
    month = "8",
    year = "1997"
}

@article{An:2012eh,
    author = "An, F. P. and others",
    collaboration = "Daya Bay",
    title = "{Observation of electron-antineutrino disappearance at Daya Bay}",
    eprint = "1203.1669",
    archivePrefix = "arXiv",
    primaryClass = "hep-ex",
    doi = "10.1103/PhysRevLett.108.171803",
    journal = "Phys. Rev. Lett.",
    volume = "108",
    pages = "171803",
    year = "2012"
}

@article{Datta:2003qg,
    author = "Datta, Aseshkrishna and Ling, Fu-Sin and Ramond, Pierre",
    title = "{Correlated hierarchy, Dirac masses and large mixing angles}",
    eprint = "hep-ph/0306002",
    archivePrefix = "arXiv",
    reportNumber = "UFIFT-HEP-03-14",
    doi = "10.1016/j.nuclphysb.2003.08.026",
    journal = "Nucl. Phys. B",
    volume = "671",
    pages = "383--400",
    year = "2003"
}

@article{Petcov:2017ggy,
    author = "Petcov, S. T.",
    title = "{Discrete Flavour Symmetries, Neutrino Mixing and Leptonic CP Violation}",
    eprint = "1711.10806",
    archivePrefix = "arXiv",
    primaryClass = "hep-ph",
    reportNumber = "SISSA-59-2017FISI, SISSA 59/2017FISI, IPMU17-0167",
    doi = "10.1140/epjc/s10052-018-6158-5",
    journal = "Eur. Phys. J. C",
    volume = "78",
    number = "9",
    pages = "709",
    year = "2018"
}

@article{KATRIN:2001ttj,
    author = "Osipowicz, A. and others",
    collaboration = "KATRIN",
    title = "{KATRIN: A Next generation tritium beta decay experiment with sub-eV sensitivity for the electron neutrino mass. Letter of intent}",
    eprint = "hep-ex/0109033",
    archivePrefix = "arXiv",
    month = "9",
    year = "2001"
}

@article{Petcov:2014laa,
    author = "Petcov, S. T.",
    title = "{Predicting the values of the leptonic CP violation phases in theories with discrete flavour symmetries}",
    eprint = "1405.6006",
    archivePrefix = "arXiv",
    primaryClass = "hep-ph",
    reportNumber = "SISSA-20-2014-FISI, IPMU14-0104",
    doi = "10.1016/j.nuclphysb.2015.01.011",
    journal = "Nucl. Phys. B",
    volume = "892",
    pages = "400--428",
    year = "2015"
}

@article{T2K:2019bcf,
    author = "Abe, K. and others",
    collaboration = "T2K",
    title = "{Constraint on the matter\textendash{}antimatter symmetry-violating phase in neutrino oscillations}",
    eprint = "1910.03887",
    archivePrefix = "arXiv",
    primaryClass = "hep-ex",
    doi = "10.1038/s41586-020-2177-0",
    journal = "Nature",
    volume = "580",
    number = "7803",
    pages = "339--344",
    year = "2020",
    note = "[Erratum: Nature 583, E16 (2020)]"
}

@article{Ganguly:2022qxj,
    author = "Ganguly, Joy and Gluza, Janusz and Karmakar, Biswajit",
    title = "{Common origin of \ensuremath{\theta}$_{13}$ and dark matter within the flavor symmetric scoto-seesaw framework}",
    eprint = "2209.08610",
    archivePrefix = "arXiv",
    primaryClass = "hep-ph",
    doi = "10.1007/JHEP11(2022)074",
    journal = "JHEP",
    volume = "11",
    pages = "074",
    year = "2022"
}

@article{Capozzi:2021fjo,
    author = "Capozzi, Francesco and Di Valentino, Eleonora and Lisi, Eligio and Marrone, Antonio and Melchiorri, Alessandro and Palazzo, Antonio",
    title = "{Unfinished fabric of the three neutrino paradigm}",
    eprint = "2107.00532",
    archivePrefix = "arXiv",
    primaryClass = "hep-ph",
    doi = "10.1103/PhysRevD.104.083031",
    journal = "Phys. Rev. D",
    volume = "104",
    number = "8",
    pages = "083031",
    year = "2021"
}

@article{Barry:2010yk,
    author = "Barry, James and Rodejohann, Werner",
    title = "{Neutrino Mass Sum-rules in Flavor Symmetry Models}",
    eprint = "1007.5217",
    archivePrefix = "arXiv",
    primaryClass = "hep-ph",
    doi = "10.1016/j.nuclphysb.2010.08.015",
    journal = "Nucl. Phys. B",
    volume = "842",
    pages = "33--50",
    year = "2011"
}

@article{King:2013psa,
    author = "King, Stephen F. and Merle, Alexander and Stuart, Alexander J.",
    title = "{The Power of Neutrino Mass Sum Rules for Neutrinoless Double Beta Decay Experiments}",
    eprint = "1307.2901",
    archivePrefix = "arXiv",
    primaryClass = "hep-ph",
    doi = "10.1007/JHEP12(2013)005",
    journal = "JHEP",
    volume = "12",
    pages = "005",
    year = "2013"
}

@article{DUNE:2020jqi,
    author = "Abi, B. and others",
    collaboration = "DUNE",
    title = "{Long-baseline neutrino oscillation physics potential of the DUNE experiment}",
    eprint = "2006.16043",
    archivePrefix = "arXiv",
    primaryClass = "hep-ex",
    reportNumber = "FERMILAB-PUB-20-251-E-LBNF-ND-PIP2-SCD, PUB-20-251-E-LBNF-ND-PIP2-SCD",
    doi = "10.1140/epjc/s10052-020-08456-z",
    journal = "Eur. Phys. J. C",
    volume = "80",
    number = "10",
    pages = "978",
    year = "2020"
}

@article{Hyper-Kamiokande:2018ofw,
    author = "Abe, K. and others",
    collaboration = "Hyper-Kamiokande",
    title = "{Hyper-Kamiokande Design Report}",
    eprint = "1805.04163",
    archivePrefix = "arXiv",
    primaryClass = "physics.ins-det",
    month = "5",
    year = "2018"
}

@article{KATRIN:2021dfa,
    author = "Aker, M. and others",
    collaboration = "KATRIN",
    title = "{The design, construction, and commissioning of the KATRIN experiment}",
    eprint = "2103.04755",
    archivePrefix = "arXiv",
    primaryClass = "physics.ins-det",
    doi = "10.1088/1748-0221/16/08/T08015",
    journal = "JINST",
    volume = "16",
    number = "08",
    pages = "T08015",
    year = "2021"
}

@article{Ballett:2014dua,
    author = "Ballett, Peter and King, Stephen F. and Luhn, Christoph and Pascoli, Silvia and Schmidt, Michael A.",
    title = "{Testing solar lepton mixing sum rules in neutrino oscillation experiments}",
    eprint = "1410.7573",
    archivePrefix = "arXiv",
    primaryClass = "hep-ph",
    reportNumber = "IPPP-14-90, DCPT-14-180, SI-HEP-2014-27, QFET-2014-20",
    doi = "10.1007/JHEP12(2014)122",
    journal = "JHEP",
    volume = "12",
    pages = "122",
    year = "2014"
}

@article{King:2007pr,
    author = "King, S. F.",
    title = "{Parametrizing the lepton mixing matrix in terms of deviations from tri-bimaximal mixing}",
    eprint = "0710.0530",
    archivePrefix = "arXiv",
    primaryClass = "hep-ph",
    doi = "10.1016/j.physletb.2007.10.078",
    journal = "Phys. Lett. B",
    volume = "659",
    pages = "244--251",
    year = "2008"
}

@article{KamLAND:2002uet,
    author = "Eguchi, K. and others",
    collaboration = "KamLAND",
    title = "{First results from KamLAND: Evidence for reactor anti-neutrino disappearance}",
    eprint = "hep-ex/0212021",
    archivePrefix = "arXiv",
    doi = "10.1103/PhysRevLett.90.021802",
    journal = "Phys. Rev. Lett.",
    volume = "90",
    pages = "021802",
    year = "2003"
}

@article{Chen:2026fod,
    author = "Chen, Mu-Chun and Matias, Harold J. and Moffett-Smith, Cameron",
    title = "{Gravitational Waves in an A4 Neutrino Mass Model}",
    eprint = "2601.14394",
    archivePrefix = "arXiv",
    primaryClass = "hep-ph",
    month = "1",
    year = "2026"
}

@article{Borah:2024gql,
    author = "Borah, Debasish and Das, Pritam and Karmakar, Biswajit and Mahapatra, Satyabrata",
    title = "{Discrete dark matter with light Dirac neutrinos}",
    eprint = "2406.17861",
    archivePrefix = "arXiv",
    primaryClass = "hep-ph",
    doi = "10.1103/PhysRevD.111.035032",
    journal = "Phys. Rev. D",
    volume = "111",
    number = "3",
    pages = "035032",
    year = "2025"
}

@article{Esteban:2024eli,
    author = "Esteban, Ivan and Gonzalez-Garcia, M. C. and Maltoni, Michele and Martinez-Soler, Ivan and Pinheiro, Jo\~ao Paulo and Schwetz, Thomas",
    title = "{NuFit-6.0: Updated global analysis of three-flavor neutrino oscillations}",
    eprint = "2410.05380",
    archivePrefix = "arXiv",
    primaryClass = "hep-ph",
    reportNumber = "IPPP/24/64, IFT-UAM/CSIC-24-140, YITP-SB-2024-24",
    month = "10",
    year = "2024"
}

@misc{NuFIT6.0data,
    title = "{NuFIT} 6.0 Data Files",
    year = 2024,
    howpublished = {\url{http://www.nu-fit.org/?q=node/294#label85}}
}

@misc{NuFIT6.1data,
    title = "{NuFIT} 6.1 Data Files",
    year = 2025,
    howpublished = {\url{http://www.nu-fit.org/?q=node/309#label85}}
}

@misc{Mathematica14,
  author = {Wolfram Research{,} Inc.},
  title = {Mathematica, {V}ersion 14.2},
  url = {https://www.wolfram.com/mathematica},
  note = {Champaign, IL, 2024}
}

@article{Super-Kamiokande:1998kpq,
    author = "Fukuda, Y. and others",
    collaboration = "Super-Kamiokande",
    title = "{Evidence for oscillation of atmospheric neutrinos}",
    eprint = "hep-ex/9807003",
    archivePrefix = "arXiv",
    reportNumber = "BU-98-17, ICRR-REPORT-422-98-18, UCI-98-8, KEK-PREPRINT-98-95, LSU-HEPA-5-98, UMD-98-003, SBHEP-98-5, TKU-PAP-98-06, TIT-HPE-98-09",
    doi = "10.1103/PhysRevLett.81.1562",
    journal = "Phys. Rev. Lett.",
    volume = "81",
    pages = "1562--1567",
    year = "1998"
}

@article{DoubleChooz:2011ymz,
    author = "Abe, Y. and others",
    collaboration = "Double Chooz",
    title = "{Indication of Reactor $\bar{\nu}_e$ Disappearance in the Double Chooz Experiment}",
    eprint = "1112.6353",
    archivePrefix = "arXiv",
    primaryClass = "hep-ex",
    doi = "10.1103/PhysRevLett.108.131801",
    journal = "Phys. Rev. Lett.",
    volume = "108",
    pages = "131801",
    year = "2012"
}

@article{Ding:2024ozt,
    author = "Ding, Gui-Jun and Valle, Jose W. F.",
    title = "{The symmetry approach to quark and lepton masses and mixing}",
    eprint = "2402.16963",
    archivePrefix = "arXiv",
    primaryClass = "hep-ph",
    doi = "10.1016/j.physrep.2024.12.005",
    journal = "Phys. Rept.",
    volume = "1109",
    pages = "1--105",
    year = "2025"
}

@article{Albright:2008rp,
    author = "Albright, Carl H. and Rodejohann, Werner",
    title = "{Comparing Trimaximal Mixing and Its Variants with Deviations from Tri-bimaximal Mixing}",
    eprint = "0812.0436",
    archivePrefix = "arXiv",
    primaryClass = "hep-ph",
    reportNumber = "FERMILAB-PUB-08-535-T",
    doi = "10.1140/epjc/s10052-009-1074-3",
    journal = "Eur. Phys. J. C",
    volume = "62",
    pages = "599--608",
    year = "2009"
}

@article{He:2011gb,
    author = "He, Xiao-Gang and Zee, A.",
    title = "{Minimal Modification to Tri-bimaximal Mixing}",
    eprint = "1106.4359",
    archivePrefix = "arXiv",
    primaryClass = "hep-ph",
    doi = "10.1103/PhysRevD.84.053004",
    journal = "Phys. Rev. D",
    volume = "84",
    pages = "053004",
    year = "2011"
}

@article{Ding:2023htn,
    author = "Ding, Gui-Jun and King, Stephen F.",
    title = "{Neutrino mass and mixing with modular symmetry}",
    eprint = "2311.09282",
    archivePrefix = "arXiv",
    primaryClass = "hep-ph",
    doi = "10.1088/1361-6633/ad52a3",
    journal = "Rept. Prog. Phys.",
    volume = "87",
    number = "8",
    pages = "084201",
    year = "2024"
}

@article{SNO:1999crp,
    author = "Boger, J. and others",
    collaboration = "SNO",
    title = "{The Sudbury neutrino observatory}",
    eprint = "nucl-ex/9910016",
    archivePrefix = "arXiv",
    reportNumber = "SNO-STR-99-025",
    doi = "10.1016/S0168-9002(99)01469-2",
    journal = "Nucl. Instrum. Meth. A",
    volume = "449",
    pages = "172--207",
    year = "2000"
}

@article{DayaBay:2012fng,
    author = "An, F. P. and others",
    collaboration = "Daya Bay",
    title = "{Observation of electron-antineutrino disappearance at Daya Bay}",
    eprint = "1203.1669",
    archivePrefix = "arXiv",
    primaryClass = "hep-ex",
    doi = "10.1103/PhysRevLett.108.171803",
    journal = "Phys. Rev. Lett.",
    volume = "108",
    pages = "171803",
    year = "2012"
}

@article{RENO:2012mkc,
    author = "Ahn, J. K. and others",
    collaboration = "RENO",
    title = "{Observation of Reactor Electron Antineutrino Disappearance in the RENO Experiment}",
    eprint = "1204.0626",
    archivePrefix = "arXiv",
    primaryClass = "hep-ex",
    doi = "10.1103/PhysRevLett.108.191802",
    journal = "Phys. Rev. Lett.",
    volume = "108",
    pages = "191802",
    year = "2012"
}

@article{Zhang:2025jnn,
    author = "Zhang, Di",
    title = "{Trimaximal mixing patterns meet the first JUNO result}",
    eprint = "2511.15654",
    archivePrefix = "arXiv",
    primaryClass = "hep-ph",
    doi = "10.1103/t3w7-y9r3",
    journal = "Phys. Rev. D",
    volume = "113",
    number = "5",
    pages = "055035",
    year = "2026"
}

@article{CUPID:2019imh,
    author = "Armstrong, W. R. and others",
    collaboration = "CUPID",
    title = "{CUPID pre-CDR}",
    eprint = "1907.09376",
    archivePrefix = "arXiv",
    primaryClass = "physics.ins-det",
    month = "7",
    year = "2019"
}

@article{nEXO:2017nam,
    author = "Albert, J. B. and others",
    collaboration = "nEXO",
    title = "{Sensitivity and Discovery Potential of nEXO to Neutrinoless Double Beta Decay}",
    eprint = "1710.05075",
    archivePrefix = "arXiv",
    primaryClass = "nucl-ex",
    reportNumber = "LLNL-JRNL-737682",
    doi = "10.1103/PhysRevC.97.065503",
    journal = "Phys. Rev. C",
    volume = "97",
    number = "6",
    pages = "065503",
    year = "2018"
}

@article{LEGEND:2017cdu,
    author = "Abgrall, N. and others",
    editor = "Civitarese, Osvaldo and Stekl, Ivan and Suhonen, Jouni",
    collaboration = "LEGEND",
    title = "{The Large Enriched Germanium Experiment for Neutrinoless Double Beta Decay (LEGEND)}",
    eprint = "1709.01980",
    archivePrefix = "arXiv",
    primaryClass = "physics.ins-det",
    doi = "10.1063/1.5007652",
    journal = "AIP Conf. Proc.",
    volume = "1894",
    number = "1",
    pages = "020027",
    year = "2017"
}

@inproceedings{Project8:2013trt,
    author = "Doe, P. J. and others",
    collaboration = "Project 8",
    title = "{Project 8: Determining neutrino mass from tritium beta decay using a frequency-based method}",
    booktitle = "{Snowmass 2013}: {Snowmass on the Mississippi}",
    eprint = "1309.7093",
    archivePrefix = "arXiv",
    primaryClass = "nucl-ex",
    month = "9",
    year = "2013"
}

@misc{GitHubCorrelations,
  author = {Gluza, Janusz and Karmakar, Biswajit and Zieba, Szymon},
  title = "{Correlation package and NuFIT data sets, Github repository}", 
  howpublished = {\url{https://github.com/szieba/correlations}},
}

@article{Chauhan:2023faf,
    author = "Chauhan, Garv and Dev, P. S. Bhupal and Dubovyk, Ievgen and Dziewit, Bartosz and Flieger, Wojciech and Grzanka, Krzysztof and Gluza, Janusz and Karmakar, Biswajit and Zi{\k{e}}ba, Szymon",
    title = "{Phenomenology of lepton masses and mixing with discrete flavor symmetries}",
    eprint = "2310.20681",
    archivePrefix = "arXiv",
    primaryClass = "hep-ph",
    doi = "10.1016/j.ppnp.2024.104126",
    journal = "Prog. Part. Nucl. Phys.",
    volume = "138",
    pages = "104126",
    year = "2024"
}

@article{Song:2020nfh,
    author = {Song, Ningqiang and Li, Shirley Weishi and Arg\"uelles, Carlos A. and Bustamante, Mauricio and Vincent, Aaron C.},
    title = "{The Future of High-Energy Astrophysical Neutrino Flavor Measurements}",
    eprint = "2012.12893",
    archivePrefix = "arXiv",
    primaryClass = "hep-ph",
    reportNumber = "FERMILAB-PUB-21-006-T",
    doi = "10.1088/1475-7516/2021/04/054",
    journal = "JCAP",
    volume = "04",
    pages = "054",
    year = "2021"
}

@article{deSalas:2020pgw,
    author = "de Salas, P. F. and Forero, D. V. and Gariazzo, S. and Mart\'\i{}nez-Mirav\'e, P. and Mena, O. and Ternes, C. A. and T\'ortola, M. and Valle, J. W. F.",
    title = "{2020 global reassessment of the neutrino oscillation picture}",
    eprint = "2006.11237",
    archivePrefix = "arXiv",
    primaryClass = "hep-ph",
    doi = "10.1007/JHEP02(2021)071",
    journal = "JHEP",
    volume = "02",
    pages = "071",
    year = "2021"
}

@Misc{10.5281/zenodo.4726908,
    author = "{P. F. De Salas} and {D. V. Forero} and {S. Gariazzo} and {P. Mart{\'\i}nez-Mirav{\'e}} and {O. Mena} and {C. A. Ternes} and {M. T{\'o}rtola} and {J. W. F. Valle}",
    title = "{Chi2 profiles from Valencia neutrino global fit}",
    year = "2021",
    doi = "10.5281/zenodo.4726908",
    howpublished = "\url{http://globalfit.astroparticles.es/}",
}
\end{document}